\documentclass[10pt,journal,compsoc]{IEEEtran}

\usepackage[numbers]{natbib}

\ifCLASSOPTIONcompsoc
  \usepackage[caption=false,font=footnotesize,labelfont=sf,textfont=sf]{subfig}
\else
  \usepackage[caption=false,font=footnotesize]{subfig}
  \fi
\usepackage{amsmath,amsfonts}
\usepackage{stmaryrd,upgreek}
\usepackage{url}
\usepackage{xcolor}
\usepackage{xspace}
\usepackage[english]{babel}
\usepackage[utf8]{inputenc}
\usepackage[T1]{fontenc}
\usepackage{graphicx}
\usepackage{tikz}
\usepackage{tkz-euclide}
\usepackage{pgfplots}
\usepgfplotslibrary{fillbetween}	
\usetikzlibrary{positioning}
\usetikzlibrary{intersections}

\usepackage{booktabs}
\usepackage[binary-units]{siunitx}
\usepackage{listings}
\usepackage{algorithm}
\usepackage[noend]{algpseudocode}
\usepackage{float}
\usepackage{hyperref}

\makeatletter

\newcount\ALG@printindent@tempcnta
\def\ALG@printindent{\ifnum \theALG@nested>0\ifx\ALG@text\ALG@x@notext \else
            \unskip
            \addvspace{-1pt}\ALG@printindent@tempcnta=1
            \loop
                \algrule[\csname ALG@ind@\the\ALG@printindent@tempcnta\endcsname]\advance \ALG@printindent@tempcnta 1
            \ifnum \ALG@printindent@tempcnta<\numexpr\theALG@nested+1\relax \repeat
        \fi
    \fi
    }\usepackage{etoolbox}
\patchcmd{\ALG@doentity}{\noindent\hskip\ALG@tlm}{\ALG@printindent}{}{\errmessage{failed to patch}}
\makeatother

\usepackage{paralist}

\newcommand{\dslname}{SB-TemPsy-DSL\xspace}
\newcommand{\checkname}{SB-TemPsy-Check\xspace}

\newcommand\real{\mathbb{R}}

\newcommand\trace{\ensuremath{\lambda}\xspace}

\newcommand{\condition}{\nonterminal{c}\xspace}
\newcommand{\property}{\ensuremath{\phi}\xspace}
\newcommand{\patternsemantics}{\ensuremath{\upzeta}}

\newcommand{\diagnosticpattern}{\ensuremath{\mathit{vc}}\xspace}

\newcommand{\afclogs}{10}
\newcommand{\afcMinRec}{22824}
\newcommand{\afcMaxRec}{23988}
\newcommand{\afcAverage}{23650}
\newcommand{\afcStd}{328}

\newcommand{\construct}{\ensuremath{\eta}\xspace}

\usepackage{fp}

\definecolor{keywordcolor}{RGB}{127,0,85}
\newcommand{\lit}[1]{\textbf{\texttt{\textcolor{keywordcolor}{#1}}}}
\newcommand\synt[1]{\textsf{#1}}
\newcommand\nonterminal[1]{\synt{#1}}
\usepackage[inline]{enumitem}

\newcommand\NAME{\emph{TD-SB-TemPsy}\xspace}

\newcommand\numsimulationtraces{361\xspace}

\newcommand{\minsimulationminutes}{25}
\newcommand{\maxsimulationhours}{23}
\newcommand{\maxsimulationminutes}{29}
\newcommand{\avgsimulationhours}{6}
\newcommand{\avgsimulationminutes}{38}
\newcommand{\stdDevsimulationhours}{7}
\newcommand{\stdDevsimulationminutes}{05}
\newcommand{\minnumberofentries}{25358}
\newcommand{\maxnumberofentries}{9328178}
\newcommand{\averageumberofentries}{438224}
\newcommand{\stdevnumberofentries}{596505}

\newcommand{\totalNumberOfCombinations}{35378}

\newcommand\numtracesremoved{13426}

\newcommand{\numpreprocessedtraces}{21952}
\newcommand{\percentagetimeout}{10.60\%}
\newcommand{\percentageNOtimeout}{89.40\%}
\newcommand{\preprocminnumberofentries}{1\xspace}
\newcommand{\preprocmaxnumberofentries}{11901}
\newcommand{\preprocaverageumberofentries}{1032\xspace}
\newcommand{\preprocstdevnumberofentries}{1471\xspace}
\newcommand{\numproperties}{98\xspace}

\newcommand{\SBTemPsyCheckTimeout}{2328\xspace}
\newcommand{\SBTemPsyCheckNoTimeout}{19624}

\newcommand{\minsimulationminutesprec}{0}
\newcommand{\maxsimulationhoursprec}{23}

\newcommand{\maxsimulationminutesprec}{28} 
\newcommand{\avgsimulationhoursprec}{4}
\newcommand{\avgsimulationminutesprec}{51}
\newcommand{\stdDevsimulationhoursprec}{6}
\newcommand{\stdDevsimulationminutesprec}{24}

\newcommand\timeoutpreprocessing{1}

\newcommand\numfinaldataset{14940}
\newcommand\percentageViolationsAmongNonTimeouts{76.13}

\newcommand\diagnosticsTimeoutInMinutes{1}
\newcommand\diagnosticsTimeouts{2889}

\newcommand\diagnosticsNoTimeouts{12051}
\newcommand\diagnosticsNoTimeoutsPercentage{80.66}
\newcommand\supportedViolationsOnly{11374}
\newcommand\supportedViolationsOnlyPercentage{94.38}
\newcommand\occurrencesOfsupportedViolationsOnly{12486}

\newcommand\mixedViolations{677}
\newcommand\mixedViolationsPercentage{5.62}
\newcommand\scopesOccurrences{4596}
\newcommand\scopesOccurrencesPercentage{36.81}
\newcommand\patternsOccurrences{7890}
\newcommand\patternsOccurrencesPercentage{63.19}

\newcommand\ordertimeout{274}
\newcommand\ordertimeoutPercentage{9.48}
\newcommand\eventBoundariestimeout{2216}
\newcommand\eventBoundariestimeoutPercentage{76.71}
\newcommand\complexPropTimeout{399}
\newcommand\complexPropTimeoutPercentage{13.81}

\newcommand\combinationsNotCoveredByRealProps{2562}

\newcommand{\violatedCombinationsWithGlobally}{2538}
\newcommand{\violatedCombinationsWithGloballyPercentage}{99.06}

\newcommand{\risesfallsCombinations}{180}

\FPeval{\totalNonTimeouts}{clip(\diagnosticsNoTimeouts+\combinationsNotCoveredByRealProps+\risesfallsCombinations)}
\FPeval{\totalCases}{clip(\numfinaldataset+\combinationsNotCoveredByRealProps+\risesfallsCombinations)}
\FPeval{\totalPercentage}{\totalNonTimeouts/\totalCases*100}
\FPeval{\totalPercentageA}{round(\totalPercentage:2)}
\FPeval{\totalSupported}{clip(\diagnosticsNoTimeouts+\violatedCombinationsWithGlobally+\risesfallsCombinations)}
\FPeval{\totalSupportedPercentage}{\totalSupported/\totalNonTimeouts*100}
\FPeval{\totalSupportedPercentageA}{round(\totalSupportedPercentage:2)} 
\usepackage{multirow}

\DeclareMathOperator{\minf}{\mathit{uni\_m\_min}}
\DeclareMathOperator{\maxf}{\mathit{uni\_m\_max}}
\DeclareMathOperator{\lminf}{\mathit{uni\_sm\_min}}
\DeclareMathOperator{\lmaxf}{\mathit{uni\_sm\_max}}
\DeclareMathOperator{\monot}{\mathit{monot}}

\begin{document}
\title{Trace Diagnostics for Signal-based   Temporal Properties}

\author{Chaima Boufaied,
  Claudio Menghi,~\IEEEmembership{Member,~IEEE,}
  	Domenico~Bianculli,~\IEEEmembership{Member,~IEEE,}
	and~Lionel~C.~Briand,~\IEEEmembership{Fellow,~IEEE}\IEEEcompsocitemizethanks{\IEEEcompsocthanksitem C.\ Boufaied
          is with the school of EECS, University of Ottawa, Ottawa, ON
          K1N 6N5, Canada (e-mail:chaima.boufaied@uottawa.ca).
          Part of this work was done when she was affiliated with the Interdisciplinary Centre for
          Security, Reliability, and Trust (SnT) of the University of
          Luxembourg.
          \IEEEcompsocthanksitem C. Menghi is with 
          University of Bergamo, Bergamo, Italy and
          McMaster
          University, Hamilton, Canada (e-mail:  claudio.menghi@unibg.it).
          Part of this work was done when he was affiliated with the Interdisciplinary Centre for
          Security, Reliability, and Trust (SnT) of the University of
          Luxembourg.
          \IEEEcompsocthanksitem D.\ Bianculli is with the Interdisciplinary Centre for
          Security, Reliability, and Trust (SnT) of the University of
          Luxembourg, Luxembourg (e-mail: domenico.bianculli@uni.lu).
          		\IEEEcompsocthanksitem L.\ Briand holds shared appointments with the Interdisciplinary Centre for Security, Reliability, and Trust (SnT) of the University of Luxembourg, Luxembourg and the school of EECS, University of Ottawa, Ottawa, ON K1N 6N5, Canada (e-mail: lionel.briand@uni.lu).}}

\IEEEtitleabstractindextext{\begin{abstract}

Trace checking is a verification technique widely used in
Cyber-physical system (CPS) development, to verify whether execution
traces satisfy or violate properties expressing system requirements.
Often these properties characterize complex signal behaviors and are
defined using domain-specific languages, such as \dslname, a
pattern-based specification language for signal-based temporal
properties.  Most of the trace-checking tools only yield a Boolean
verdict. However, when a property is violated by a trace, engineers
usually inspect the trace to understand the cause of the violation;
such manual diagnostic is time-consuming and
error-prone. Existing approaches that complement trace-checking tools
with diagnostic capabilities either produce low-level explanations
that are hardly comprehensible by engineers or do not support complex
signal-based temporal properties.

In this paper, we propose \NAME, a trace-diagnostic approach for
properties expressed using \dslname. Given a property and a trace that violates the property, \NAME determines the
root cause of the property violation. \NAME relies on the concepts of
\emph{violation cause}, which characterizes one of the behaviors of
the system that may lead to a property violation, and
\emph{diagnoses}, which are associated with violation causes and
provide additional information to help engineers understand the
violation cause.  As part of \NAME, we propose a language-agnostic methodology
to define violation causes and diagnoses. In our context, its application resulted in a catalog of 34
violation causes, each associated with one diagnosis, tailored to
properties expressed in \dslname. 

We assessed the applicability of \NAME \ on two datasets, including one based on a complex industrial case study.
The results show that \NAME could finish within a timeout of \SI{1}{\minute}
for $\approx\totalPercentageA\%$ of the trace-property combinations in the industrial dataset,
yielding a diagnosis in $\approx\totalSupportedPercentageA\%$ of these cases; 
moreover, it also yielded a diagnosis for all the trace-property combinations in the other dataset.
These  results suggest that our tool is applicable and efficient in most cases.

 \end{abstract}

}

\maketitle

\IEEEdisplaynontitleabstractindextext
\IEEEpeerreviewmaketitle

\section{Introduction}
\label{sec:introduction}
\IEEEPARstart{C}{yber-physical}
system (CPS) development requires engineers to verify whether the system meets its requirements.
In industrial contexts, verification is often performed through \emph{trace-checking} tools (e.g.,~\cite{boufaied2020trace,icse2021,FSE2019,gorostiaga2018striver,convent2018tessla,10.1007/978-3-319-46982-9_10}).
Engineers collect traces, sequences of records 
representing the behavior of the system, and use trace-checking tools to check whether the traces satisfy or violate properties expressing the system requirements. If properties are violated, the system has faults that need to be identified and corrected.

In the case of \emph{pattern-based} trace-checking tools, properties are expressed using pattern-based languages. 
Pattern-based languages contain domain-specific constructs to express complex requirements~\cite{8859226} that increase their usability in industrial contexts~\cite{fowler2010domain,8859226}.
In this work, we consider requirements expressed in \dslname~\cite{boufaied2020trace}, a pattern-based language that can express complex signal behaviors based on a recent taxonomy~\cite{DBLP:journals/corr/abs-1910-08330}.
This language enables engineers to write properties describing important types of requirements for industrial CPSs, through constructs that express complex signal behaviors, such as spikes and oscillations.

When a property is checked on a trace, trace-checking tools usually provide a \emph{Boolean verdict}: \emph{true} if the trace satisfies the property, \emph{false} otherwise.
When the property is violated by a trace, engineers usually inspect the trace to understand the cause of the violation, leading to the analysis of a high number of records. 
For example, in our industrial case study in the satellite domain, the average number of records included in \numsimulationtraces  traces is \averageumberofentries.
Inspecting a large number of records, and checking the causes of property violations, requires in general significant time. 
Additionally, this activity is error-prone and engineers may fail to identify the actual cause of the property violation. 
Therefore, they need automated tools that can explain the reasons leading to the violation of the properties. These tools should provide diagnostic information enabling engineers to understand the cause of violations.

Two complementary strategies were proposed in the literature to help engineers in these activities: (i)~isolating \emph{slices of traces} that explain the property violation;
and (ii)~checking whether traces show common \emph{behaviors that lead to the property violation}. These two complementary strategies are discussed in the following.

Approaches that isolate \emph{slices of the trace} that explain the property violation (e.g.,~\cite{ferrere2015trace,mukherjee2012computing,beer2009explaining,10.1007/978-3-319-89963-3_18}) usually assume that the properties are specified using a logical formula. 
To explain the property violation, these approaches iteratively analyze the sub-formulae of the logical formula and identify minimal slices of the trace that explain the satisfaction or violation of each sub-formula.
Using this approach, the size of the explanation increases with the number of sub-formulae of the logical formula expressing the property.
For properties expressed using pattern-based languages, which provide domain-specific constructs encoding complex logical formulae, using such an approach is likely to produce large explanations that are hardly comprehensible by engineers.
Besides, none of these approaches was implemented and evaluated on realistic case studies.

Approaches that check for the presence of common \emph{behaviors leading to the property violation} (e.g.,~\cite{dawes2019explaining,dou2018model,10.1007/978-3-319-11164-3_24}), assume that such behaviors  correspond to common causes of such violation.
Each cause therefore encodes one of the behaviors, observed in the trace, that may lead to a property violation and help explain it. 
However, existing approaches do not support complex signal-based temporal properties of CPS, such as the one expressed using \dslname.
Besides,
it is unclear how to extend these approaches  to support signal-based temporal properties,
since such approaches do not come with a precise methodology that describes how to add new causes that support more complex properties.

In this work, we propose \NAME, a trace-diagnostic approach for signal-based temporal properties. 
\NAME takes as input a trace and a property expressed using \dslname and violated by the trace; it provides as output an explanation that describes why the property is violated on that trace. 

To detect the source of the property violation, we define the notions of \emph{violation cause} and \emph{diagnosis}. 
A violation cause characterizes one of the behaviors of the system that may lead to a property violation.
For example, for a property requiring a signal to show a spike with an amplitude and a width lower than specific thresholds, the absence of any  spike behavior in a signal is a violation cause. 
Diagnoses are associated with violation causes and provide additional information to help engineers understand such causes.
For example, a diagnosis for the previous violation cause, for the case in which the value of the signal is increasing over time, contains two records (timestamps and signal values) where the signal shows its minimum and maximum values, while increasing. These values
allow engineers to understand the range of values taken by the signal while it exhibits an increasing behavior.

We propose a novel  \emph{methodology to define violation causes and diagnoses} (Section~\ref{sec:methodology}). 
Our methodology provides formal guarantees of the soundness of the proposed violation causes: if a violation cause holds on a trace, the corresponding property is violated.
Though we applied our methodology to define violation causes for
properties expressed using \dslname, our methodology is
language-agnostic and can therefore be applied to other
pattern-based specification languages such as
TemPsy~\cite{dou2017model} and FRETISH~\cite{giannakopoulou20:_gener_formal_requir_struc_natur_languag}.
To further support this claim, we also sketch how to apply our methodology to one construct supported by the latter.

We present a \emph{catalog of 34 violation causes, each associated with one diagnosis,} for signal-based temporal properties expressed in \dslname (Section~\ref{sec:diag-patterns-def}).
These violation causes are not complete as they do not encode all the possible reasons that may lead to a property violation, but are the results of applying our methodology in the context of our industrial case study. 
Indeed, such a catalogue of violation causes and diagnoses has been defined (and validated) together with a group of system and software engineers of our industrial partner, with the goal of maximizing the usefulness of a diagnosis for a certain violation cause.
However, following the same methodology, users can add new violation causes depending on their specific needs or on the requirements of particular domains. 

We implemented \NAME as a plugin for \checkname~\cite{boufaied2020trace}, a trace-checking tool for \dslname.
We assessed the applicability of \NAME  on a large, proprietary industrial dataset from the satellite domain (PROP-SAT), 
as well as a smaller dataset (AFC) generated from a benchmark model  used in the ARCH competition~\cite{ernst2021arch}.
\NAME could finish within a timeout of \SI{1}{\minute}
for $\approx\totalPercentageA\%$ of the trace-property combinations in the PROP-SAT dataset,
yielding a diagnosis in $\approx\totalSupportedPercentageA\%$ of these cases; 
moreover, it also yielded a diagnosis for all the trace-property combinations in the AFC dataset.

\emph{Significance.}
Since diagnoses were provided in $\approx\totalSupportedPercentageA\%$ of the cases for which no timeout occurred in the PROP-SAT dataset, and in the totality of the trace-property combinations in the AFC one, \NAME was deemed widely applicable across trace-property combinations in both datasets. Given the high expressiveness of \dslname, the many violation causes and related diagnoses we have defined in \NAME, and the run-time performance of our tool, we expect significant impact for this technology across many CPS domains.
Moreover, the methodology for defining violation causes and diagnoses can be adopted by other researchers working on the problem of trace diagnostics in the context of run-time verification.

To summarize, the main contributions of this paper are:
\begin{compactitem}
\item \NAME, a trace-diagnostic approach for signal-based temporal properties expressed in \dslname, based on the concepts of \emph{violation cause} and \emph{diagnosis};
\item a language-agnostic methodology for defining violation causes
  and diagnoses, with formal guarantees of the soundness of the
  proposed violation causes, and its application to \dslname;
\item a catalog of 34 violation causes, each associated with one diagnosis, for signal-based temporal properties expressed in \dslname;
\item a comprehensive evaluation of the applicability of \NAME on two datasets, including one based on a complex industrial case study.
\end{compactitem}

\emph{Paper structure.} This paper is organized as follows. 
Section~\ref{sec:motivating} introduces our case study from the satellite domain and identifies concrete motivations for our work.
Section~\ref{sec:background} illustrates the syntax and semantics of \dslname.
Section~\ref{sec:approach} presents \NAME, our pattern-based trace-diagnostic approach.
Section~\ref{sec:methodology} describes our methodology to define violation causes and diagnoses.
Section~\ref{sec:diag-patterns-def} presents the violation causes and diagnoses proposed in this work.
Section~\ref{sec:evaluation} 
reports on the evaluation of the applicability of \NAME on two datasets.
Section~\ref{sec:practical} discusses the practical implications of our approach.
Section~\ref{sec:related} surveys related work.
Section~\ref{sec:conclusion} concludes the paper, providing directions for future work.

\section{Case Study and Motivations}
\label{sec:motivating}
Our case study is a satellite developed by our industrial partner.
This is a representative CPS as it contains many complex software components that interact with actuators and sensors of the satellite. 

During the satellite development, and after its deployment, engineers collect traces that describe the behavior of the satellite.  
A fragment of one of these traces is depicted in \figurename~\ref{fig:traceExample} and plotted in \figurename~\ref{fig:tracegraphical}.
A trace is a sequence of records that describe how the values of some signals change over time.
For example, the fragment of the trace in \figurename~\ref{fig:traceExample} contains eight records.
Each record contains a timestamp, identifying the time at which the record was collected, and the values assumed by some variables, each recording the values of one of the monitored signals at that time.
In the example, the variables $\beta$ and $\rho$ record respectively the signals representing the beta angle~\cite{beta} and the pointing error~\cite{ott2011esa} of the satellite.
For example, for record $r_3$ the timestamp is $0.9$, and the values of the variables $\beta$ and $\rho$ are respectively \SI{55.0}{\degree} and \SI{125.0}{\degree}.
The recording interval of a trace is the difference between the maximum and the minimum timestamps. For example, the recording interval of the trace in \figurename~\ref{fig:traceExample} is 
\SI{6}{\s}.

\begin{figure}
\begin{tikzpicture}
	\pgfmathsetmacro{\ilocationangularrate}{0.25}
	\pgfmathsetmacro{\ilocationrho}{-0.25}
	\pgfmathsetmacro{\ilocationtimestamp}{-0.75}
	\pgfmathsetmacro{\ilocationindex}{-1.2}
	\draw[dashed] (-2,0.5) -- (6.2,0.5);
		\draw node at (-1.5,\ilocationangularrate) {\footnotesize $\beta$};
	\draw node at (-0.5,\ilocationangularrate) {\small $2.0$};
	\draw node at (0.4,\ilocationangularrate) {\small $153.5$};
	\draw node at (1.3,\ilocationangularrate) {\small $55.0$};
	\draw node at (2.2,\ilocationangularrate) {\small $0.5$};
	\draw node at (3.1,\ilocationangularrate) {\small $80.0$};
	\draw node at (4.0,\ilocationangularrate) {\small $203.5$};
	\draw node at (4.9,\ilocationangularrate) {\small $20.0$};
	\draw node at (5.8,\ilocationangularrate) {\small $0.5$};
		\draw[dashed] (-2,-0) -- (6.2,-0);
\draw node at (-1.5,\ilocationrho) {\footnotesize $\rho$};
	\draw node at (-0.5,\ilocationrho) {\small $1.0$};
	\draw node at (0.4,\ilocationrho) {\small $52.5$};
	\draw node at (1.3,\ilocationrho) {\small $125.0$};
	\draw node at (2.2,\ilocationrho) {\small $125.5$};
	\draw node at (3.1,\ilocationrho) {\small $25.0$};
	\draw node at (4.0,\ilocationrho) {\small $75.5$};
	\draw node at (4.9,\ilocationrho) {\small $35.0$};
	\draw node at (5.8,\ilocationrho) {\small $200.5$};

		\draw node at (-1.5,\ilocationtimestamp) {\small timestamp};
	\draw node at (-0.5,\ilocationtimestamp) {\small $0.0$};
	\draw node at (0.4,\ilocationtimestamp) {\small $0.2$};
	\draw node at (1.3,\ilocationtimestamp) {\small $0.9$};
	\draw node at (2.2,\ilocationtimestamp) {\small $1.8$};
	\draw node at (3.1,\ilocationtimestamp) {\small $3.0$};
	\draw node at (4.0,\ilocationtimestamp) {\small $4.9$};
	\draw node at (4.9,\ilocationtimestamp) {\small $5.7$};
	\draw node at (5.8,\ilocationtimestamp) {\small $6.0$};
	\draw[dashed] (-2,-0.5) -- (6.2,-0.5);
	\draw[dashed] (-2,-1) -- (6.2,-1);
	\pgfmathsetmacro{\llocation}{-1.5}
	\pgfmathsetmacro{\dlocation}{-2}
	\pgfmathsetmacro{\xlocation}{-2.5}
	\pgfmathsetmacro{\edgelocation}{-3.5} ;
		\draw (0.9,0.6) -- (1.7,0.6) -- (1.7,-1.1) -- (0.9,-1.1) -- (0.9,0.6);
		\draw node at (1.5,-1.3) {\small Record $r_3$};
	\end{tikzpicture}
	
	 \caption{A fragment of a trace from our case study.}
\label{fig:traceExample}
\end{figure}

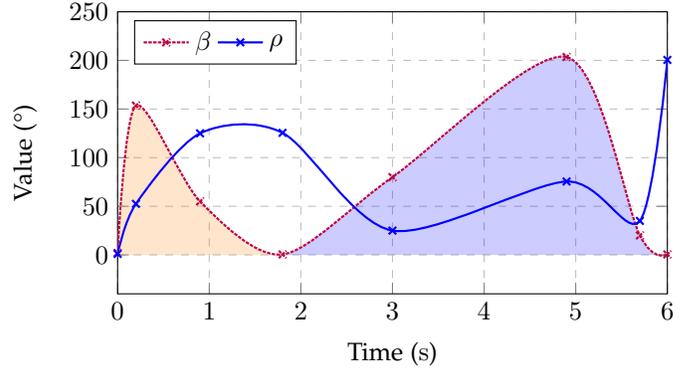
\begin{figure}
 \begin{tikzpicture}[domain=0:6] 
 \begin{axis}[
  width=\columnwidth,
       height=0.6\columnwidth,
    xlabel={Time (\si{\s})},
    ylabel={Value (\si{\degree})},
    xmin=0, xmax=6,
    ymin=-40, ymax=250,
    xtick={0,1,2,3,4,5,6},
    ytick={0,50,100,150,200,250},
     ymajorgrids=true,
      xmajorgrids=true,
    grid style=dashed,
      legend pos=north west,
      legend style={legend columns=-1}
]

 \path[name path=axis] (axis cs:0,0) -- (axis cs:6,0);
 
    \addplot[color=purple,smooth,thick,mark=x,name path=s,dash pattern=on 1.3pt off 0.5pt on 1.3pt off 0.5pt] plot coordinates {
          (0,2)
          (0.2,153.5)
          (0.9,55.0)
          (1.8,0.5)
          (3.0,80.0)
          (4.9,203.5)
          (5.7,20)
          (6.0,0.5)
      };
     
            \addplot[color=blue,smooth,thick,mark=x,name path=g,smooth] plot coordinates {
          (0,1.0)
          (0.2,52.5)
          (0.9,125.0)
          (1.8,125.5)
          (3.0,25.0)
          (4.9,75.5)
          (5.7,35)
          (6.0,200.5)
      };
       \addplot[color=orange,opacity=0.2] fill between[ 
    of = s and axis,      
    soft clip={domain=0:1.8}
  ];
    \addplot[color=blue,opacity=0.2] fill between[ 
    of = s and axis,      
    soft clip={domain=1.8:6.0}
  ];
\legend{$\beta$,$\rho$}
  \end{axis}

\end{tikzpicture}

 \caption{Graphical representation of the trace in
  Figure~\ref{fig:traceExample}.}
\label{fig:tracegraphical}
\end{figure}

After the traces are collected, engineers analyze whether the behaviors recorded in the traces satisfy the CPS requirements. An example of a requirement (inspired by the ones from the case study) is the following:
\begin{center}
    R1: \emph{``Within the trace, the beta angle shall contain at least one spike with an amplitude lower than \SI{90}{\degree}  
    and a width less than \SI{0.5}{\second}''.} 
\end{center}
 The beta angle is the angle between the orbital plane of the satellite and the vector of the Sun (i.e., the direction from which the Sun is shining). After deployment, the satellite aligns its orbital plane. Therefore, $\beta$ shall contain a spike with an amplitude lower than \SI{90}{\degree}.
The trace shown in \figurename~\ref{fig:traceExample} violates
the requirement R1.
As we will discuss in the next section, automated trace-checking tools, such as \checkname~\cite{boufaied2020trace} (see section~\ref{sec:background}), can verify whether a trace satisfies or violates a requirement.
However, they do not provide any additional information to help engineers understand the cause of the violation.
 This means that engineers have to manually inspect the values of the variables recorded in the trace records and check why these values led to the violation of the requirement. In our example, looking at the plot in \figurename~\ref{fig:tracegraphical}, one can see 
that the two spikes (i.e., \emph{spike1} and \emph{spike2} defined over the
time intervals $[0,1.8]$ and $[1.8,6.0]$) of signal $\beta$ have an amplitude value ($A_1=\SI{153.5}{\degree}$ and $A_2=\SI{203.5}{\degree}$) greater
than \SI{90}{\degree} and show a width ($w_1=\SI{1.8}{\s}$ and $w_2=\SI{4.2}{\s}$) greater than \SI{0.5}{\second}.
Our pattern-based diagnostic approach (see section~\ref{sec:approach}) aims to automatically detect the causes of requirement violation.

 \section{Background: \dslname}
\label{sec:background}

\begin{figure}[t]
\begin{footnotesize}
\begin{tabular}{l@{\ }l@{\ }p{63mm}}
\toprule
\textbf{Property} & $  \property::=$ &$
\property_1\ \mathbin{\lit{or}}\ \property_2  \ |\ \delta$ \\\midrule
\textbf{Clause} & $  \delta ::=$ &$
\delta_1\  \mathbin{\lit{and}}\ \delta_2 \ |\ \alpha$ \\\midrule
\textbf{Atom} &  $\alpha::= $ & $\lit{not}\ \synt{sc}\ |\ 
\synt{sc} $ \\\midrule
\textbf{Scope} &$\synt{sc}::= $&
$  \lit{globally}\ \synt{p}\ |\ \lit{before}\ \synt{t}\ \synt{p} \ |\
                                 \lit{after}\ \synt{t}\ \synt{p}  \ |\
                                 \lit{at}\ \synt{t}\ \synt{p}\ |$ \\
 & & $\lit{before}\ \synt{p}\textsubscript{1}\ \synt{p}$
$|\ \lit{after}\ \synt{p}\textsubscript{1}\
 \synt{p}\ |$
  \\
&& $\lit{between}\ \synt{t}\textsubscript{1} \ \mathbin{\lit{and}} \ \synt{t}\textsubscript{2}\ \synt{p}\ |\
    \lit{between}\ \synt{p}\textsubscript{1} \ \mathbin{\lit{and}} \ \synt{p}\textsubscript{2}\ \synt{p} $
   \\ \midrule
\textbf{Pattern} & $\synt{p}::=$  & 
$  \lit{assert}\ \synt{c} \ |\ \synt{s} \mathbin{\lit{becomes} \sim} \synt{v}\ |$\\
&& $\lit{if}\
 \synt{p}\textsubscript{1}\ \lit{then}\  [\lit{within} \bowtie \synt{t}]\
 \synt{p}\textsubscript{2} \ |$\\
                  & & $\lit{exists}\  \lit{spike}\ \lit{in}\ \synt{s}$\\
&& $ \phantom{\lit{ex}} \left[\lit{with}\
    [\lit{width}\ \sim_1 \synt{v}\textsubscript{1}] \ | [\lit{amplitude}\ \sim_2 \synt{v}\textsubscript{2}]\right] \ |\ $ \\
                  & & $ \lit{exist}\ \lit{oscillation}\ \lit{in}\ \synt{s}$\\
&& $ \phantom{\lit{ex}} \left[\lit{with}\ 
 [\lit{p2pAmp}\ \sim_1 \synt{v}\textsubscript{1}] [\lit{period}\ \sim_2 \synt{v}\textsubscript{2}]\right]\ |\ $\\
&&  $ \synt{s}\  \lit{rises} \ [\lit{monotonically}] \ \lit{reaching} \ \synt{v}\ |\ $\\
&&   $\synt{s}\  \lit{falls} \ [\lit{monotonically}] \ \lit{reaching} \ \synt{v}\ |\  $\\
&&  $\synt{s}\ \lit{overshoots}\ [\lit{monotonically}]\ \synt{v}\textsubscript{1}\ \lit{by}\ \synt{v}\textsubscript{2}\  |\ $\\ 
&&  $ \synt{s}\ \lit{undershoots}\
 [\lit{monotonically}]\ \synt{v}\textsubscript{1}\ \lit{by}\
 \synt{v}\textsubscript{2}\   $\\
& $\bowtie\ ::=$ & \lit{exactly} | \lit{at most} | \lit{at least}
\\ \midrule
\textbf{Condition} & $\synt{c}::=$ &
 $ \synt{c}\textsubscript{1}\ \mathbin{\lit{and}}\ \synt{c}\textsubscript{2}\ |\ \synt{c}\textsubscript{1}\ \mathbin{\lit{or}}\ \synt{c}\textsubscript{2}\ |\  \synt{s}\ \sim \synt{v}$ \\
\bottomrule
\end{tabular}\\[5pt]
$\synt{t}, \synt{t}\textsubscript{1},\synt{t}\textsubscript{2} \in \real$;
  $\synt{v}, \synt{v}\textsubscript{1}, \synt{v}\textsubscript{2} \in \real$;
  $\sim \in \{ <,>,=,<>, <=, >= \}$; \\$\synt{s}$ is a signal or a mathematical expression over the signals $S$ defined in property $\phi$.
\end{footnotesize}

 \caption{\dslname  syntax.}
\label{tab:syntax}
\end{figure}

\begin{figure*}[t]
\begin{footnotesize}
\begin{tabular}{p{\linewidth}}
\toprule
$  \trace \models  \property_1 \mathbin{\lit{or}}   \property_2   $ \emph{iff} $  
 (\trace \models \property_1) \vee (\trace \models \property_2)  $; \hfill
$  \trace  \models \ \delta_1 \mathbin{\lit{and}} \ \delta_2  $ \emph{iff} $   (\trace  \models \delta_1) \wedge 
 (\trace \models \delta_2)  $ ; \hfill
 $  \trace \models  \lit{not}\   \synt{sc}   $ \emph{iff} $   (\trace
                                \not \models \synt{sc} )  $\\
\midrule
$  \trace \models \lit{before}\ \synt{t}\ \synt{p}  $ \emph{iff} $   t_i < \synt{t} \leq t_e \wedge \trace, [t_i,\synt{t}] \models \synt{p}$ ;\hfill
 $\trace \models \lit{between}\ \synt{n} \ \lit{and} \
\synt{m}\ \synt{p}  $ \emph{iff} $   t_i \leq \synt{n} < \synt{m} \leq t_e \wedge \trace, [\synt{n},\synt{m}] \models \synt{p} $\\
$ \trace \models \lit{globally}\ \synt{p} $ \emph{iff} $
  \trace, [t_i,t_e] \models \synt{p} $ ;\hfill $ \trace \models \lit{at}\ \synt{t}\ \synt{p} $ \emph{iff} $    t_i \leq \synt{t} \leq t_e \wedge \trace, [\synt{t},\synt{t}] \models \synt{p}$  ;\hfill
 $\trace \models \lit{after}\ \synt{t}\ \synt{p}   $ \emph{iff} $    t_i \leq \synt{t} < t_e \wedge \trace, [\synt{t},t_e] \models \synt{p}$   
 \\
$  \trace \models \lit{before}\ \synt{p}_1\ \synt{p}  $ \emph{iff} $   \forall t_1,t_2, ( (t_i<t_1<t_2 \leq t_e \wedge \trace, [t_1,t_2] \models
 \synt{p}_1) \Rightarrow \exists t_3,t_4, (t_i \leq t_3<t_4<t_1 \wedge \trace, [t_3,t_4] \models \synt{p} ))$\\
$ \trace \models \lit{after}\ \synt{p}_1\ \synt{p}  $ \emph{iff} $   
 \forall t_1,t_2, ( (t_i \leq t_1<t_2 < t_e \wedge \trace, [t_1,t_2] \models \synt{p}_1) \Rightarrow \exists t_3,t_4, ( t_2<t_3<t_4 \leq t_e \wedge \trace, [t_3,t_4] \models \synt{p}) )$\\
$\trace \models \lit{between}\ \synt{p}_1 \
     \lit{and} \ \synt{p}_2\ \synt{p}  $ \emph{iff} $  \forall
     t_1,t_2,t_3,t_4, ( (t_i\leq t_1<t_2<t_3<t_4 \leq t_e \wedge \trace, [t_1,t_2] \models \synt{p}_1 \wedge \trace, [t_3,t_4] \models \synt{p}_2) \Rightarrow  \trace, [t_2,t_3] \models \synt{p} ) $
\\
\midrule
$  \trace,[t_l,t_u] \models  \lit{assert}\ \condition    $ \emph{iff} $ \forall t \in [t_l,t_u], (\trace,t \models \condition )$. For every time instant $t$ within $[t_l,t_u]$, condition \condition holds.
\\
$ \trace, [t_l,t_u] \models \synt{s}\ \lit{becomes}  \sim \synt{v}   $ \emph{iff} $ \exists t \in (t_l,t_u], 
 (\synt{s}(t) \sim  \synt{v}
 \wedge  \forall t_1 \in [t_l,t),  (\synt{s}(t_1) \not \sim  \synt{v})
  )$. Formula  $\synt{s}(t)\sim \synt{v}$ is \emph{true} for some $t$,
  and for any time instant $t_1$ before $t$,  $\synt{s}(t)\sim \synt{v}$ is \emph{false}.
\\
 $ \trace,[t_l,t_u] \models 
 \lit{exists}\  \lit{spike}\ \lit{in}\ \synt{s}\ [\lit{with}\ [\lit{width} \sim_1 \synt{v}_1]_\beta  [\lit{amplitude} \sim_2 \synt{v}_2]_\gamma]_\alpha $ \emph{iff} $  \exists t_1,t_2,t_3, t_4, t_5 \in [t_l,t_u],   (
t_1 < t_2 < t_2 < t_3 < t_4 < t_5 \wedge
  \maxf (s,t_2,[t_1,t_3] ) \wedge \lminf (s,t_3,[t_2,t_4] )  \wedge
 \maxf(s,t_4,[t_3,t_5])
[[\wedge ( t_3-t_1) \sim_1
\synt{v}_1]_\beta
[\wedge \max(
(
\synt{s}(t_2)-\synt{s}(t_3)
),
( \synt{s}(t_4)-\synt{s}(t_3))
)\sim_2 \synt{v}_2 ]_\gamma]_\alpha )$. Signal \synt{s} has a strict maximum within two (non strict) minima.
The values $\synt{v}_1$ and $ \synt{v}_2$ constrain the width and
the amplitude of the spike.$^\ast$
\\
$ \trace,[t_l,t_u]  \models
\lit{exist}\ \lit{oscillation}\ \lit{in}\ \synt{s}\ [\lit{with}\ 
[\lit{period}
 \sim_1 \synt{v}_1]_\zeta
 [\lit{p2pAmp} \sim_2 \synt{v}_2]_\epsilon ]_\delta
  $ \emph{iff} $ \exists t_1,t_2,t_3, t_4, t_5 \in [t_l,t_u],
(
t_1 < t_2 < t_2 < t_3 < t_4 < t_5 \wedge
\lmaxf(s,t_2,[t_1,t_3]) \wedge  \lminf(s,t_3,[t_2,t_4]) \wedge
 \lmaxf(s,t_4,[t_3,t_5])[
[\wedge (t_4-t_2) \sim_1
      \synt{v}_1]_\zeta 
[
\wedge
( \synt{s}(t_2)-\synt{s}(t_3)) \sim_2 \synt{v}_2
\wedge
( \synt{s}(t_4)-\synt{s}(t_3))
 \sim_2 \synt{v}_2]_\epsilon ]_\delta
)$. Signal \synt{s} shows  a strict maximum within two strict minima.
The values  $ \synt{v}_1$ and $ \synt{v}_2$ constrain the period and the  amplitude of the oscillation.$^\ast$ 
\\
$ \trace, [t_l, t_u]  \models \synt{s}\  \lit{rises}\  [\lit{monotonically}]_\alpha\ \lit{reaching}\ \synt{v} $ \emph{iff} $ \exists t \in (t_l, t_u], ( \synt{s}(t) \ge \synt{v} \wedge \forall t_1 \in [t_l,t), (
 \synt{s}(t_1) < \synt{v} )
  [\wedge \monot(\synt{s},t_l,t)]_\alpha )$. There exists a time instant $t$ where $\synt{s}(t)\geq \synt{v}$,  and for any time instant $t_1$ before $t$, $\synt{s}(t_1) < \synt{v}$.
The character $\alpha$ labels the formula indicating that the signal shall  rise monotonically.
\\ 
$\trace,[t_l,t_u]  \models  \synt{s}\ \lit{overshoots}\ [\lit{monotonically}]_\alpha\
 \synt{v}_1\ \lit{by}\ \synt{v}_2 $ \emph{iff} $  \exists t \in (t_l, t_u],  ( \synt{s}(t) \geq \synt{v}_1 \wedge \forall t_1 \in [t,t_u], ( \synt{s}(t_1) \leq \synt{v}_1+ \synt{v}_2 ) 
\wedge \forall t_2 \in [t_l,t) (\synt{s}(t_2) < \synt{v}_1 )  [\wedge \monot(\synt{s},t_l,t)]_\alpha  ) $. Signal \synt{s} is initially lower than $\synt{v}_1$. It then exceeds  $\synt{v}_1$ at time instant $t$ by remaining below $\synt{v}_1+\synt{v}_2$. 
The character $\alpha$ labels the formula indicating that the signal shall overshoot monotonically.
\\
$ \trace,[t_l,t_u]  \models  \lit{if}\ \synt{p}_1\ \lit{then}\ [\lit{within}\ \bowtie\ \synt{d}]_\alpha\ \synt{p}_2 $ \emph{iff} $ 
\forall t_{1}, t_{2} \in [t_l,t_u),  ((t_1 < t_2 \wedge \trace, [t_{1}, t_{2}]  \models \synt{p}_1) \Rightarrow
\exists t_{3}, t_{4} \in [t_2, t_u], (t_3< t_4 \wedge \trace, [t_3,t_4] \models \synt{p}_2 [\wedge (t_3-t_2)
 \llbracket\bowtie\rrbracket\synt{d}]_\alpha  )  )$  where $\llbracket \bowtie \rrbracket$ is such that $\llbracket \lit{exactly}\rrbracket \equiv \text{`='}$,
    $\llbracket \lit{at most}\rrbracket \equiv \text{`\textless ='}$,
    $\llbracket \lit{at least}\rrbracket \equiv  \text{`\textgreater ='}$. If pattern $\synt{p}_1$ holds in an interval $[t_1,t_2]$, then pattern $\synt{p}_2$ holds in a subsequent interval $[t_3,t_4]$.\\
\midrule
 $  \trace,t \models \condition_1 \mathbin{\lit{and}}  \condition_2   $ \emph{iff} $  
(\trace,t \models \condition_1) \wedge
(\trace,t \models \condition_2)  $ ;\hfill  
 $\trace,t  \models  \condition_1 \mathbin{\lit{or}}  \condition_2   $ \emph{iff} $   (\trace,t  \models \condition_1) \vee
 (\trace,t  \models \condition_2)  $ ;\hfill
 $ \trace,t  \models \synt{s} \sim \synt{v}$ \emph{iff} $s(t)\sim  \synt{v} $ 
\\
\bottomrule
\end{tabular}\\[5pt]
$\synt{t}, \synt{t}\textsubscript{1},\synt{t}\textsubscript{2} \in \real$;
  $\synt{v}, \synt{v}\textsubscript{1}, \synt{v}\textsubscript{2} \in \real$;
  $\sim \in \{ <,>,=,\neq,\leq, \geq \}$; $\synt{s}$ is a signal in $S$ or a mathematical expression over the signals in $S$.\\
  $\monot(\synt{s},t_1,t_2)::=  \forall t_3 \in [t_1,t_2), \forall t_4 \in (t_3,2], ( \synt{s}(t_3) < \synt{s}(t_4) ).$
  \newline
$\maxf(\synt{s},t,[t_a,t_b])::=$ 
$\synt{s}(t)=x \text{ and } \forall t_1 \in [t_a,t_b],
 \synt{s}(t_1)<x \text{ and }  \forall t_1,t_2 \in [t_a,t], \text{ if } t_1<t_2 \text{ then }
 \synt{s}(t_1) \leq \synt{s}(t_2) \text{ and } \forall t_1,t_2 \in [t_a,t], \text{ if } t_1<t_2 \text{ then } \synt{s}(t_1) \geq \synt{s}(t_2) $ \newline
$\lmaxf(\synt{s},t,[t_a,t_b])::=$ 
$\synt{s}(t)=x \text{ and } \forall t_1 \in [t_a,t_b],
 \synt{s}(t_1)<x \text{ and }  \forall t_1,t_2 \in [t_a,t], \text{ if } t_1<t_2 \text{ then }
 \synt{s}(t_1) < \synt{s}(t_2) \text{
   and } \forall t_1,t_2 \in [t_a,t], \text{ if } t_1<t_2 \text{ then } \synt{s}(t_1) > \synt{s}(t_2) $ \newline
  $^\ast$ We present the case  where  a (strict) minimum is followed by a strict maximum followed by 
    a (strict) minimum. The dual case can be derived from our
    formulation. Similarly, we present the predicates that
    characterize a local (strict) maximum ($\maxf$ and
    $\lmaxf$). Their dual case, i.e., the predicates that characterize
    a local (strict) minimum ($\minf$ and $\lminf$) can be derived from the above formulations.  
\end{footnotesize}

 \caption{\dslname  formal semantics (based on~\cite{boufaied2020trace})}
\label{tab:propertypatterns}
\end{figure*}

\dslname~\cite{boufaied2020trace} is a domain-specific language for expressing requirements that concern signal-based temporal properties.
The syntax  of \dslname is shown in \figurename~\ref{tab:syntax}; optional
items are enclosed in square brackets; the symbol `$|$' separates
alternatives.\footnote{The grammar of  \dslname considered in this
  paper is slightly different from the original one~\cite{boufaied2020trace}. Any \dslname property can be rewritten following this grammar by using standard rewriting rules~\cite{robinson2001handbook}.}
A \emph{property} is a disjunction of clauses.
A \emph{clause} is a conjunction of atoms.
An \emph{atom} is defined in terms of a \emph{scope} (non-terminal~\synt{sc}) or the negation operator~(\lit{not})  applied to constructs of type scope~(\synt{sc}).
A scope operator constrains a \emph{pattern} (non-terminal~\synt{p}) to hold within a given time interval. 
There are two types of scope operators: absolute scopes and event scopes.
Absolute scopes are delimited by
absolute time instants (e.g.,  \lit{before}\ \synt{t}\ \synt{p}).
Event scopes are delimited by other patterns (e.g., \lit{before}~\synt{p}\textsubscript{1}~\synt{p}). 
A pattern (e.g., $\lit{exists}\  \lit{spike}\ \lit{in}\ \synt{s}\ [...]$)  specifies a constraint on the behavior of one or more
signals. 
A \emph{condition}, which is used within the \lit{assert}~\synt{c}
pattern, is a comparison ($\synt{s} \sim \synt{v}$)  between the value
of a signal \synt{s} and the value \synt{v}, or a combination of two
conditions with the $\mathbin{\lit{and}}$ and $\mathbin{\lit{or}}$
logical operators.

\dslname supports the following patterns:
\begin{itemize}
\item \lit{assert} indicates an event-based data assertion. It specifies a constraint on the value of a signal. 
A requirement with the \lit{assert}  construct is as follows:\\ $R2$: The beta angle shall vary between  \SI{90}{\degree} and  \SI{-90}{\degree}.
The corresponding \dslname specification of the pattern is:
$\synt{p}_2 \equiv \lit{assert} \ \beta \ \lit{<=}\  90 \ \lit{and} \ \beta \ \lit{>=}  -90$
\item \lit{becomes} represents a state-based data assertion. It specifies a state of the signal, within a specific time interval, that satisfies a condition, which was not satisfied before that interval. A requirement with the \lit{becomes}  construct is as follows:\\
$R3$: the value of signal $\mathit{RWS\_command}$
shall become greater than 0.
The corresponding \dslname specification of the pattern is: \newline
$\synt{p}_3 \equiv \mathit{RWS\_command} \ \lit{becomes} \ > 0$
\item \lit{rises} indicates a constraint on the transient behavior of a signal, while it reaches, possibly monotonically, a target value. Its dual behavior is called \lit{falls}. \newline
A requirement with the \lit{rises} construct is as follows:  \\
$R4$: The\textit{ X\_cur} signal of the sun sensor shall rise monotonically reaching the value of \SI{3650}{\micro\ampere}.
The corresponding \dslname specification of the pattern is: \newline
$\synt{p}_4 \equiv \mathit{X\_cur} \ \lit{rises} \ \lit{monotonically} \ \lit{reaching} \ 3650$
\item \lit{overshoots} specifies a maximum value (i.e., above the target value) that a signal can reach when overshooting (i.e., when it exceeds the target value). The pattern can possibly be defined with a monotonicity constraint that requires a monotonic increase of the signal prior to reaching its target value.
The dual behavior of this pattern is called \emph{undershoot} and is expressed with the keyword \lit{undershoots}.
A requirement with the \lit{overshoots} construct is as follows:\\  
$R5$: The\textit{ X\_cur} signal of the sun sensor shall monotonically overshoot the value of \SI{3650}{\micro\ampere} by at most \SI{50}{\micro\ampere}.
The corresponding \dslname specification of the pattern is: \newline
$\synt{p}_5 \equiv \mathit{X\_cur} \ \lit{overshoots} \ \lit{monotonically} \ 3650 \ \lit{by} \ 50$.
\item \lit{spike} specifies a large increase (or decrease) of the value of a signal. 
A spike is characterized by three extrema (one strict maximum surrounded by two local minima if it represents an increase of the signal, or one strict minimum surrounded by two local maxima if it represents a decrease of the signal).
A requirement with the \lit{spike} construct is as follows:\\
$R6$: The \textit{beta\_angle} signal shall show a spike with an amplitude less than \SI{90}{\degree}. 
The corresponding \dslname specification of the pattern is: \newline
$\synt{p}_6 \equiv \lit{exists} \ \lit{spike} \ \lit{in} \ \mathit{beta\_angle} \\ \ \lit{with} \ \lit{amplitude} \ < 90$.
\item \lit{oscillation} specifies a repeated variation, over time, of the signal value. During an oscillatory behavior, the signal value swings from one extremum to the adjacent extremum of the same type (i.e., maximum or minimum) by traversing an extremum of the other type. 
A requirement with the \lit{oscillation} construct is as follows:\\ 
$R7$: The velocity of the satellite along the \textit{X\_axis} signal shall oscillate with a maximum amplitude of \SI{8000}{\km} per hour and a maximum period of \SI{180}{\min}. 
The corresponding \dslname specification of the pattern is: \newline
$\synt{p}_7 \equiv \lit{exist} \ \lit{oscillation} \ \lit{in} \ \mathit{X\_axis} \\ \lit{with} \ \lit{p2pAmp} \ <= 8000 \\ \lit{with} \ \lit{period} \ <= \ 180 $.
\item \lit{if}-\lit{then} represents a constraint on a response behavior of one or two signals, where a pattern (i.e., effect pattern) shall hold some time after a trigger pattern (i.e., cause pattern) has held in the past.
A requirement with the \lit{if}-\lit{then} construct is as follows:\\ 
$R8$: If the value of signal \textit{not\_Eclipse} is equal to 0, then the value of signal \textit{sun\_currents} should eventually be equal to 0.
The corresponding \dslname specification is: \newline
$\synt{p}_8 \equiv \ \lit{if} \ \mathit{not\_Eclipse}  \ = 0 \  \lit{then} \ \mathit{sun\_currents}  \ = 0$
\end{itemize} 

The syntax of \dslname enables engineers to define the property $\phi_1$ expressing requirement R1 (see Section~\ref{sec:motivating}) as:
$\phi_1 \equiv \ $ \lit{globally}\ \lit{exists}\  \lit{spike}\ \lit{in}\ $\beta$ 
\lit{with}\ \lit{width} <0.5 \lit{amplitude} < 90. 
Note that this property is made by a single atom (represented by the
\lit{globally} scope construct, which is applied to a pattern \synt{p}).  

The semantics of each construct $\construct$ of \dslname is shown in
\figurename~\ref{tab:propertypatterns}, which is divided into four parts.
The first part contains the semantics of properties, clauses, and atoms.
The other three parts address the semantics of scopes, patterns, and conditions.
The semantics of a construct $\construct$ is the formula $\patternsemantics(\construct)$ (in first-order logic)  written on the right side of the \emph{iff} (if and only if) sign.

Recall from section~\ref{sec:motivating} that a trace $\trace$ is a sequence of
records that describe how the values of one or more signals in the set $S=\{s_1,s_2,\dots,s_n\}$
change over time. More precisely, each record contains a timestamp, identifying the time at which the record was collected, and the values assumed by some variables, each recording the values of one of the monitored signals in $S$ at that time.
For properties, the semantics specifies the conditions that make a property~$\phi$ satisfied by the trace \trace, i.e., $\trace \models \phi$.
For example, the semantics of $ \property_1 \mathbin{\lit{or}}
\property_2$ requires at least one of them to hold on trace $\trace$.
For scopes, the semantics specifies the conditions that make a scope~$\synt{sc}$ satisfied on the trace \trace, i.e., $\trace \models \synt{sc}$.
For example, the semantics of the \lit{globally} scope indicates that a pattern \synt{p} scoped by the  \lit{globally} operator holds on the trace \trace if the pattern \synt{p} holds on the interval of the trace \trace delimited by the timestamps $t_i$ and $t_e$.
The timestamps $t_i$ and $t_e$ indicate the initial and the last timestamps of the trace.
For patterns, the semantics specifies the conditions that make a pattern~\synt{p}  satisfied on the interval of the trace \trace delimited by the timestamps $t_l$ and $t_u$ defined by a given scope, i.e., $\trace, [t_l,t_u] \models \synt{p}$. 
\figurename~\ref{tab:propertypatterns} also includes an informal description of the pattern semantics, after the formal definition. 
For example, the semantics of pattern ``\lit{exists}\ \lit{spike}\
\lit{in}\ \synt{s}\ \lit{with}\ \lit{width}\
$\sim_1$ \synt{v}\textsubscript{1} \lit{amplitude}\
$\sim_2$ \synt{v}\textsubscript{2}''  specifies that signal \synt{s} shows a spike behavior with a width satisfying the constraint ``$\sim_1$ \synt{v}\textsubscript{1}'' and an amplitude satisfying the constraint ``$\sim_2$ \synt{v}\textsubscript{2}''. 
A spike informally denotes a temporary (large) increase (or decrease) of the value of a signal. It occurs when the signal has a strict maximum  surrounded by two minima (or a strict minimum surrounded by two maxima).
This behavior can be subjected to additional constraints on the width (i.e., the difference between the time instants at which the two minima --- or the two maxima occur) and on the amplitude (i.e., the difference between the maximum and minimum values of the signal).
Finally, the semantics of conditions specifies how to satisfy a condition \synt{c} for a trace \trace at time instant~$t$, i.e., $\trace, t \models \synt{c}$.
For example, the semantics of $ \condition_1 \mathbin{\lit{and}}   \condition_2$ requires both $\condition_1$ and $\condition_2$ to hold on the trace $\trace$ at timestamp~$t$.

\dslname is supported by \checkname~\cite{boufaied2020trace}, an automated, model-driven trace-checking tool that verifies whether a property is satisfied or violated by a given trace.
\checkname yields a Boolean verdict: \emph{true} if the property is satisfied, \emph{false} otherwise.
For example, when property $\phi_1$ is checked on the trace shown in \figurename~\ref{fig:traceExample}, \checkname returns the \emph{false} verdict.
However, \checkname does not provide any additional information to help engineers understand the cause of the violation.
Our trace diagnostic approach aims to solve this problem.

\newcommand*{\algrule}[1][\algorithmicindent]{\makebox[#1][l]{\hspace*{.5em}\vrule height .75\baselineskip depth .25\baselineskip}}

\algrenewcommand\algorithmicindent{1em}

\section{Pattern-based trace diagnostic}
\label{sec:approach}
This section describes \NAME, our trace-diagnostic approach.
At the core of the approach, there is the computation of
\emph{violation causes} and \emph{diagnoses}.
A violation cause characterizes one of the possible behaviors of the system that may lead to the property violation.
An example of violation cause for property~$\phi_1$ and the
trace in \figurename~\ref{fig:tracegraphical} is that all the spikes have an amplitude greater than or equal to~\SI{90}{\degree}. 
A diagnosis provides additional information to explain the violation cause.
For example, a diagnosis for the previous violation cause is the amplitude
$A_1=\SI{150}{\degree}$ and the time interval $[\SI{0}{\s},\SI{1.8}{\s}]$ of spike
 \emph{spike1}, that is the closest (among those
contained in the trace) to satisfy
the amplitude constraint of property~$\phi_1$.

Algorithm~\ref{tdApp} shows the main steps of \NAME. The inputs
of  \NAME are a trace $\lambda$ and a property $\phi$ violated by
$\lambda$.
Trace $\lambda$ is a set of consecutive records that contain the values of the variables at different time instants, such as the trace depicted in \figurename~\ref{fig:traceExample}.
 Property $\phi$ is a specification of a requirement in \dslname
defined according to the grammar presented in
\figurename~\ref{tab:syntax}.

The algorithm relies on the following intuition.
Based on the \dslname grammar shown in \figurename~\ref{tab:syntax}, property $\phi$  is specified as a disjunction of clauses.
Since the property is violated,  the disjunction evaluates to false; this
means that all its clauses must be violated. 
To be violated, each clause must contain one or more violated atoms (since a clause is a conjunction of atoms).
Therefore, to explain the violation of property~$\phi$, we return the diagnoses (if available\footnote{As we will discuss later on, it is possible for a violated atom not to have any diagnosis.}) for all the violated atoms of~$\phi$. 
Each diagnosis explains why the corresponding atom is violated.

\begin{algorithm}[t]
\footnotesize
\begin{flushleft}
\textbf{Inputs}. $\trace$: trace \hfill \\
 \hspace{0.8cm} $\phi$: violated property \\
 \textbf{Outputs}. \textit{diags}: set of diagnoses instances
\end{flushleft}
	\caption{\NAME} \label{tdApp}
	\begin{algorithmic}[1]
	\Function{TD-SB-Tempsy}{$\trace$, $\phi$} \label{alg:inputs}
		\State \textit{diags}=$\{\}$; \label{alg:init}
		\State \textit{PropertyAtoms}=\Call{getAtoms}{$\phi$}; \label{alg:getsubproperties}
		\For {$\alpha$ \textbf{in} \textit{PropertyAtoms}} \label{alg:getsubpropertiesloop}
				 \If {\Call{checkAtomOnTrace}{$\trace$,$\alpha$}==\texttt{false}}  \label{alg:subpropviolation}
					 \State \textit{diags}.add(\Call{TD-Atom}{$\trace$,$\alpha$})  \label{alg:diagnosticinf}
				\EndIf
		\EndFor
		\State \textbf{return} \textit{diags};\label{alg:ret}	
		\EndFunction
	\end{algorithmic} 
\end{algorithm}

Algorithm~\ref{tdApp} works as follows. After initializing a set of
diagnoses instances to be returned (line~\ref{alg:init}),
it extracts all the atoms from property~$\phi$ by analyzing its
abstract syntax tree
(line~\ref{alg:getsubproperties}). Then, it iteratively analyzes each
atom (line~\ref{alg:getsubpropertiesloop}).
It first checks, by calling a trace checker for \dslname like \checkname~\cite{boufaied2020trace},
if the atom is violated by the trace
(line~\ref{alg:subpropviolation}). If it is the case, the algorithm computes (through
algorithm \Call{TD-Atom}{} described below) the  diagnosis (if
available) that explains why the atom is violated (line~\ref{alg:diagnosticinf}).
Finally, the algorithm returns the set of the computed diagnoses instances (line~\ref{alg:ret}).
\NAME returns a diagnosis if the set of the computed diagnoses instances is not empty.

\begin{algorithm}[t]
\footnotesize
\begin{flushleft}
\textbf{Inputs}. $\trace$: trace \hfill \\
 \hspace{0.8cm} $\alpha$: violated atom \\
 \textbf{Outputs}. \textit{diag}: diagnosis for $\alpha$ (if available)
\end{flushleft}
	\caption{\NAME\ - Atoms} \label{tdAtom}
	\begin{algorithmic}[1]
		\Function{TD-Atom}{$\trace$,$\alpha$} \label{alg:inputsAtom}
                \State vcs=\Call{getViolationCauses}{$\alpha$}; \label{alg:getdp}
				\For {i=0; i$<$vcs.size(); i++} \label{alg:analyzeddp}
				\If{\Call{checkViolationCause}{$\trace$,vcs[i]}==\texttt{true}}  \label{alg:analyzeddpcheck}
						\State \textbf{return} \Call{getDiagnosis}{$\trace$,vcs[i]}; \label{alg:analyzeddpcheckviolinfo}
					 \EndIf
				\EndFor
			    \State \textbf{return} \texttt{null};\label{alg:null}
		\EndFunction
	\end{algorithmic} 
\end{algorithm}

\emph{Computing the diagnosis.}
Algorithm~\ref{tdAtom} describes how to compute the diagnosis for an atom  of an \dslname property.
The inputs of  \Call{TD-Atom}{} are a trace $\lambda$ and an atom $\alpha$ violated by $\lambda$.
 
To compute the diagnosis for atom~$\alpha$, Algorithm~\ref{tdAtom} extracts the \emph{violation causes} associated with
the atom (line~\ref{alg:getdp}) by calling the auxiliary function
\textsc{getViolationCauses},
which relies on a predefined mapping associating each
type of atom of \dslname with one or more violation causes (see
section~\ref{sec:diag-patterns-def}). 
A violation cause encodes a behavior that may lead
to the violation of an atom.  
For example, a violation cause for the atom of property~$\phi_1$
defined in Section~\ref{sec:motivating} is the following:
\begin{center}
    $\lit{c\_{spike}}_{1}$: \emph{all the spikes in signal $\beta$ violate the amplitude constraint}.
\end{center}
If the behavior captured by this violation cause holds, the atom of formula
$\phi_1$ is violated.
The violation of an atom can be caused by a violation of the scope
used in the atom (or its negation), a violation of the pattern constrained by the scope,
or both. Function \textsc{getViolationCauses} returns a list of
violation causes, sorted such that violation causes of the scope
precede the violation causes of the pattern\footnote{The priority of
  the different violation causes is application-specific. Our current
  implementation is based on the feedback received by the engineers of
  our industrial partners}.
Then, the algorithm loops through
this list of violation causes (line~\ref{alg:analyzeddp}).
The loop body includes a check that determines whether the violation cause holds on the trace
(line~\ref{alg:analyzeddpcheck}); this is achieved through the call of
the auxiliary function \textsc{checkViolationCause}.
If the violation cause holds, the algorithm stops, returning the corresponding diagnosis  (line~\ref{alg:analyzeddpcheckviolinfo})
using the auxiliary function \textsc{getDiagnosis},
which relies on a predefined mapping of
diagnoses for each type of violation cause (see section~\ref{sec:diag-patterns-def}).
A diagnosis is relevant information that enables engineers to understand \emph{why a violation cause holds on a trace}. 
For example, the diagnosis for the violation cause $\lit{c\_{spike}}_{1}$ is the following
\begin{center}
	$\lit{d\_spike}_1$: \emph{the amplitude $\mathrm{a}$ and
          the time interval $[\mathrm{t_1,t_2}]$ of the spike that is
          the closest to satisfy the amplitude constraint}.
      \end{center}
\label{txt:spk-dg}      
The amplitude value of the spike that is the closest to satisfy the amplitude constraint enables engineers to determine how close is the atom to be satisfied.
The time interval $[\mathrm{t_1,t_2}]$ enables engineers to isolate the portion of the trace containing that spike and to inspect the values assumed by the variables within the records included in this portion of the trace.
The amplitude $A_1=\SI{150}{\degree}$ and the time interval $[0, 1.8]$ of \emph{spike1} in \figurename~\ref{fig:tracegraphical} is an instance of the $\lit{d\_spike}_1$ diagnosis for our case study. Diagnoses like $\lit{d\_spike}_1$ help engineers
understand why an atom is violated by a trace. 
If all the violation causes are checked and none of them led to the computation of diagnoses, a \texttt{null} value is returned (line~\ref{alg:null}).

In the following, we present our methodology to define violation causes and diagnoses.

 \section{Methodology for Defining 
Violation Causes and Diagnoses}
\label{sec:methodology}
This section describes our methodology to define violation causes and diagnoses by using \dslname as an example.
Our methodology considers each construct~$\construct$ used to define an atom~$\alpha$ of \dslname and follows three steps: \emph{behavior analysis},
\emph{definition of violation causes}, 
and \emph{definition of diagnoses}.

\subsection{Behavior Analysis} 
\begin{figure}[htb]
 \begin{tikzpicture}[domain=0:6] 
 \begin{axis}[
  width=\columnwidth,
       height=0.6\columnwidth,
    xlabel={\texttt{Timestamp}},
    ylabel={\texttt{Value}},
    xmin=0, xmax=6,
    ymin=0, ymax=250,
    xtick={0,1,2,3,4,5,6},
    ytick={0,50,100,150,200,250},
     ymajorgrids=true,
      xmajorgrids=true,
    grid style=dashed,
      legend pos=north west,
      legend style={legend columns=-1}
]
    \addplot[color=purple,smooth,thick,mark=x,name path=$s_1$,dash pattern=on 1.3pt off 0.5pt on 1.3pt off 0.5pt] plot coordinates {
          (0,2)
          (0.2,153.5)
          (0.9,20.0)
          (1.8,0.5)
          (3.0,80.0)
          (4.9,203.5)
          (5.7,20)
          (6.0,0.5)
      };
            \addplot[color=orange,smooth,thick,mark=x,name path=$s_1$,dash pattern=on 8.5pt off 8.5pt on 8.5pt off 8.5pt] plot coordinates {
          (0,100)
          (0.2,100)
          (0.9,100)
          (1.8,100)
          (3.0,100)
          (4.9,100)
          (5.7,100)
          (6.0,100)
      };

      \addplot[color=green,smooth,thick,mark=x,name path=$s_4$,dash pattern=on 4.5pt off 4.5pt on 4.5pt off 4.5pt] plot coordinates {
          (0,200)
          (0.2,180)
          (0.9,160)
          (1.8,100)
          (3.0,90)
          (4.9,70)
          (5.7,60)
          (6.0,55)
      };
        \addplot[color=blue,smooth,thick,mark=x,name path=$s_3$,dash pattern=on 2.3pt off 2.5pt on 1.3pt off 0.5pt] plot coordinates {
          (0,30)
          (0.2,35)
          (0.9,50)
          (1.8,125)
          (3.0,150)
          (4.9,160)
          (5.7,170)
          (6.0,190)
      };    
\legend{$\beta_1$,$\beta_2$,$\beta_3$,$\beta_4$}
\end{axis}
\end{tikzpicture}
\caption{A trace with signals violating the expression ``\lit{exists}\
  \lit{spike}\ \lit{in}\ $\beta$  \lit{with}\ \lit{width} <
  0.5 \lit{amplitude} < 90''.}
\label{fig:spk}
\end{figure}
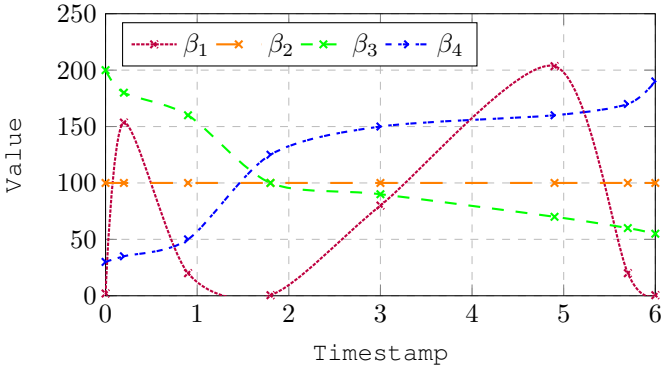

It identifies traces capturing relevant behaviors that violate the semantics of  construct \construct\  as follows.
\begin{enumerate}
    \item It considers an instance of construct $\construct$ obtained
      by selecting some values for its parameters. This instance is a
      concrete example utilized to identify the relevant behaviors that
      violate \construct. For example, for the \lit{spike} construct
      of \dslname (see \figurename~\ref{tab:syntax}), we considered
      the instance ``\lit{exists}\  \lit{spike}\ \lit{in}\ $\beta$
      \lit{with}\ \lit{width}  < 0.5 \lit{amplitude} < 90 '', which sets the parameters  \synt{s},  $\sim_1$,
      \synt{v}\textsubscript{1}, $\sim_2$, and
      \synt{v}\textsubscript{2} to the values $\beta$, ``<'', 0.5, ``<'', and
      90 respectively.
  \item It considers the logical formula $\patternsemantics(\construct)$ describing the semantics of the construct $\construct$.
For example, for the \lit{spike} construct the logical formula describing its semantics is reported in \figurename~\ref{tab:propertypatterns} (on the right side of the \emph{iff} operator).
\item It identifies traces capturing relevant behaviors that violate the instance of construct~$\construct$ (i.e., that make formula~$\patternsemantics(\construct)$ evaluate to \emph{false}).

For example, \figurename~\ref{fig:spk} shows a trace with four signals ($\beta_1$, $\beta_2$, $\beta_3$, and $\beta_4$) that violate the instance we considered for the \lit{spike} construct;  
for instance, signal $\beta_4$ does not contain a strict maximum.
\end{enumerate}

\subsection{Definition of Violation Causes} It characterizes each of the traces identified by the behavior analysis step through a violation cause.
Each violation cause $\diagnosticpattern$ is defined by writing a
logical formula~$\patternsemantics(\diagnosticpattern)$ that specifies
its semantics; the formula~$\patternsemantics(\diagnosticpattern)$ is
true when trace \trace satisfies violation cause~$\diagnosticpattern$.

For example, the shape of signal $\beta_3$ can be defined, with
respect to a trace \trace delimited by timestamps $t_i$ and $t_e$, through the
logical formula:
$$\lit{c\_spike}_4  \equiv \forall t_1 \in [t_l,t_u), (
\forall t_2 \in (t_1,t_u],
(\synt{s}(t_1) \geq \synt{s}(t_2))
).$$
This formula characterizes the behavior
for which a signal \texttt{s} decreases.
More precisely,  $\lit{c\_spike}_4$ holds on a trace $\trace$ if, for any timestamp $t_1$ within $t_l$ and $t_u$,
the value $\synt{s}(t_1)$ of signal $\synt{s}$ at  timestamp $t_1$ is
greater than or equal to the value $\synt{s}(t_2)$ of signal
$\synt{s}$ for  any timestamp $t_2$ that follows $t_1$.

Violation causes for \dslname are illustrated in
section~\ref{sec:diag-patterns-def} (and summarized in \figurename~\ref{tab:diagnosticPatterns}). The formula
$\patternsemantics(\diagnosticpattern)$ characterizing the semantics
of each violation cause is reported on the right side of the
\emph{iff} operator.

Since violation causes should encode root causes leading to property violations, a violation cause $\diagnosticpattern$ for a construct $\construct$ should satisfy the following relation:
\begin{equation*}
\textbf{if } \patternsemantics(\diagnosticpattern)=
\texttt{true}, \textbf{ then }  \patternsemantics(\construct)=\texttt{false}
\end{equation*}
Intuitively,  if violation cause $\diagnosticpattern$ holds on trace \trace, i.e., formula  $\patternsemantics(\diagnosticpattern)$ is true, the trace \trace should violate  construct~$\construct$, i.e., formula  $\patternsemantics(\construct)$ should be false.
To check the satisfaction of this relation, we automatically verified
that the formula $$\Psi \equiv
\patternsemantics(\diagnosticpattern)\Rightarrow \neg
\patternsemantics(\construct)$$
holds.
Formula  $\Psi$ holds if, whenever the violation cause holds ($\patternsemantics(\diagnosticpattern)$ is true), the construct $\construct$  does not hold ($\patternsemantics(\construct)$ is false).
To check if formula $\Psi$ holds, we verified
whether the formula $\neg \Psi$ is unsatisfiable. 
If the  formula $\neg \Psi$ is unsatisfiable,  $\Psi$ always holds.
We used Microsoft Z3~\cite{10.5555/1792734.1792766} ---  an industry-strength tool --- to check if $\neg \Psi$ is unsatisfiable.
For example, Z3 confirmed that the formula $\neg \Psi$ obtained by considering the $\lit{c\_spike}_4$ violation cause and the \lit{spike} construct is unsatisfiable. 
Therefore, whenever violation cause $\lit{c\_spike}_4$ holds, the \lit{spike} construct is violated.

\subsection{Definition of diagnoses} It defines a diagnosis for each of the violation causes.
To define each diagnosis, we (i)~analyzed the semantics of the corresponding violation cause, and (ii)~identified minimum relevant information that enables engineers to understand why a violation cause holds on a trace. 

For example, the diagnosis for violation cause $\lit{c\_spike}_4$, denoted by $\lit{d\_spike}_4$, includes the lowest and the highest values of signal~$s$ within the trace.
These values allow engineers to understand the range of values of signal~$s$ while it exhibits a decreasing behavior. 
For example, for signal $\beta_3$ in \figurename~\ref{fig:spk}, the
diagnosis for violation cause $\lit{c\_spike}_4$ is $\langle \langle
0, 200  \rangle, \langle 6, 55 \rangle \rangle$ showing that signal
$\beta_3$ reaches its maximum value ($200$) at timestamp $0$ and its
minimum value ($55$) at timestamp~$6$. In this particular case, when
the signal does not show any spike, this is some information
required for the engineers to understand why this particular violation
was caused.

The diagnoses associated with the violation causes shown in
\figurename~\ref{tab:diagnosticPatterns} are  illustrated in
section~\ref{sec:diag-patterns-def} (and summarized in Figures~\ref{tab:diagnosticInformationAtomsScopes}--\ref{tab:diagnosticInformationPatterns}).

\subsection{Properties of the methodology} Our methodology provides formal guarantees
of the soundness of the proposed violation causes: if a violation
cause holds on a trace, the corresponding property is violated.

We remark that the methodology relies mostly on manual steps (except for proving, using a solver like Z3, that the proposed violation causes can lead to unsatisfiable properties). Moreover, the methodology cannot guarantee the completeness of the set of violation causes resulting from step ``Definition of violation causes''.

Though we applied our methodology to define violation causes for
properties expressed using \dslname, our methodology is
\emph{language-agnostic}, 
in the sense that \emph{it can be applied to other pattern-based languages}
like TemPsy~\cite{dou2017model},
FRETISH~\cite{giannakopoulou20:_gener_formal_requir_struc_natur_languag}
and, more in general, languages based on specification pattern
catalogues (such as those for robotic missions~\cite{8859226}).
More specifically, this claim is
supported by the fact that, in the definition of the methodology, we
only refer to generic syntactic constructs of the specification
language and generic logical formulae capturing their semantics (which
we assume to be available in formal syntax and semantics definitions),
as well as generic logical formulae of violation causes (which have to
be defined from scratch). 
The restriction to pattern-based languages is due to the fact that we assume the existence of some pattern-based structure in the  syntax of the specification language.
We also remark that the methodology would not work for specification languages in which the Boolean and temporal operators are less constrained (e.g., STL).

For the sake of illustration, in the rest of this subsection, we show how to apply the
methodology to one construct supported by FRETISH.

\subsubsection*{Behavior Analysis}
Let us consider the FRETISH scope construct \lit{only in
  mode}~\cite{giannakopoulou2021automated}; it informally indicates
that a requirement shall only hold within the time interval delimited
by the scope boundaries (also called endpoints). More specifically, it
restricts the satisfaction of that requirement to only the time
interval delimited by the scope boundaries. In other words, this
implies that the requirement shall be violated within the remaining
time interval(s) outside the interval determined by the scope boundaries.  
Let us consider the following requirement 
\begin{center}
RF: \emph{``Only in mode M, signal s shall be less than or equal to 3''.} 
\end{center}
The corresponding FRETISH specification is the following: 
\[
 \begin{array}{ll}
     \phi_{f} \equiv & \lit{only in} \ M \  \lit{mode}\ s \ \lit{shall} \ \lit{satisfy} \ s < 3
  \end{array}
\]
\begin{figure}[!htbp]
\begin{tikzpicture}[domain=0:6] 
 \begin{axis}[
  width=\columnwidth,
       height=0.6\columnwidth,
    xlabel={\texttt{Timestamp}},
    ylabel={\texttt{Value}},
    xmin=0, xmax=7,
    ymin=0, ymax=7,
    xtick={0,1,2,3,4,5,6,7},
    ytick={0,1,2,3,4,5},
     ymajorgrids=true,
      xmajorgrids=true,
    grid style=dashed,
      legend pos=north west,
      legend style={legend columns=-1}
]
   \addplot[color=purple,smooth,thick,mark=x,name path=$s_1$,dash pattern=on 1.3pt off 0.5pt on 1.3pt off 0.5pt] plot coordinates {
          (0,1)
          (1,1.5)
          (2,0.5)
          (3,2)
          (4,0.5)
          (5,4.8)
          (6,4.5)
          (7,3.8)
      };
     \addplot[color=orange,smooth,thick,mark=x,name path=$s_1$,dash pattern=on 1.3pt off 0.5pt on 1.3pt off 0.5pt] plot coordinates {
          (0,4.5)
          (1,4)
          (2,3.5)
          (3,2.7)
          (4,2.5)
          (5,1.5)
          (6,2)
          (7,0.7)
      };
     \draw[] [name path = vertical, dashed, black, thick] (axis cs:0,3) -- (axis cs:7,3) ;
     \draw[] [name path = horizental, dashed, green, thick] (axis cs:2,0) -- (axis cs:2,5) ;
     \draw[] [name path = horizental, dashed, green, thick] (axis cs:4,0) -- (axis cs:4,5) ;
\legend{$\beta_1$,$\beta_2$}
\end{axis}
\end{tikzpicture}
\caption{A trace with signals violating the expression ``\lit{only in} \ M \  \lit{mode}\ s \ \lit{shall} \ \lit{satisfy} \ s < 3".}
\label{figures/fig:fretish}
\end{figure}
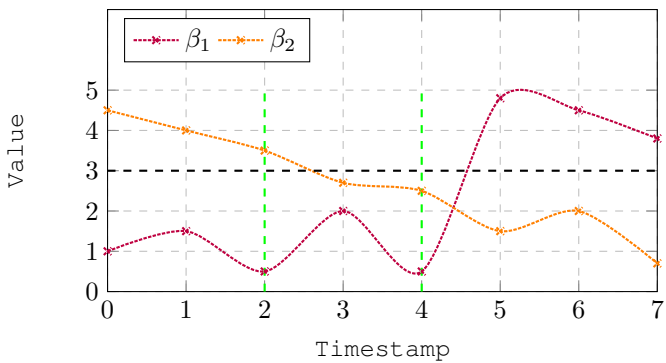

The logical formula describing the formal semantics of a property with
the \lit{only in mode} construct is defined as follows:
\\
$  \trace \models \lit{only in}\ \synt{M} \ \lit{mode}\ \synt{s}(t) \sim \synt{v}  $ \  \emph{iff}$ \ \forall t_1 \in [\mathit{ftp},\mathit{fim}), \synt{s}(t) \not \sim \synt{v} \ \land \ \forall t_2 \in (\mathit{lim}, \mathit{ltp}], \synt{s}(t) \not  \sim \synt{v} $.
Informally, given a trace length delimited by the
$\mathit{ftp}$ and $\mathit{ltp}$ time points where $\mathit{ftp} < \mathit{ltp}$, and a time interval 
$[\mathit{fim},\mathit{lim}]$ in which mode  \synt{M} holds, where $\mathit{fim} \ge \mathit{ftp}$ and $\mathit{lim} \le \mathit{ltp}$, the definition of the semantics of the \lit{only in mode}
construct indicates that the property condition ($s \sim \synt{v}$) shall not be satisfied outside of the scope interval $[\mathit{fim},\mathit{lim}]$.

We identify two possible relevant behaviors that violate an instance
of the \lit{only in mode} construct (i.e., that make formula $\phi_{f}$
evaluate to \emph{false}).

As depicted in~\figurename~\ref{figures/fig:fretish}, let us consider
a trace  defined within the time horizon delimited by the first time
point $\mathit{ftp}$ and the last time point $\mathit{ltp}$ ($0$ and
$7$ in the figure, respectively). \synt{M} holds within the time
interval $[2,4]$ (delimited by the dashed green lines in the figure)  where timestamp $2$ is referred to as $\mathit{fim}$ (first in mode) and timestamp $4$ represents the last timestamp in mode $M$ ($\mathit{lim}$), such that $ \mathit{ftp} \le \mathit{fim} < \mathit{lim} \le \mathit{ltp}$.
For property $\phi_f$ \emph{to be violated}, $s < 3$ shall be
satisfied outside the interval $[\mathit{fim},\mathit{lim}]$; this
means that $s < 3$ is satisfied either within the time interval $[\mathit{ftp},\mathit{fim})$ or within the time interval $(\mathit{lim},\mathit{ltp}]$).
For instance, signal $\beta_1$ in the figure satisfies the condition,
since its value is less than $3$ (i.e., ranging between $1$ and
$1.5$) in the time interval $[0,1]$ before mode \synt{M} holds (i.e., in
the time interval $[2,4]$). Similarly, signal $\beta_2$ in the figure
satisfies the condition, since its value ranges between $0.7$ and $2$ ($1.5$, $2$ and $0.7$)
in the time interval $[5,7]$, which is an interval in which mode \synt{M} does not hold.
\subsubsection*{Definition of Violation Causes}
\label{sec:definition-violation-causes}

In the following, we define and formalize two violation causes that, when satisfied, lead to the violation of property $\phi_f$.

\begin{itemize}
\item \lit{c\_only\_in\_mode}$_1$: Some time before mode \synt{M}
  starts holding, the signal satisfies the property condition. However, the signal
  does not satisfy the constraint after mode \synt{M}  stops holding.
Formally, the violation cause is defined through the following logical formula: \\
$  \trace \models \lit{c\_only\_in\_mode}_1 \ \emph{iff} \ \exists t_1
\in [\mathit{ftp},\mathit{fim}), \synt{s}(t) \sim \synt{v} \ \land \
\forall t_2 \in (\mathit{lim}, \mathit{ltp}], \synt{s}(t) \not  \sim
\synt{v} $.

For instance, signal $\beta_1$ in the figure takes values ranging between $1$ and $1.5$ in the time interval
$[0,1]$ before mode $M$ started holding. However, $\beta_1$ violates the constraint (i.e., $\beta_1 \ge 3$) by taking values ranging between $3.8$ and $4.8$ ($4.8, 4.5$ and $3.8$) within the time interval $[5,7]$, after mode $M$ stopped holding.

\item \lit{c\_only\_in\_mode}$_2$: Some time after mode $M$ stops holding, the
  signal satisfies the property condition. However, the signal does not satisfy the constraint before mode $M$ starts holding.
Formally, the violation cause is defined through the following logical formula: \\
$  \trace \models \lit{c\_only\_in\_mode}_2 \ \emph{iff} \ \exists  t_2 \in (\mathit{lim},\mathit{ltp}], \synt{s}(t) \sim \synt{v} \ \land \ \forall t_1 \in [\mathit{ftp}, \mathit{fim}), \synt{s}(t) \not  \sim \synt{v}$.
For instance, signal $\beta_2$ in the figure takes values ranging between $0.7$ and $2$ ($1.5,2$ and $0.7$) in the time interval $[5,7]$. However, $\beta_2$ violates the constraint (i.e., $\beta_2 \ge 3$) within the time interval $[0,1]$ (showing values $4.5$ and $4$), right before mode $M$ starts holding.
\end{itemize}

Similar to what we did before with \dslname violation causes, we used
Z3 to confirm that the formulae $\neg \Psi$ obtained from the two 
violation causes \lit{c\_only\_in\_mode}$_1$  and 
\lit{c\_only\_in\_mode}$_2$ were unsatisfiable. 

\subsubsection*{Definition of diagnoses}
We propose the following diagnoses related to the satisfaction of the violation causes  \lit{c\_only\_in\_mode}$_1$ and \lit{c\_only\_in\_mode}$_2$:
\begin{itemize}
    \item \lit{d\_only\_in\_mode}$_1$ includes the first record
      (timestamp and the corresponding signal value) in which the
      signal satisfies the property condition, right before mode \synt{M}
      starts holding.
      The choice of this diagnosis is motivated by the fact that we are interested in reporting the root cause of the property violation.
    More formally, we have: \\ 
   $ \lit{d\_only\_in\_mode}_1= \langle t, \synt{s}(t) \rangle \mid t \in [\mathit{ftp},\mathit{fim}), \synt{s}(t) \sim \synt{v} \wedge \forall t_1 \in (\mathit{lim},\mathit{ltp}], \synt{s}(t_1) \not \sim \synt{v} \wedge \forall t_2 \in [\mathit{ftp},t), \synt{s}(t_2) \not \sim \synt{v}$, where $t$ represents the first timestamp in which the signal satisfies the property condition (i.e., $\synt{s}(t) \sim \synt{v}$) before mode $M$ starts holding and $\synt{s}(t)$ denotes the corresponding signal value. \\
For instance, for signal $\beta_1$ in the figure, the
         diagnosis for violation cause $\lit{d\_only\_in\_mode}_1$ is
         $\langle (0,1) \rangle$, showing that the signal first
         satisfies the property condition outside the time interval in
         which mode M holds (i.e., the signal takes value $1$, which is
         less than $3$, at timestamp $0$).
    \item  \lit{d\_only\_in\_mode}$_2$ includes the first record
      in which the
      signal satisfies the property condition, right after mode \synt{M}
      stops holding.
    Formally, the diagnosis is defined as follows: $ \lit{d\_only\_in\_mode}_2= \langle t, \synt{s}(t) \rangle \mid t \in (\mathit{lim},\mathit{ltp}], \synt{s}(t) \sim \synt{v} \ \wedge \ \forall t_1 \in [\mathit{ftp},\mathit{fim}), \synt{s}(t_1) \not \sim \synt{v} \wedge \forall t_2 \in (\mathit{lim},t),\synt{s}(t_2) \not \sim \synt{v}$ where $t$ represents the first timestamp in which the signal satisfies the property condition (i.e., $\synt{s}(t) \sim \synt{v}$) after mode $M$ stops holding and $\synt{s}(t)$ denotes the corresponding signal value. \\
For instance, for signal $\beta_2$ in the figure, the diagnosis for
violation cause $\lit{d\_only\_in\_mode}_2$ is $\langle (5,1.5)
\rangle$, showing that the signal first satisfies the property
condition (i.e., the signal value is less than $3$) at timestamp $5$, taking value $1.5$.
\end{itemize}

\section{Violation Causes and Diagnoses for \dslname}
\label{sec:diag-patterns-def}
In this section, we describe the \emph{violation causes} and the
corresponding \emph{diagnoses} for each \emph{construct} supported by
\dslname. We first provide a high-level overview through
\figurename~\ref{tab:diagnosticPatterns} (for \emph{violation causes})
and Figures~\ref{tab:diagnosticInformationAtomsScopes}--\ref{tab:diagnosticInformationPatterns} (for
\emph{diagnoses}); the remaining subsections discuss in detail the violation
causes and diagnoses for each main construct of \dslname.

We remark that the violation causes and corresponding diagnoses for \dslname have been defined (and validated) together with a group of system and software engineers of our industrial partner, with the goal of maximizing the usefulness of a diagnosis for a certain violation cause.
Overall, we spent 20 hours (over three business days) to define the catalogue of violation causes and diagnoses, following the methodology described in section~\ref{sec:methodology}.

\figurename~\ref{tab:diagnosticPatterns} presents the \emph{violation causes} for the constructs of \dslname that can be used in the definition of an \emph{atom} $\alpha$. 
It is divided into three parts that respectively contain the violation causes for the \dslname atoms, scopes, and patterns.
Each violation cause has a name that identifies the construct of \dslname the violation cause refers to, and an incremental index that distinguishes violation causes that refer to the same construct; for example, 
\lit{c\_becomes}$_1$, \lit{c\_becomes}$_2$, and \lit{c\_becomes}$_3$ are the three violation causes that refer to the \lit{becomes}  construct  of \dslname.
Each violation cause is parameterized with the \emph{same} parameters as the corresponding construct. For example, the parameters of the \lit{c\_becomes}$_1$ violation cause ($\sim$ and $\synt{v}$) are the same as those of the \lit{becomes} construct in \figurename~\ref{tab:syntax}.
For conciseness, in \figurename~\ref{tab:diagnosticPatterns}, we omit the parameters of the violation causes. 

The semantics of each violation cause is the (first-order logic) formula $\patternsemantics(\diagnosticpattern)$ on the right side of the \emph{iff} operator; it is followed by an informal description of the semantics in English.
The semantics of the violation causes specifies the conditions that make the violation causes satisfied by trace \trace.
For example, the semantics of the violation cause \lit{c\_becomes}$_1$ specifies that, for every timestamp $t$, the value $\synt{s}(t)$ does not satisfy $\synt{s}(t) \sim \synt{v}$.
Note that the parameters of the \dslname constructs associated with the violation causes, e.g., the value of~$\synt{v}$, are used to define the semantics of the corresponding violation cause.

\begin{figure*}[htp]
\footnotesize
\begin{tabular}{p{\linewidth} }
\toprule
  $\trace \models \lit{c\_not}_1\ \synt{sc}$  \emph{iff} $\trace \models \synt{sc}$. The atom $\synt{sc}$ is satisfied. \\
\midrule 
  $\trace \models \lit{c\_a\_at}_1$  \emph{iff} $t < t_i \vee
  t_e<t$. The value of $t$ is not within the time interval $[t_i,t_e]$.\\
 $\trace \models \lit{c\_a\_bef}_1$  \emph{iff} $t \le t_i \vee
  t_e<t$. The value of $t$ is not within the time interval $[t_i,t_e]$.  \\
  $\trace \models \lit{c\_a\_aft}_1$  \emph{iff} $t < t_i \vee
  t_e \leq t$. The value of $t$ is not within the time interval $[t_i,t_e]$\\
$\trace \models \lit{c\_a\_bet}_1$ \emph{iff} $n < t_i \vee  t_e<m
  \vee m \leq n$. Either the value of $n$ or $m$ is not within $[t_i,t_e]$,
  or the value of $n$ is not smaller than $m$.\\
 $\trace \models \lit{c\_e\_bef}_1$  \emph{iff} $\exists t_1,t_2, ( 
 t_i<t_1< t_2 \le t_e \wedge
\trace, [t_1,t_2] \models p_1 \wedge 
\forall t_3,t_4, ( t_i \le t_3 < t_4 < t_1 \Rightarrow \trace, [t_3,t_4] \not\models p
 ) ) $.  Pattern $p_1$ holds within  $[t_1,t_2]$. Pattern $p$ is violated before $p_1$.\\ 
$\trace \models \lit{c\_e\_aft}_1$  \emph{iff} $\exists t_1,t_2, ( 
t_i \le t_1 < t_2  < t_e \wedge
\trace, [t_1,t_2] \models p_1 \wedge
\forall t_3,t_4, (t_2 < t_3 < t_4  \leq t_e \Rightarrow \trace, [t_3,t_4] \not\models p
  ) ) $. Pattern $p_1$ holds within $[t_1,t_2]$. Pattern $p$ is violated after $p_1$.\\
 $\trace \models \lit{c\_e\_bet}_1$  \emph{iff} $\exists t_1,t_2,t_3,t_4, ( 
 t_i\leq t_1  < t_2 < t_3  < t_4 \leq t_e \wedge
\trace, [t_1,t_2] \models p_1 \wedge
\trace, [t_3,t_4] \models p_2 \wedge 
\trace, [t_2,t_3] \not\models p
) $. Pattern $p_1$ holds within $[t_1,t_2]$ and pattern $p_2$ holds within $[t_3,t_4]$, but pattern $p$ does not hold between $p_1$ and $p_2$.\\  
\midrule
 $\trace, [t_l,t_u] \models \lit{c\_assert}_1$  \emph{iff}
$\exists t \in [t_l,t_u], \left(\trace,t \not\models \condition
  \right)$. There exists a  timestamp $t$ within $[t_l,t_u]$ in which
  condition  \condition is violated  \\
 $\trace, [t_l,t_u] \models \lit{c\_becomes}_1$  \emph{iff} $\forall t
  \in (t_l,t_u], \big(\synt{s}(t) \not\sim \synt{v} )$ .The signal values violate the pattern
  constraint $\sim \synt{v}$ throughout the time interval, delimited by $t_l$ and $t_u$, over which the pattern is evaluated. \\
  $\trace, [t_l,t_u] \models \lit{c\_becomes}_2$  \emph{iff} $\forall
  t \in (t_l,t_u], \big( \exists t_1 \in [t_l,t), (\synt{s}(t_1) \sim
  \synt{v} ) \big) $. All the signal values observed within the time interval $[t_l,t_u]$ satisfy the pattern constraint $\sim \synt{v}$.\\
    $\trace, [t_l,t_u]\models \lit{c\_becomes}_3$  \emph{iff}   $\exists t \in (t_l,t_u), (
    \forall t_1 \in [t_l,t), (\synt{s}(t_1) \sim \synt{v})  \wedge \forall t_2 \in (t,t_u], 
   (\synt{s}(t_2) \not \sim \synt{v})  
    )
$.  The signal satisfies the semantics of the
  pattern instance in which the constraint  $\sim \synt{v}$ is
  negated (i.e., $ \trace, [t_l,t_u] \models \synt{s}\
  \lit{becomes} \not \sim \synt{v}   $ holds).  \\
 $\trace, [t_l,t_u]\models \lit{c\_spike}_1$ \emph{iff} $
\forall t_1, t_2, t_3, t_4, t_5  \in [t_l,t_u],
(  
		(
		t_1 < t_2 < t_3 < t_4< t_5 \wedge
 					\minf (\synt{s},t_2,[t_1,t_3]) \wedge  
		        \lmaxf(\synt{s},t_3,[t_2,t_4])\wedge 
					\minf (\synt{s},t_4,[t_3,t_5])
		)
		\Rightarrow
		\neg (
		\mathit{amp}(\synt{s},t_1,t_2,t_3) \sim_2 \synt{v}_2
		)
)
		$. All the spike instances violate the amplitude constraint$^\ast$. \\
 $\trace, [t_l,t_u]\models \lit{c\_spike}_2$ \emph{iff} $
\forall t_1, t_2, t_3, t_4, t_5  \in [t_l,t_u],
(  
		(
		t_1 < t_2 < t_3 < t_4< t_5 \wedge
 					\minf (\synt{s},t_2,[t_1,t_3]) \wedge  
					\lmaxf(\synt{s},t_3,[t_2,t_4])\wedge 
					\minf (\synt{s},t_4,[t_3,t_5])
		)
		\Rightarrow
		\neg (\mathit{width}(t_2,t_4) \sim_1 \synt{v}_1)
)
		$. All the spike instances violate the  width constraint$^\ast$.\\
 $\trace, [t_l,t_u]\models \lit{c\_spike}_3$ \emph{iff} $\forall t \in [t_l,t_u], ( \synt{s}(t)=\synt{s}(t_l))$. The signal \synt{s} is constant.\\
 $\trace, [t_l,t_u] \models \lit{c\_spike}_4$ \emph{iff} $\forall t_1 \in [t_l,t_u), (
\forall t_2 \in (t_1,t_u],
(\synt{s}(t_1) \geq \synt{s}(t_2))
)$. The signal \synt{s} decreases.\\
$\trace, [t_l,t_u] \models \lit{c\_spike}_5$ \emph{iff} $\forall t_1 \in [t_l,t_u), (
\forall t_2 \in (t_1,t_u], 
(\synt{s}(t_1) \leq \synt{s}(t_2))
)$. The signal \synt{s}  increases.\\
$\trace, [t_l,t_u] \models \lit{c\_oscillation}_1$ \emph{iff} $
\forall t_1, t_2, t_3, t_4, t_5 \in [t_l,t_u],
(  (
	t_1 < t_2 < t_3 < t_4 < t_5 \wedge
 				\lminf (\synt{s},t_2,[t_1,t_3]) \wedge  
					\lmaxf(\synt{s},t_3,[t_2,t_4])\wedge 
					\lminf (\synt{s},t_4,[t_3,t_5])
			)		
		\Rightarrow
		\neg (
 \mathit{p2p}(\synt{s}, t_2,t_3) \sim_2  \synt{v}_2
		) \wedge \neg (
 \mathit{p2p}(\synt{s}, t_3,t_4)  \sim_2  \synt{v}_2
		)
)
		$. All the oscillation instances violate the  amplitude constraint.\\
$\trace, [t_l,t_u] \models \lit{c\_oscillation}_2$ \emph{iff} $\forall t_1, t_2, t_3, t_4, t_5 \in [t_l,t_u],
(  (t_1 < t_2 < t_3 < t_4 < t_5 \wedge	
 					\lminf (\synt{s},t_2,[t_1,t_3]) \wedge  
					\lmaxf(\synt{s},t_3,[t_2,t_4])\wedge 
					\lminf (\synt{s},t_4,[t_3,t_5])
		)
		\Rightarrow
		\neg (
		 \mathit{width}(t_2,t_4)  \sim_1 \synt{v}_1 
		)
)
		$. All the oscillation instances violate the period
  constraint.\\
$\trace, [t_l,t_u] \models \lit{c\_oscillation}_3$ \emph{iff} $\exists t_1, t_2,t_3 \in [t_l,t_u], (
			t_1 < t_2 < t_3 \wedge
	\mathit{ext}(\synt{s},t_2,[t_1,t_3])
		\wedge 
		\forall t_4,t_5,t_6 \in [t_l,t_u],
		(
			(t_5 \neq t_2 \wedge t_4<t_5<t_6) \Rightarrow \neg 
	\mathit{ext}(\synt{s},t_5,[t_4,t_6])
	)
	)
$. The signal \synt{s}  contains  only one strict local extremum (minimum or maximum).\\
$\trace, [t_l,t_u] \models \lit{c\_oscillation}_4$ \emph{iff} $
\exists t_1, t_2,t_3 \in [t_l,t_u], (
			t_1 < t_2 < t_3 \wedge	\mathit{ext}(\synt{s},t_2,[t_1,t_3]) 
		\wedge 
		\exists t_4,t_5,t_6 \in [t_l,t_u],
		( 
		t_5 \neq t_2 \wedge t_4<t_5<t_6 \wedge \mathit{ext}(\synt{s},t_5,[t_4,t_6]) 
		\wedge
		\forall t_7,t_8,t_9 \in [t_l,t_u],
		(
			t_2 \neq t_8 \neq t_5 \wedge t_7<t_8<t_9
			\wedge \neg \mathit{ext}(\synt{s},t_8,[t_7,t_9])
		)))
$. The signal \synt{s} shows only two local extrema.\\
$\trace, [t_l,t_u] \models \lit{c\_oscillation}_5$ \emph{iff} $\forall t \in [t_l,t_u], ( \synt{s}(t)=\synt{s}(t_l))$. The signal \synt{s} is constant. \\
$\trace, [t_l,t_u] \models \lit{c\_oscillation}_6$ \emph{iff}
$\forall t_1 \in [t_l,t_u), (
\forall t_2 \in (t_1,t_u],
(\synt{s}(t_1) \geq \synt{s}(t_2))
)$. The signal \synt{s}  decreases. \\
$\trace, [t_l,t_u] \models \lit{c\_oscillation}_7$ \emph{iff} $\forall t_1 \in [t_l,t_u), (
\forall t_2 \in (t_1,t_u],
(\synt{s}(t_1) \leq \synt{s}(t_2))$. The signal \synt{s}   increases.\\
$\trace, [t_l,t_u] \models \lit{c\_rises}_1$ \emph{iff} $\forall t \in [t_l,t_u], (\synt{s}(t)<\synt{v} )$. The signal value is always below \synt{v}.\\
$\trace, [t_l,t_u] \models \lit{c\_rises}_2$ \emph{iff} $\forall t \in [t_l,t_u], (\synt{s}(t)\geq \synt{v})$. The signal value is always greater than or equal to \synt{v}.\\
$\trace, [t_l,t_u] \models \lit{c\_rises}_3$ \emph{iff} $\exists t \in (t_l,t_u], (\synt{s}(t) \geq \synt{v}
\wedge 
\forall t_1 \in [t_l,t), (\synt{s}(t_1)<v)
\wedge
	\neg (
		mon(\synt{s},t_l,t)
    	)
	)
)
$. The signal rises at timestamp $t$, reaching value $v$. However, it violates the monotonicity constraint defined in the pattern.\\
$\trace, [t_l,t_u] \models \lit{c\_rises}_4$ \emph{iff} $\exists t \in (t_l,t_u), (
    \forall t_1 \in [t_l,t), (\synt{s}(t_1) \geq \synt{v})  \wedge
    \forall t_2 \in[t,t_u], (\synt{s}(t_2)<\synt{v} ) 
    )
$.  The value of the signal is initially above the threshold value \synt{v}. The signal then drops and remains below that value. \\
$\trace, [t_l,t_u] \models \lit{c\_overshoots}_1$ \emph{iff} $\forall t \in [t_l,t_u], (\synt{s}(t) <\synt{v}_1)$. The signal \synt{s} is always below $\synt{v}_1$.\\
$\trace, [t_l,t_u] \models \lit{c\_overshoots}_2$ \emph{iff} $\exists t \in [t_l,t_u], ( \synt{s}(t)>\synt{v}_1+\synt{v}_2	\wedge
        \forall t_1 \in (t,t_u], (\synt{s}(t_1)>\synt{v}_1+\synt{v}_2)
  )$. The signal \synt{s} exceeds (and remains above) the value $\synt{v}_1+\synt{v}_2$.\\
$\trace, [t_l,t_u] \models \lit{c\_overshoots}_3$ \emph{iff} $\exists t \in (t_l, t_u],
	(	\synt{s}(t) \geq \synt{v}_1 \wedge \synt{s}(t) \leq \synt{v}_1+\synt{v}_2 
		\wedge
		\forall t_2 \in [t_l,t), (\synt{s}(t_2) \leq  \synt{v}_1 )
		\wedge
		\forall t_1 \in (t,t_u], (\synt{s}(t_1) \leq  \synt{v}_1+\synt{v}_2 )
	    \wedge
    	\neg(
    		mon(\synt{s},t_l,t)
    		)
    	)
    )
$.  The signal overshoots value $\synt{v}_1$, without exceeding the maximum threshold set to $\synt{v}_1$+ $\synt{v}_2$, but it violates the monotonicity constraint.\\
    $\trace, [t_l,t_u] \models \lit{c\_overshoots}_4$ \emph{iff} $\exists t \in (t_l,t_u], (
                \forall t_1 \in [t_l,t], (\synt{s}(t_1) \geq \synt{v}_1 \wedge \synt{s}(t_1) \leq \synt{v}_1+\synt{v}_2) \wedge
                \forall t_2 \in (t,t_u], (\synt{s}(t_2)<\synt{v}_1)
            )
    )$.  The signal undershoots, going below $\synt{v}_1$ after timestamp $t$, and remains below that value instead of overshooting.\\
$\trace, [t_l,t_u] \models \lit{c\_if-then}_1$ \emph{iff}
 $
 \exists t_1,t_2 \in [t_l, t_u),
(
 \trace, [t_1,t_2] \models \synt{p}_1
\wedge
(\forall t_3,t_4 \in [t_2, t_u],
(
	  \trace, [t_3,t_4] \not \models \synt{p}_2
)
)
)
$. Pattern $\synt{p}_1$ holds within the time interval $[t_1,t_2]$. Pattern $\synt{p}_2$ never holds after the satisfaction of pattern $\synt{p}_1$, until the end of the time interval, right-bounded by value $t_u$.\\
$\trace, [t_l,t_u] \models \lit{c\_if-then}_2$ \emph{iff} $
\exists t_1,t_2 \in [t_l,t_u), (
	\trace, [t_1,t_2] \models \synt{p}_1
	\wedge
		\forall t_3, t_4 \in [t_2,t_u], 
		( \trace, [t_3,t_4] \models \synt{p}_2 \Rightarrow \neg  ((t_3-t_2)
 \left\llbracket\bowtie\right\rrbracket\synt{d}) ))
$  where $\llbracket \bowtie \rrbracket$ is such that $\llbracket \lit{exactly}\rrbracket \equiv \text{`='}$,
    $\llbracket \lit{at most}\rrbracket \equiv \text{`\textless ='}$,
    $\llbracket \lit{at least}\rrbracket \equiv  \text{`\textgreater ='}$. Pattern $\synt{p}_1$ is  satisfied within the time interval $[t_1,t_2]$.
 Any time interval $[t_3,t_4]$ satisfying pattern $\synt{p}_2$ violates the time distance constraint on the size of $t_3-t_2$.\\
\midrule
\end{tabular}
\begin{tabular}{p{0.5cm} p{16.7cm}}
\multicolumn{2}{p{16.9cm}}{$^\ast$ We present the case  where  a (strict) minimum is followed by a strict maximum followed by
a (strict) minimum. The dual case can be derived from our formulation. \newline
$\mathit{ext}(\synt{s},t_2,[t_1,t_3])=(\lmaxf(\synt{s},t_2,[t_1,t_3])
  \vee \lminf(\synt{s},t_2,[t_1,t_3]))$; \quad
  $\mathit{p2p}(\synt{s},t_1,t_2)=
\lvert \synt{s}(t_1)-\synt{s}(t_2)) \rvert$;\newline
$\mathit{amp}(\synt{s},t_1,t_2,t_3) = \max\left(
\lvert \synt{s}(t_2)-\synt{s}(t_1) \rvert,
\lvert \synt{s}(t_2)-\synt{s}(t_3) \rvert
  \right)$;  \quad
  $\mathit{width}(t_1,t_2) = (\lvert t_{2}-t_{1}\rvert)$ 
}
\end{tabular}

 \caption{Violation causes for the constructs of \dslname}
\label{tab:diagnosticPatterns}
\end{figure*}

Figures~\ref{tab:diagnosticInformationAtomsScopes}--\ref{tab:diagnosticInformationPatterns} present the \emph{diagnoses} for the \emph{violation causes} in \figurename~\ref{tab:diagnosticPatterns}. Figure~\ref{tab:diagnosticInformationAtomsScopes} contains the diagnoses related to the violation causes for \dslname atoms and scopes,  while \figurename~\ref{tab:diagnosticInformationPatterns} contains the diagnoses related to violation causes for patterns.

The name of the diagnosis is obtained by replacing the string ``$\lit{c\_}$'' with ``$\lit{d\_}$'' from the name of the corresponding violation cause.
For example, diagnosis $\lit{d\_becomes}_1$ refers to violation cause \lit{c\_becomes}$_1$.

The formal definition of the diagnosis is reported on the right side of the symbol ``=''.
For example, the definition of the diagnosis $\lit{d\_becomes}_1$ is
the tuple containing the maximum and the minimum values (as well as
their timestamps) of signal\footnote{To minimize cluttering, hereafter we omit to indicate that each signal value is associated with a signal name.} \synt{s}.
Violation causes sharing the same diagnosis are separated by the symbol~``$/$''.
For example, $\lit{d\_a\_at}_1$/$\lit{d\_a\_bef}_1$/ $\lit{d\_a\_aft}_1$ is the diagnosis associated with violation causes $\lit{c\_a\_at}_1$, $\lit{c\_a\_bef}_1$, and $\lit{c\_a\_aft}_1$.
The \emph{informal definition} provides a high-level description of the diagnosis.
\subsection*{Implementation}

We implemented \NAME as an OCL~\cite{iso19507ocl} plugin for  \checkname~\cite{boufaied2020trace}.
The plugin
contains the definitions of OCL constraints that encode the violation causes
(see \figurename~\ref{tab:diagnosticPatterns}) as well as OCL functions that compute the diagnoses (see
Figures \ref{tab:diagnosticInformationAtomsScopes}--\ref{tab:diagnosticInformationPatterns}) associated with the
violation causes. The full OCL encoding is available at  \url{https://figshare.com/s/50f355f84a28fcbcc153}.

\begin{figure*}[htp]
\footnotesize
\begin{tabular}{p{\linewidth}}
\toprule
$\lit{d\_not\_assert}$=$\langle t, \synt{s}_1(t), \synt{s}_2(t), \dots , \synt{s}_n(t) \rangle \mid  \left(\trace,t \models \condition \right)$. One timestamp and all the corresponding signal values where condition $c$ is satisfied.\\
$\lit{d\_not\_becomes}$=$ \langle t, \synt{s}(t) \rangle \mid  t \in (t_l,t_u], 
 (\synt{s}(t) \sim  \synt{v}
 \wedge  \forall t_1 \in [t_l,t),  (\synt{s}(t_1) \not \sim  \synt{v}) )$. The first record that satisfies  $\synt{s}(t)\sim \synt{v}$, such that $\synt{s}(t_1) \not \sim \synt{v}$ for any time $t_1$ before $t$.\\
 $ \lit{d\_not\_spike} =\langle ( t_1,\synt{s}(t_1) ), ( t_5,\synt{s}(t_5) ) \mid \exists t_2,t_3, t_4 \in [t_l,t_u],   (
t_l < t_1 < t_2 < t_3 < t_4 < t_5 \wedge
  \maxf (\synt{s},t_2,[t_1,t_3] ) \wedge \lminf (\synt{s},t_3,[t_2,t_4] )  \wedge
 \maxf(\synt{s},t_4,[t_3,t_5])
[[\wedge ( t_3-t_1) \sim_1
\synt{v}_1]_\beta
[\wedge \max(
(
\synt{s}(t_2)-\synt{s}(t_3)
),
( \synt{s}(t_4)-\synt{s}(t_3))
)\sim_2 \synt{v}_2 ]_\gamma]_\alpha )$. The first 
$\langle t_1,\synt{s}(t_1) \rangle$ 
and the last $\langle t_5,\synt{s}(t_5) \rangle$ records that show an occurrence of a spike$^\ast$.\\
$ \lit{d\_not\_oscillation} = \langle ( t_1,\synt{s}(t_1) ), ( t_5,\synt{s}(t_5) )  \mid 
\exists t_2,t_3, t_4 \in [t_l,t_u],
(
t_1 < t_2 < t_2 < t_3 < t_4 < t_5 \wedge
\lmaxf(\synt{s},t_2,[t_1,t_3]) \wedge \lminf(\synt{s},t_3,[t_2,t_4]) \wedge
 \lmaxf(\synt{s},t_4,[t_3,t_5])[
[\wedge (t_4-t_2) \sim_1
      \synt{v}_1]_\zeta 
[
\wedge
( \synt{s}(t_2)-\synt{s}(t_3)) \sim_2 \synt{v}_2
\wedge
( \synt{s}(t_4)-\synt{s}(t_3))
 \sim_2 \synt{v}_2]_\epsilon ]_\delta
 )$. The first 
$\langle t_1,\synt{s}(t_1) \rangle$ 
and the last $\langle t_5,\synt{s}(t_5) \rangle$ records that show an occurrence of  oscillations$^\ast$. 
\\
$\lit{d\_not\_rises}= \langle t, \synt{s}(t)\rangle \mid t \in (t_l, t_u], ( \synt{s}(t) \ge \synt{v} \wedge \forall t_1 \in [t_l,t), (
 \synt{s}(t_1) < \synt{v} )
  [\wedge \emph{mon}(\synt{s},t_l,t)]_\alpha )$. The first record $\langle t, \synt{s}(t)\rangle$ at which the signal becomes greater than or equal to $\synt{v}$, where the optional monotonicity constraint is satisfied, if defined in the property.
\\ 
$\lit{d\_not\_overshoots} = \langle t, \synt{s}(t) \rangle \mid t \in (t_l, t_u],  ( \synt{s}(t) \geq \synt{v}_1 \wedge \forall t_1 \in [t,t_u], ( \synt{s}(t_1) \leq \synt{v}_1+ \synt{v}_2 ) 
\wedge \forall t_2 \in [t_l,t), (\synt{s}(t_2) < \synt{v}_1 )  [\wedge \emph{mon}(\synt{s},t_l,t)]_\alpha  ) $. The first record at which signal \synt{s} reaches  value ~$\synt{v}_1$. The signal never goes above the maximum allowed amplitude of $\synt{v}_1+\synt{v}_2$ and satisfies the monotonicity constraint, if defined in the property.
\\
$\lit{d\_not\_if-then}=\langle [t_{1}, t_{2}], [t_3,t_4] \rangle \mid  (t_l<t_1 < t_2 <t_3< t_4<t_u \wedge \trace, [t_{1}, t_{2}]  \models \synt{p}_1  \wedge \trace, [t_3,t_4] \models \synt{p}_2 [\wedge (t_3-t_2)
 \llbracket\bowtie\rrbracket\synt{d}]_\alpha)$. 
 An interval $[t_1,t_2]$ where pattern $\synt{p}_1$ holds and a subsequent interval $[t_3,t_4]$ where pattern $\synt{p}_2$ holds.\\
\midrule
$\lit{d\_a\_at}_1$/$\lit{d\_a\_bef}_1$/$\lit{d\_a\_aft}_1=\langle
  [t_i,t_e], t \rangle$. The time interval $[t_i, t_e]$ and the absolute boundary $t$, that is not within that interval. \\
  $\lit{d\_a\_bet}_1=\langle [t_i,t_e], n,m \rangle$. Values $n$ and $m$ and the interval $[t_i,t_e]$.\\
$\lit{d\_e\_bef}_1=\langle [t_1,t_2] \rangle \mid ( 
 t_l<t_1< t_2 \le t_u \wedge
\trace, [t_1,t_2] \models p_1 
\wedge 
\forall t_3,t_4, (  t_l \le t_3   < t_4 < t_1 \Rightarrow \trace, [t_3,t_4] \not\models p
 ) ) $. The interval $[t_1,t_2]$ where $p_1$ holds, and before which the property pattern $p$ failed to hold. \\
 $\lit{d\_e\_aft}_1=\langle [t_1,t_2] \rangle \mid ( 
t_l \le t_1 < t_2  < t_u \wedge
\trace, [t_1,t_2] \models p_1 
\wedge
\forall t_3,t_4, (t_2 < t_3 < t_4  \leq t_u \Rightarrow \trace, [t_3,t_4] \not\models p
 ) ) $. The interval $[t_1,t_2]$, where $p_1$ holds and after which the property pattern $p$ failed to hold.\\
 $\lit{d\_e\_bet}_1=\langle [t_2,t_3] \rangle \mid ( 
\exists t_1 \in [t_l,t_u), 
 t_l\leq t_1  < t_2 < t_3  < t_u \wedge
\trace, [t_1,t_2] \models p_1 \wedge
\exists t_4 \in (t_3,t_u], 
t_3 < t_4 \leq t_u \wedge 
\trace, [t_3,t_4] \models p_2 \wedge 
\trace, [t_2,t_3] \not\models p
) $. The time interval $[t_2,t_3]$, where $t_2$ is the last timestamp in which pattern $p_1$ held  and $t_3$ is the first timestamp in which pattern $p_2$ held.\\
\bottomrule

\end{tabular}

 \caption{Diagnoses associated with the violation causes of atoms and scopes in \figurename~\ref{tab:diagnosticPatterns}.}
\label{tab:diagnosticInformationAtomsScopes}
\end{figure*}

\begin{figure*}[htp]
\footnotesize
\begin{tabular}{p{\linewidth}}
\toprule
$\lit{d\_assert}_1=\langle t, \synt{s}_1(t), \synt{s}_2(t), \dots,
  \synt{s}_n(t) \rangle \mid  \left(\trace,t \not\models \condition
  \right) \wedge \forall t_1 \in [t_l,t), \left(\trace,t_1 \models \condition
  \right) $. The first timestamp $t$ and the values, taken in correspondence of
  $t$, of the signals that lead to the violation of condition $c$.\\
$\lit{d\_becomes}_1 / \lit{d\_becomes}_2=\left\langle \left(t_1,  \synt{s}(t_1)\right), \left(t_2, \synt{s}(t_2)\right) \right\rangle  \mid 
		 \forall t \in [t_l,t_u], ( \synt{s}(t) \leq \synt{s}(t_1)  \wedge  
		 \synt{s}(t) \geq \synt{s}(t_2) )$. The maximum and the minimum  values (and the corresponding timestamps) of signal $\synt{s}$.\\ 
$\lit{d\_becomes}_3=\langle \left(t_1,  \synt{s}(t_1)\right),
  \left(t_2, \synt{s}(t_2)\right) \rangle \mid   t_l \le t_1 < t_2 \le t_u  \wedge \synt{s}(t_1) \sim  \synt{v} \wedge \synt{s}(t_2) \not \sim  \synt{v} \wedge
		\nonumber \neg \exists t_3 \in [t_{l},t_{u}], (t_1<t_3<t_2)$. The last time instant $t_1$ (and the corresponding value) at which the signal \synt{s} satisfies the predicate  $\synt{s}(t_1) \sim \synt{v}$, exactly followed by the next time instant $t_2$ (and the corresponding value) at which
  the signal value satisfies the predicate  $\synt{s}(t_2) \not \sim \synt{v}$.
\\
$\lit{d\_spike}_1=\langle [t_1,t_2],a \rangle \mid 
 (
 	\exists t_3,t_4,t_5 \in [t_l,t_u], 
 	(
 			\mathit{spk}(\synt{s},t_3,t_1,t_4,t_2,t_5)
			\wedge \neg
  (\mathit{amp}(\synt{s},t_1,t_4,t_2)\sim_2 \synt{v}_2)
			 \wedge a=\mathit{amp}(\synt{s},t_1,t_4,t_2)
			\wedge 
			\forall 	t_6, t_7, t_8, t_9, t_{10} \in [t_l,  t_u],
			(	
					(	
						t_7 \neq t_1 \wedge t_8 \neq t_4 \wedge t_9 \neq t_2 \wedge
						\mathit{spk}(\synt{s},t_6,t_7,t_8,t_9,t_{10})
					)
					\Rightarrow 
\lvert a - \synt{v}_2 \rvert  <
						\mathit{ampv}(\synt{s},t_7,t_8,t_9,\synt{v}_2)
			)				
 	)		
)$. The amplitude $a$ and the interval $[t_1,t_2]$ of the spike that
  is the closest to satisfy the amplitude constraint.  \\
$\lit{d\_spike}_2=	\langle [t_1,t_2],w \rangle \mid 
 (
 	\exists t_3,t_4,t_5 \in [t_l,t_u], 
 	(
 			\mathit{spk}(\synt{s},t_3,t_1,t_4,t_2,t_5)
			\wedge \neg (\mathit{width}(t_1,t_2) \sim_1 \synt{v}_1) 
			\wedge w=\mathit{width}(t_1,t_2)
			\wedge 
			\forall 	t_6, t_7, t_8, t_9, t_{10} \in [t_l,  t_u],
			(	
					(	
						t_7 \neq t_1 \wedge t_8 \neq t_4 \wedge t_9 \neq t_2 \wedge
						\mathit{spk}(\synt{s},t_6,t_7,t_8,t_9,t_{10})
					)
					\Rightarrow 
\lvert w - \synt{v}_1 \rvert <
						\mathit{widthv}(t_7,t_9,\synt{v}_1)
			)				
 	)		
)$. The width $w$ and the time interval $[t_1,t_2]$ of the spike that
  is the closest to satisfy the 
width constraint. 
\\
$\lit{d\_spike}_3= \langle [t_l,t_u], s(t_l) \rangle$. The first and
  the last timestamps ($t_l$ and $t_u$) delimiting the interval
  throughout which signal \synt{s} is constant, and the signal value. \\
 $\lit{d\_spike}_4$/$\lit{d\_spike}_5= \langle 
 			  ( t_1, \synt{s}(t_1) ), ( t_2, \synt{s}(t_2) ) \rangle  \mid 
		 \forall t \in [t_l,t_u], ( \synt{s}(t) \leq \synt{s}(t_1)  \wedge  
		 \synt{s}(t) \geq \synt{s}(t_2) )$. The maximum and the minimum values (and their timestamps) taken by signal \synt{s}. 
 \\
 $\lit{d\_oscillation}_1=\langle [t_1,t_5],a \rangle \mid 
 (
 	\exists t_2,t_3,t_4 \in [t_l,t_u], 
 	(
 			\mathit{osc}(\synt{s},t_1,t_2,t_3,t_4,t_5)
			\wedge \neg
(\mathit{p2p}(\synt{s},t_2,t_3) \sim_2 \synt{v}_2 \vee \mathit{p2p}(\synt{s},t_3,t_4) \sim_2 \synt{v}_2 )
			\wedge 
			a= \max(\mathit{p2p}(\synt{s},t_2,t_3),\mathit{p2p}(\synt{s},t_3,t_4)) \wedge
			\forall 	t_6, t_7, t_8, t_9, t_{10} \in [t_l, t_u],
			(	
					(	
						t_8 \neq t_2 \wedge t_9 \neq t_3 \wedge t_{10} \neq t_4 \wedge
						\mathit{osc}(\synt{s},t_6,t_8, t_9, t_{10}, t_7)
					)
					\Rightarrow 
					(
\lvert a - \synt{v}_2 \rvert \leq
					   \mathit{p2pv}(\synt{s},t_8,t_9,\synt{v}_2) \wedge 
					   \lvert a - \synt{v}_2 \rvert \leq
					   \mathit{p2pv}(\synt{s},t_9,t_{10},\synt{v}_2) 
					)
			)				
 	)		
)$. The amplitude $a$ and the time interval $[t_1,t_5]$ of the closest
  oscillation instance to satisfy the amplitude constraint. 
\\
 $\lit{d\_oscillation}_2=\langle [t_1,t_5],w \rangle \mid 
 (
 	\exists t_2,t_3,t_4 \in [t_l,t_u], 
 	(
 			\mathit{osc}(\synt{s},t_1,t_2,t_3,t_4,t_5)
			\wedge \neg (
\mathit{width}(t_2,t_4) \sim_1  \synt{v}_1
			)
			\wedge w=width(t_2,t_4) \wedge
			\forall 	t_6, t_7, t_8, t_9, t_{10} \in [t_l, t_u],
			(	
					(	
						t_8 \neq t_2 \wedge t_9 \neq t_3 \wedge t_{10} \neq t_4 \wedge
						\mathit{osc}(\synt{s},t_6,t_8, t_9, t_{10}, t_7)
					)
					\Rightarrow 
					(\lvert w - \synt{v}_1 \rvert <
					      \ widthv(t_{8},t_{10},\synt{v}_1)
					)
			)				
 	)		
)$. The period $w$ and the interval $[t_1,t_2]$ of the closest
  oscillations instance to satisfy the period constraint.\\

$\lit{d\_oscillation}_3=	\langle t_1,\synt{s}(t_1)\rangle \mid ~
\exists t_2,t_3 \in [t_l,t_u],
t_l \le t_2 < t_1 ~\wedge~ t_1 < t_3 \le t_u \wedge 
(\lminf(\synt{s},t_1,[t_2,t_3]) \vee \lmaxf(\synt{s},t_1,[t_2,t_3])
)$. The record 
at which the only seen strict extremum occurs in the signal, within the time interval $[t_l,t_u]$. 
\\ 
$\lit{d\_oscillation}_4=\langle ( t_1,\synt{s}(t_1) ), ( t_4,\synt{s}(t_4) ) \rangle \mid ~
\exists t_2,t_3 \in [t_l,t_u],
t_l \le t_2 < t_1 \wedge t_1 < t_4 \wedge t_4 < t_3 \le t_u \wedge
(\lminf(\synt{s},t_1,[t_2,t_4]) \vee \lmaxf(\synt{s},t_1,[t_2,t_4]))  \wedge
(\lminf(\synt{s},t_4,[t_1,t_3]) \vee \lmaxf(\synt{s},t_4,[t_1,t_3])) 
  \wedge t_4 \neq t_1 $. The two records 
at which the strict maximum and the strict minimum occur in the signal, within the time interval $[t_l,t_u]$. 
\\ 
$\lit{d\_oscillation}_5=\langle [t_l,t_u], \synt{s}(t_l) \rangle$. The first and
  the last timestamps ($t_l$ and $t_u$) delimiting the interval
  throughout which signal \synt{s} is constant, and the signal value. \\
$\lit{d\_oscillation}_6$/$\lit{d\_oscillation}_7=  \langle 
 			  ( t_1, \synt{s}(t_1) ), ( t_2, \synt{s}(t_2) ) \rangle  \mid 
		 \forall t \in [t_l,t_u], ( \synt{s}(t) \leq \synt{s}(t_1)  \wedge  
		 \synt{s}(t) \geq \synt{s}(t_2) )$.
The maximum and the minimum values (and their timestamps) taken by the signal $s$. \\	
$\lit{d\_rises}_1/\lit{d\_rises}_2=\langle (
t_1,  \synt{s}(t_1) ), ( t_2, \synt{s}(t_2)) \rangle \mid 
		 \forall t \in [t_l,t_u], ( \synt{s}(t) \leq \synt{s}(t_1)  \wedge  
		 \synt{s}(t) \geq \synt{s}(t_2) )$.  The  maximum and the minimum  values (and their timestamps) of signal $\synt{s}$.\\
$\lit{d\_rises}_3=\langle ( t_{1},  \synt{s}(t_{1}) ), ( t_{2}, \synt{s}(t_{2}) ) \rangle \mid 
\neg (\exists t \in [t_{l},t_{u}], (t_1<t<t_2))
\wedge
\exists t \in (t_l,t_u], (\synt{s}(t) \geq \synt{v}
\wedge 
\forall t_3 \in [t_l,t), (\synt{s}(t3)<v)
\wedge  t_1<t_2<t 
\wedge \synt{s}(t_1)>\synt{s}(t_2)
)
$. Two signal values that violate the monotonicity constraint and the corresponding consecutive timestamps $t_1$ and $t_2$.\\
$\lit{d\_rises}_4= \langle (
t_1,  \synt{s}(t_1) ), ( t_2, \synt{s}(t_2) ) \rangle \mid  t_l \le t_1 < t_2 \le t_u  \wedge \synt{s}(t_1) \ge  \synt{v} \wedge \synt{s}(t_2) <  \synt{v} \wedge
		\nonumber \neg \exists t \in [t_{l},t_{u}], (t_1<t<t_2) $. The record 
at which the signal \synt{s} is greater than or equal to value $\synt{v}$, followed by the record at which the signal falls, going below $\synt{v}$. \\
$\lit{d\_overshoots}_1/\lit{d\_overshoots}_2=\langle (
t_1,  \synt{s}(t_1) ), ( t_2, \synt{s}(t_2\rangle) )  \mid 
		 \forall t \in [t_l,t_u], ( \synt{s}(t) \leq \synt{s}(t_1)  \wedge  
		 \synt{s}(t) \geq \synt{s}(t_2) )$.  The  maximum and the minimum  values (and timestamps) of signal~$\synt{s}$.\\
$\lit{d\_overshoots}_3=\langle  t_{1}, \synt{s}(t_{1}),  t_{2}, \synt{s}(t_{2}) \rangle \mid 
\neg (\exists t \in [t_{l},t_{u}], (t_1<t<t_2))
\wedge
\exists t \in (t_l, t_u],
	(	\synt{s}(t) \geq \synt{v}_1
		\wedge
		\forall t_4 \in [t_l,t), (\synt{s}(t_4) \leq  \synt{v}_1 )
		\wedge
		\forall t_5 \in [t,t_u], (\synt{s}(t_5) \leq  \synt{v}_1+\synt{v}_2 )
        \wedge 
        (t_l<t_1<t_2<t)
\wedge
        \synt{s}(t_{1}) \geq \synt{s}(t_{2})
    )
$. Two consecutive records of a signal that overshoots, but does not satisfy the monotonicity constraint.\\
$\lit{d\_overshoots}_4= \langle (
t_1,  \synt{s}(t_1) ), ( t_2, \synt{s}(t_2) \rangle \mid  t_l \le t_1 < t_2 \le t_u  \wedge \synt{s}(t_1) \ge \synt{v}_1 \wedge \synt{s}(t_1) \le \synt{v}_1+ \synt{v}_2 \wedge \synt{s}(t_2) <  \synt{v}_1 \wedge
\nonumber \neg \exists t \in [t_{l},t_{u}], (t_1<t<t_2) $.
The record at which the signal $\synt{s}$ is greater than or equal to value $\synt{v}_1$ and less than or equal to $\synt{v}_1$+$\synt{v}_2$,
followed by the record at which the signal undershoots, going below $\synt{v}_1$. \\
$\lit{d\_if-then}_1=\langle [ t_2, t_u ] \rangle \mid   (
\exists t_1 \in [t_l, t_2), 
\trace, [t_1,t_2] \models \synt{p}_1
\wedge
\forall t_3, t_4 \in (t_2,t_u], 
( \trace, [t_3,t_4] \not \models \synt{p}_2) 
)$. The time interval delimited by $t_2$ (the last time instant of the last occurrence of pattern $p_1$) up to the last time instant ($t_u$) of the trace. \\
$\lit{d\_if-then}_2= \langle [t_2,t_3], t_3-t_2 \rangle \mid 
		 (
	\exists t_1 \in [t_l, t_2), 
	\trace, [t_1,t_2] \models \synt{p}_1
	\wedge
		\exists t_4 \in (t_3,t_u], 
		( \trace, [t_3,t_4] \models \synt{p}_2 \Rightarrow \neg  ((t_3-t_2)
 \left\llbracket\bowtie\right\rrbracket\synt{d} )
\wedge \forall t_5,t_6 \in (t_2, t_3), (\trace, [t_5,t_6] \not \models \synt{p}_2 )) 
 )$. The time interval $[t_2,t_3]$ representing the time distance between patterns $p_1$ and $p_2$ hold, and the exact value of that violated time distance $(t_3-t_2)$.   \\
\bottomrule
\multicolumn{1}{p{16.9cm}}{
$^\ast$ $\mathit{spk}(\synt{s},t_1,t_2,t_3,t_4,t_5)=\minf (\synt{s},t_2,[t_1,t_3] ) \wedge  \lmaxf (\synt{s},t_3,[t_2,t_4]) \wedge   \minf (\synt{s},t_4,[t_3,t_5] ) 
 $\newline
$^\ast$ $\mathit{osc}(\synt{s},t_1,t_2,t_3,t_4,t_5)=\lminf (\synt{s},t_2,[t_1,t_3] ) \wedge  \lmaxf (\synt{s},t_3,[t_2,t_4]) \wedge   \lminf (\synt{s},t_4,[t_3,t_5] ) 
$\newline
$\mathit{ampv}(\synt{s},t_1,t_2,t_3,\synt{v})= \lvert 
\mathit{amp}(\synt{s},t_1,t_2,t_3)- \synt{v}\rvert $;
  $\quad
  \mathit{p2pv}(\synt{s},t_1,t_2,\synt{v})=
		\lvert \mathit{p2p}(\synt{s},t_1,t_2) - \synt{v} \rvert$;
	$\quad\mathit{widthv}(t_1,t_2,\synt{v})= \lvert \mathit{width}(t_1,t_2) - \synt{v} \rvert$ \newline
}
\end{tabular}

 \caption{Diagnoses associated with the violation causes of patterns in \figurename~\ref{tab:diagnosticPatterns}.}
\label{tab:diagnosticInformationPatterns}
\end{figure*}

\subsection{\emph{Patterns}}\label{patternAssertion}

\subsubsection{\lit{assert}: \textbf{Event-based Data Assertion}}
\paragraph*{\textbf{Violation cause}}
This pattern is violated if there exists at least one record in the trace that violates the condition used in the assertion. 
Recall that a record is used to represent a timestamp and a signal value observed in that timestamp.
Therefore,
the corresponding violation cause $\lit{c\_assert}_1$ checks for the
presence of a timestamp in which the assertion condition \synt{c} is violated.

\begin{figure}[ht]

 \begin{tikzpicture}[domain=0:6] 
 \begin{axis}[
  width=\columnwidth,
       height=0.6\columnwidth,
    xlabel={\texttt{Timestamp}},
    ylabel={\texttt{Value}},
    xmin=0, xmax=7,
    ymin=0, ymax=7,
    xtick={0,1,2,3,4,5,6,7},
    ytick={0,1,2,3,4,5,6},
     ymajorgrids=true,
      xmajorgrids=true,
    grid style=dashed,
      legend pos=north west,
      legend style={legend columns=-1}
]
    \addplot[color=purple,smooth,thick,mark=x,name path=$s_1$,dash pattern=on 1.3pt off 0.5pt on 1.3pt off 0.5pt] plot coordinates {
          (1,1.5)
          (2,2)
          (3,3)
          (4,5)
          (5,2.5)
          (6,4.3)
          (7,3.5)
      };
      \draw[] [name path = vertical, dashed, black, thick] (axis cs:0,4) -- (axis cs:7,4) ;
\legend{$\beta_1$}
\end{axis}
\end{tikzpicture}

 \caption{A trace violating the expression ``\lit{assert}\ $\beta$ < 4''.}
\label{figures/fig:eDA}
\end{figure}
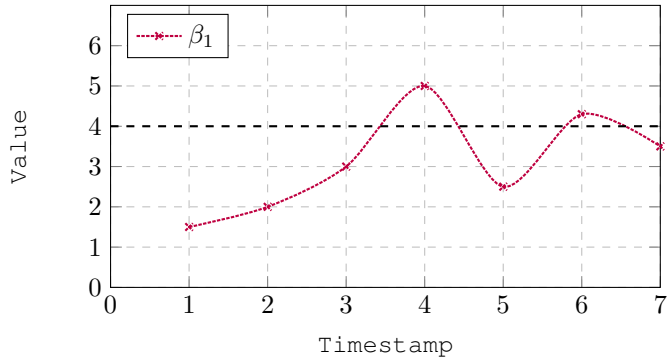

For example, the trace shown in \figurename~\ref{figures/fig:eDA}
violates the expression ``$\lit{assert}\ \beta_{1} < 4$'' because
signal $\beta_1$ shows  a value equal to 5 at timestamp 4, satisfying
the violation cause $\lit{c\_assert}_1$ on the interval $[0,7]$.

\paragraph*{\textbf{Diagnoses}}

The diagnosis $\lit{d\_assert}_1$ associated with violation cause
$\lit{c\_assert}_1$ includes the first timestamp $t$ at which one or more
signals ($\synt{s}_1, \synt{s}_2, \dots, \synt{s}_n$) violate the assertion condition
\synt{c}, as well as the values taken by these signals at $t$. This diagnosis 
allows engineers to identify the root cause of the violation of the assertion condition by looking at the first timestamp in which this violation was observed.

For instance, in the case of the trace shown in
\figurename~\ref{figures/fig:eDA}, the diagnosis is the tuple
containing timestamp $4$ and the value of $\beta_1(4)=5$ taken by signal $\beta_1$.

\subsubsection{\lit{becomes}: \textbf{State-based Data Assertion}}
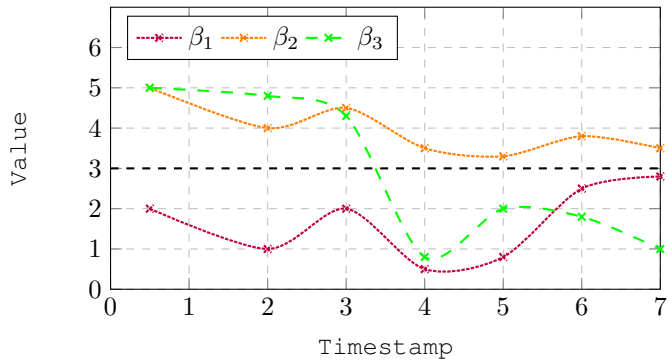
\begin{figure}[ht]
  
 \begin{tikzpicture}[domain=0:6] 
 \begin{axis}[
  width=\columnwidth,
       height=0.6\columnwidth,
    xlabel={\texttt{Timestamp}},
    ylabel={\texttt{Value}},
    xmin=0, xmax=7,
    ymin=0, ymax=7,
    xtick={0,1,2,3,4,5,6,7},
    ytick={0,1,2,3,4,5,6},
     ymajorgrids=true,
      xmajorgrids=true,
    grid style=dashed,
      legend pos=north west,
      legend style={legend columns=-1}
]
   \addplot[color=purple,smooth,thick,mark=x,name path=$s_1$,dash pattern=on 1.3pt off 0.5pt on 1.3pt off 0.5pt] plot coordinates {
          (0.5,2)
          (2,1)
          (3,2)
          (4,0.5)
          (5,0.8)
          (6,2.5)
          (7,2.8)
      };
         \addplot[color=orange,smooth,thick,mark=x,name path=$s_1$,dash pattern=on 1.3pt off 0.5pt on 1.3pt off 0.5pt] plot coordinates {
          (0.5,5)
          (2,4)
          (3,4.5)
          (4,3.5)
          (5,3.3)
          (6,3.8)
          (7,3.5)
      };

      \addplot[color=green,smooth,thick,mark=x,name path=$s_4$,dash pattern=on 4.5pt off 4.5pt on 4.5pt off 4.5pt] plot coordinates {
          (0.5,5)
          (2,4.8)
          (3,4.3)
          (4,0.8)
          (5,2)
          (6,1.8)
          (7,1)
      };
      \draw[] [name path = vertical, dashed, black, thick] (axis cs:0,3) -- (axis cs:7,3) ;
\legend{$\beta_1$,$\beta_2$,$\beta_3$}
\end{axis}
\end{tikzpicture}

 \caption{A trace with signals violating the expression ``$\beta\ \lit{becomes} > 3$''.}
\label{figures/fig:sDA}
\end{figure}

\paragraph*{\textbf{Violation causes}}
This pattern can be violated in at least three ways, as illustrated
with different signal behaviors in \figurename~\ref{figures/fig:sDA}
using the expression ``$\beta\ \lit{becomes} > 3$'':
\begin{itemize}
\item \lit{c\_becomes}$_1$: The signal value violates the pattern constraint $\sim \synt{v}$ throughout the time interval over which the pattern is evaluated.
For instance, signal $\beta_1$ in the figure is never greater than 3.
\item \lit{c\_becomes}$_2$:  The signal value satisfies the pattern
   constraint $\sim \synt{v}$ throughout the time interval over which the pattern is evaluated. This violation cause is the dual
  of the previous case. For instance, signal $\beta_2$ in the figure
  is always greater than value 3. 
\item \lit{c\_becomes}$_3$: The signal violates the semantics of the pattern by satisfying the negation of the pattern constraint (i.e., $ \trace, [t_l,t_u] \models \synt{s}\ \lit{becomes} \not \sim \synt{v}   $ holds).

For instance, signal $\beta_3$ in the figure becomes less than or equal to 3 (instead of becoming greater than 3). More precisely, it goes below value $3$ at timestamps $4$, and remains below that value until the end of the time interval, delimited by timestamp $7$. 
\end{itemize}

\paragraph*{\textbf{Diagnoses}}
The diagnoses associated with the three violation causes above are the following:
\begin{itemize} 
\item \lit{d\_becomes}$_1$ and \lit{d\_becomes}$_2$
include two records from the signal showing a minimum and a maximum value.
In this way, we show the range of values over which the signal changes.
In the example shown in \figurename~\ref{figures/fig:sDA}, we report records
$\langle (7, 2.8), (4, 0.5)\rangle$ for  signal $\beta_1$ and records $\langle
(0.5,5), (5, 3.3) \rangle$ for signal $\beta_2$.
\item \lit{d\_becomes}$_3$ includes the last-seen record at which the signal value satisfies the constraint $\sim \synt{v}$, followed by
  the next-seen record at which the signal value satisfies $\not \sim
  \synt{v}$. Through this diagnosis, we want to capture the exact time interval, delimited by two consecutive timestamps, within $[t_l,t_u]$, in which
  the signal exhibits a behavior compatible with the negation of the constraint
  specified in the \lit{becomes} expression. 
  For instance, for signal
  $\beta_3$ in \figurename~\ref{figures/fig:sDA}, the diagnosis is $\langle (3,
  4.3), (4, 0.8)\rangle$.
\end{itemize}

\subsubsection{\textbf{Spike}}

\paragraph*{\textbf{Violation causes}}
This pattern can be violated in at least five ways, as illustrated with different signal behaviors in \figurename~\ref{fig:spk} using the expression ``\lit{exists} \ \lit{spike} \ \lit{in} $\beta$ \ \lit{with} \ \lit{amplitude} < 90  \lit{width} < 0.5''.
These alternatives are the following:
\begin{itemize} 
\item \lit{c\_spike}$_1$: All spike instances in the signal violate the amplitude constraint.
For instance, signal $\beta_1$ in the figure shows two spike amplitude values greater than $90$ ($150$ and $200$, respectively).
\item \lit{c\_spike}$_2$: All spike instances in the signal violate the width constraint. For example, signal $\beta_1$ shows two spike width values greater than $0.5$ ($1.8$ and $4.2$, respectively).
\item \lit{c\_spike}$_3$: The signal is constant throughout the time interval over which the pattern is evaluated. For example, the constant signal $\beta_2$ in the figure always takes the value $100$ within the time interval $[0,6]$.
\item  \lit{c\_spike}$_4$: The signal decreases,within the time interval over which the pattern is evaluated, without showing any
  spike behavior. For example, signal $\beta_3$ in the figure
  decreases within the time interval $[0,6]$, going from value $190$ to $30$.
\item  \lit{c\_spike}$_5$: The signal increases within the time interval over which the pattern is evaluated, without showing any
  spike behavior. For instance, signal $\beta_4$ increases within the time interval $[0,6]$, going from
  value $30$ to $190$. 
\end{itemize}
\paragraph*{\textbf{Diagnoses}}
The diagnoses associated with the five violation causes above are the following:
\begin{itemize} 
\item \lit{d\_spike}$_1$ includes the time interval in which the spike
  with the closest amplitude to satisfy the amplitude constraint
  occurs, as well as the amplitude value of that spike instance (see page~\pageref{txt:spk-dg} for a detailed explanation).
  The intuition behind this diagnosis is that when a spike property with an amplitude constraint is violated, the engineers are interested in knowing the amplitude value of the spike that is the closest to satisfy the amplitude constraint, to assess how close the signal behavior was to satisfy the property .
 For instance, for signal $\beta_1$ in \figurename~\ref{fig:spk}, the diagnosis is
$\langle [0,1.8], 150 \rangle$.
\item \lit{d\_spike}$_2$ is defined in a similar way, but with respect
  to the width constraint. It includes the time interval in which the
  spike with the closest width to satisfy the width constraint occurs, as well as the width value of that spike instance. 
  Similar to \lit{d\_spike}$_1$, the choice of this specific width value enables engineers to determine the closest value of a spike width to the satisfaction of the width constraint defined in the pattern.
  For instance, for signal $\beta_1$ in \figurename~\ref{fig:spk}, the diagnosis is $\langle [0,1.8], 1.8 \rangle$. 
\item \lit{d\_spike}$_3$ includes the time interval [$t_l,t_u$] over which the property pattern is evaluated, as well as the value taken by the constant signal throughout that interval. 
This diagnosis shows that the signal is constant (i.e., it shows a single value) throughout the full time interval $[t_l,t_u]$, over which the pattern is evaluated.  
For instance, for signal $\beta_2$ in \figurename~\ref{fig:spk}, the diagnosis is $\langle [0,6], 100 \rangle$.
\item \lit{d\_spike}$_4$ and \lit{d\_spike}$_5$
include two records from the signal corresponding to its minimum and
maximum values (and the timestamps at which these values occur). In this way, we show the range of values over which the signal changes
(decreasing or increasing).
In the example shown in \figurename~\ref{fig:spk}, we report records $\langle
(0, 200), (6, 55)\rangle$ for the decreasing signal $\beta_3$  and records $\langle
(0,30), (6, 190) \rangle$ for the increasing  signal $\beta_4$.
\end{itemize}

\subsubsection{\textbf{Oscillation}}
\paragraph*{\textbf{Violation causes}}
This pattern can be violated in at least seven ways, as illustrated
with different signal behaviors in \figurename~\ref{figures/fig:osc}
using the expression ``\lit{exist} \ \lit{oscillation} \
\lit{in} $\beta$ \ \lit{with} \ \lit{p2pAmp} < 90 \ \lit{period} <
0.5''. These alternatives are the following:
\begin{itemize} 
\item \lit{c\_oscillation}$_1$: All oscillation instances in the
  signal violate the amplitude constraint. For instance,
  signal $\beta_1$ in the figure shows two oscillation instances, both having an amplitude value greater than $90$ ($125$ and $200$, respectively).
\item \lit{c\_oscillation}$_2$: All oscillation instances in the
  signal violate the period constraint. For instance, signal $\beta_1$
  shows two oscillation instances whose period value is greater than
  $0.5$: the first oscillation has a period of $0.8$ (i.e., the time
  difference between timestamps $0.2$ and $1$), while the second
  oscillation has a period of $1$ (i.e., the time difference between timestamps $3.5$ and $4.5$).
\item \lit{c\_oscillation}$_3$: The signal  does not show any
  oscillation; instead, it shows only one strict local extremum (a maximum or a minimum).
  This is the case, for instance, of signal $\beta_2$ in the figure, that exhibits a strict local maximum (reaching the value of $150$ at timestamp $1.5$). 
\item  \lit{c\_oscillation}$_4$: The signal does not show any
  oscillation; instead, it shows only two strict local extrema. For
  instance, signal $\beta_3$ in the figure exhibits a strict local minimum (taking value $80$ at timestamp $1.5$), followed by a strict local maximum (taking value $150$ at timestamp $2$).
\item  \lit{c\_oscillation}$_5$: The signal is constant throughout the time interval $[t_l,t_u]$ (see, for example, signal $\beta_4$ in the figure).  
\item  \lit{c\_oscillation}$_6$: The signal  decreases without showing
  any oscillatory behavior. For instance, signal $\beta_5$ in the
  figure decreases, going from value $180$ at timestamp $0.1$ to value  $20$ at timestamp $5.8$. 
\item  \lit{c\_oscillation}$_7$: The signal  increases without showing
  any oscillatory behavior.
For example, signal $\beta_6$ in the figure increases, going from
value $40$ at timestamp $0.2$ to value $150$ at timestamp $5.8$. 
\end{itemize}
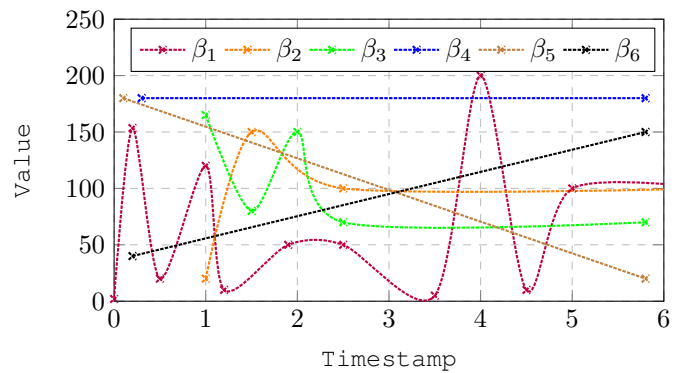
\begin{figure}[!htbp]
 \begin{tikzpicture}[domain=0:6] 
 \begin{axis}[
  width=\columnwidth,
       height=0.6\columnwidth,
    xlabel={\texttt{Timestamp}},
    ylabel={\texttt{Value}},
    xmin=0, xmax=6,
    ymin=0, ymax=250,
    xtick={0,1,2,3,4,5,6},
    ytick={0,50,100,150,200,250},
     ymajorgrids=true,
      xmajorgrids=true,
    grid style=dashed,
      legend pos=north west,
      legend style={legend columns=-1}
]
    \addplot[color=purple,smooth,thick,mark=x,name path=$s_1$,dash pattern=on 1.3pt off 0.5pt on 1.3pt off 0.5pt] plot coordinates {
          (0,2)
          (0.2,153.5)
          (0.5,20.0)
          (1,120)
          (1.2,10.0)
          (1.9,50)
          (2.5,50)
          (3.5,5)
          (4,200)
          (4.5,10)
          (5,100)
          (6.5,100)
      };
    \addplot[color=orange,smooth,thick,mark=x,name path=$s_1$,dash pattern=on 1.3pt off 0.5pt on 1.3pt off 0.5pt] plot coordinates {
          (1,20)
          (1.5,150)
          (2.5,100)
          (6.5,100)
      }; 
    \addplot[color=green,smooth,thick,mark=x,name path=$s_1$,dash pattern=on 1.3pt off 0.5pt on 1.3pt off 0.5pt] plot coordinates {
      (1,165)
      (1.5,80)
      (2,150)
      (2.5,70)
      (5.8,70)
    };   
    
     \addplot[color=blue,smooth,thick,mark=x,name path=$s_1$,dash pattern=on 1.3pt off 0.5pt on 1.3pt off 0.5pt] plot coordinates {
      (0.3,180)
      (5.8,180)
    };   
      
      \addplot[color=brown,smooth,thick,mark=x,name path=$s_1$,dash pattern=on 1.3pt off 0.5pt on 1.3pt off 0.5pt] plot coordinates {
      (0.1,180)
      (5.8,20)
    };   
      
      \addplot[color=black,smooth,thick,mark=x,name path=$s_1$,dash pattern=on 1.3pt off 0.5pt on 1.3pt off 0.5pt] plot coordinates {
      (0.2,40)
      (5.8,150)
    };   
          
\legend{$\beta_1$,$\beta_2$,$\beta_3$,$\beta_4$,$\beta_5$,$\beta_6$}
\end{axis}
\end{tikzpicture}
\caption{A trace with signals violating the expression ``\lit{exist}\  \lit{oscillation}\ \lit{in}\ $\beta$ \lit{with}\ \lit{p2pAmp} < 90 \lit{period} < 0.5 ''.}
\label{figures/fig:osc}
\end{figure}
 \paragraph*{\textbf{Diagnoses}}
The diagnoses associated with the seven violation causes above are the following:
\begin{itemize} 
\item \lit{d\_oscillation}$_1$ includes the time interval in which the
  oscillation with the closest amplitude to satisfy the amplitude
  constraint occurs, as well as the amplitude value of that oscillation instance.
The choice of this diagnosis enables engineers to determine the oscillation instance with the closest amplitude to the satisfaction of the amplitude constraint defined in the pattern.
For instance, for signal $\beta_1$ in \figurename~\ref{figures/fig:osc}, the diagnosis is $\langle [0,1.9], 125 \rangle$. 
\item \lit{d\_oscillation}$_2$ includes the time interval in which the
  oscillation with the closest period to satisfy the period
  constraint occurs, as well as the period of that oscillation instance.
For instance, for signal $\beta_1$ in \figurename~\ref{figures/fig:osc}, the diagnosis is $\langle [0,1.9], 0.8 \rangle$.
Similar to \lit{d\_oscillation}$_1$, we allow engineers to identify the oscillation instance that shows
the closest period value to the satisfaction of the period constraint defined in the pattern.
\item \lit{d\_oscillation}$_3$ includes the timestamp (and the
  corresponding signal value) in which the signal exhibits a strict extremum. 
  The reported diagnosis allows engineers to identify the first time in which the signal exhibited a considerable deviation, leading to a change of the sign of its derivative.
  For instance, for signal $\beta_2$ in \figurename~\ref{figures/fig:osc}, the diagnosis is
$\langle 1.5,150 \rangle$.
\item \lit{d\_oscillation}$_4$ includes the two records 
from the signal in which the strict  maximum and the strict minimum occur. 
By considering this diagnosis, engineers are able to see a considerable change of the signal shape, showing two different consecutive strict extrema.
For instance, for signal $\beta_3$ in \figurename~\ref{figures/fig:osc}, the diagnosis is 
$\langle (1.5,80), (2,150) \rangle$.
\item \lit{d\_oscillation}$_5$ includes the time interval $[t_l,t_u]$ throughout which the signal \synt{s} is constant, as well as the value taken by that signal.
 Similar to \lit{d\_spike}$_3$, this diagnosis shows that the signal is constant throughout the full time interval $[t_l,t_u]$ over which the pattern is evaluated. 
For instance, for signal $\beta_4$ in \figurename~\ref{figures/fig:osc}, the diagnosis is $\langle [0,6], 180 \rangle$.
\item \lit{d\_oscillation}$_6$ and \lit{d\_oscillation}$_7$ include the records in which the maximum and the minimum values of the signal were observed.
In this way, we show the range of values over which the signal changes (i.e., decreases or increases).
In the example shown in \figurename~\ref{figures/fig:osc}, we report records $\langle
(0.1, 180), (5.8, 20) \rangle$ for signal $\beta_5$ and records $\langle
(0.2, 40), (5.8, 150) \rangle$ for signal $\beta_6$.
\end{itemize}

\subsubsection{\textbf{Rise time}}
\begin{figure}[!htbp]
 \begin{tikzpicture}[domain=0:6] 
 \begin{axis}[
  width=\columnwidth,
       height=0.6\columnwidth,
    xlabel={\texttt{Timestamp}},
    ylabel={\texttt{Value}},
    xmin=0, xmax=7,
    ymin=0, ymax=7,
    xtick={0,1,2,3,4,5,6,7},
    ytick={0,1,2,3,4,5,6},
     ymajorgrids=true,
      xmajorgrids=true,
    grid style=dashed,
      legend pos=north west,
      legend style={legend columns=-1}
]
   \addplot[color=purple,smooth,thick,mark=x,name path=$s_1$,dash pattern=on 1.3pt off 0.5pt on 1.3pt off 0.5pt] plot coordinates {
          (0.6,2)
          (2,2.4)
          (3,2)
          (4,1.5)
          (5,0.8)
          (6,1.5)
          (6.7,2.5)
      };
         \addplot[color=orange,smooth,thick,mark=x,name path=$s_1$,dash pattern=on 1.3pt off 0.5pt on 1.3pt off 0.5pt] plot coordinates {
          (0.5,4)
          (2,5)
          (3,5)
          (4,4.5)
          (5,5.4)
          (6,5)
          (7,6)
      };

      \addplot[color=green,smooth,thick,mark=x,name path=$s_4$,dash pattern=on 4.5pt off 4.5pt on 4.5pt off 4.5pt] plot coordinates {
          (1,0.5)
          (2,2)
          (3,0.5)
          (4,4)
          (5,4.8)
          (6,3.5)
          (7,3.2)
      };
        \addplot[color=black,smooth,thick,mark=x,name path=$s_4$,dash pattern=on 4.5pt off 4.5pt on 4.5pt off 4.5pt] plot coordinates {
          (0.5,4.5)
          (2,3.8)
          (3,3.1)
          (4,0.5)
          (5,1.8)
          (6,1)
          (7,2.2)
      };
      
    \draw[] [name path = vertical, dashed, black, thick] (axis cs:0,3) -- (axis cs:7,3) ;  
\legend{$\beta_1$,$\beta_2$,$\beta_3$,$\beta_4$}
\end{axis}
\end{tikzpicture}
\caption{A trace with signals violating the expression ``$\beta$ \lit{rises}\ \lit{monotonically}\ \lit{reaching} 3''.}
\label{figures/fig:rt}
\end{figure}
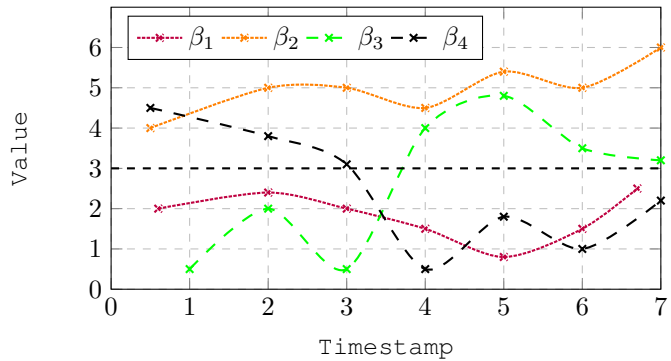

\paragraph*{\textbf{Violation causes}}
This pattern can be violated in at least four ways, as illustrated with different signal behaviors in \figurename~\ref{figures/fig:rt} using the expression ``$\beta$ \ \lit{rises} \ \lit{monotonically} \ \lit{reaching 3}". 

\begin{itemize}
\item \lit{c\_rises}$_1$: The signal is always
below the threshold value $v$ defined in the pattern constraint. For instance, signal $\beta_1$ is always below the value of $3$, showing values ranging between $0.8$ and $2.5$.
\item \lit{c\_rises}$_2$: The signal is always greater than or equal to the threshold value $v$. For instance, signal $\beta_2$ is always above the value  $3$, showing values ranging between $4$ and $6$.
\item  \lit{c\_rises}$_3$: The signal 
shows a rising behavior, but violates the monotonicity constraint defined in the pattern. For instance, signal
$\beta_3$ rises reaching the target value (showing a value of $4$ at timestamp $4$), but it violates the monotonicity constraint since its value
decreases from $2$ (at timestamp $2$) to $0.5$ (at timestamp $3$).
\item \lit{c\_rises}$_4$: The signal is initially above the threshold value $\synt{v}$. It then falls (and remains) below that value, instead of rising. 
For instance, signal $\beta_4$ falls (and remains) below the target value of $3$ (starting from timestamp $4$, up to timestamp $7$, showing values ranging within the interval $[0.5,2]$) instead of rising.
\end{itemize}
\paragraph*{\textbf{Diagnoses}}
The diagnoses associated with the four violation causes above are the following:
\begin{itemize}
\item \lit{d\_rises}$_1$ and \lit{d\_rises}$_2$ include the records in which the signal shows a maximum and a minimum value. 
In this way, we show the range of values the signal takes.
For instance, for signal $\beta_1$ in \figurename~\ref{figures/fig:rt}, the diagnosis is 
$\langle (5,0.8), (6.7, 2.5)\rangle$.
Similarly, diagnosis for signal $\beta_2$ is $\langle (0.5,4), (7,6)\rangle$.
\item \lit{d\_rises}$_3$ includes two consecutive records 
where the monotonicity constraint is violated. 
In this way, we show the exact interval over which the signal deviated from the last time it exhibited an increasing behavior, showing a negative derivative.
For instance, for signal $\beta_3$ in \figurename~\ref{figures/fig:rt}, the diagnosis is 
$\langle (2,2), (3,0.5)\rangle$.
\item \lit{d\_rises}$_4$ includes two consecutive records in which the signal shows a dual behavior (i.e., it falls instead of rising). More precisely, the signal value in the first record is above the threshold value $\synt{v}$. The signal value in the second reported record is, however, below that value. 
This diagnosis determines the interval over which the signal shows a dual behavior, within the time interval $[t_l,t_u]$ over which the pattern is evaluated.
For instance, the diagnosis of signal $\beta_4$ in \figurename~\ref{figures/fig:rt} is
$\langle (3,3.1), (4,0.5)\rangle$ .
\end{itemize}

\subsubsection{\textbf{Overshoot}}
\begin{figure}[!htbp]
 \begin{tikzpicture}[domain=0:6] 
 \begin{axis}[
  width=\columnwidth,
       height=0.6\columnwidth,
    xlabel={\texttt{Timestamp}},
    ylabel={\texttt{Value}},
    xmin=0, xmax=7,
    ymin=0, ymax=7,
    xtick={0,1,2,3,4,5,6,7},
    ytick={0,1,2,3,4,5,6},
    ymajorgrids=true,
    xmajorgrids=true,
    grid style=dashed,
    legend pos=north west,
    legend style={legend columns=-1}
]
   \addplot[color=purple,smooth,thick,mark=x,name path=$s_1$,dash pattern=on 1.3pt off 0.5pt on 1.3pt off 0.5pt] plot coordinates {
          (0.6,2)
          (2,2.4)
          (3,2)
          (4,1.5)
          (5,0.8)
          (6,1.5)
          (6.7,2.5)
      };
         \addplot[color=orange,smooth,thick,mark=x,name path=$s_1$,dash pattern=on 1.3pt off 0.5pt on 1.3pt off 0.5pt] plot coordinates {
         (0.6,0.5)
          (2,4.5)
          (3,4.9)
          (4,4.5)
          (5,4.8)
          (6,4.1)
          (6.7,4.5)
      };

      \addplot[color=green,smooth,thick,mark=x,name path=$s_4$,dash pattern=on 4.5pt off 4.5pt on 4.5pt off 4.5pt] plot coordinates {
          (1,0.5)
          (2,2)
          (3,0.5)
          (4,3.8)
          (5,3.5)
          (6,3.3)
          (7,3.5)
      };
        \addplot[color=black,smooth,thick,mark=x,name path=$s_4$,dash pattern=on 4.5pt off 4.5pt on 4.5pt off 4.5pt] plot coordinates {
          (0.5,4)
          (2,3.8)
          (3,2.1)
          (4,0.5)
          (5,1.8)
          (6,1)
          (7,0.9)
      };
      \draw[] [name path = vertical, dashed, black, thick] (axis cs:0,3) -- (axis cs:7,3) ;
      \draw[] [name path = vertical, dashed, black, thick] (axis cs:0,4) -- (axis cs:7,4) ;
\legend{$\beta_1$,$\beta_2$,$\beta_3$,$\beta_4$}
\end{axis}
\end{tikzpicture}
\caption{A trace with signals violating the expression ``$\beta$ \lit{overshoots}\ \lit{monotonically} \ 3 \ \lit{by} 1''.}
\label{fig:osh}
\end{figure}
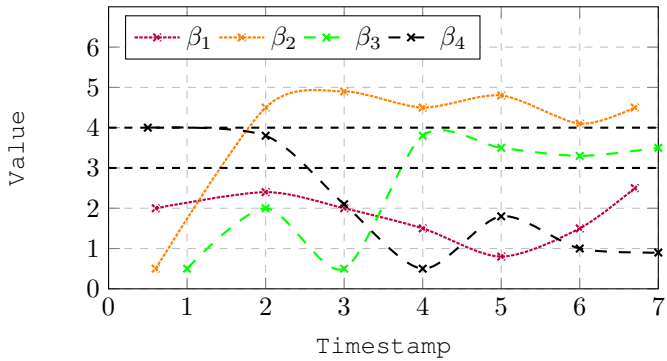

\paragraph*{\textbf{Violation causes}}
This pattern can be violated in at least four ways, as illustrated with different signal behaviors in \figurename~\ref{fig:osh} using the expression 
``$\beta \ \lit{overshoots} \ \lit{monotonically} \ 3\ \lit{by} \ 1 $”:
\begin{itemize}
\item \lit{c\_overshoots}$_1$: The signal violates the pattern constraint, by always showing values below the threshold value $\synt{v}_1$. For example, signal $\beta_1$ is always below the value of $3$.
\item \lit{c\_overshoots}$_2$: The signal
goes beyond the maximum allowed value, which consists of the sum of the target value $\synt{v}_1$ and the maximum threshold value $\synt{v}_2$ ($\synt{v}_{1}+\synt{v}_{2}$), and remains above that value. For instance, signal $\beta_2$ exceeds $4$ (showing a value of $4.5$ at timestamp $2$) and remains above the value of $4$, ranging over $[4.1,4.9]$.
\item  \lit{c\_overshoots}$_3$: The signal
overshoots the threshold value $\synt{v}_1$, without going beyond the maximum allowed value (delimited by $\synt{v}_{1}+\synt{v}_{2}$ defined in the pattern). However, it violates the monotonicity constraint. For instance, signal 
$\beta_3$ overshoots, reaching the value of $3.8$ at timestamp $4$, without going beyond the value of $4$ after then. It violates the monotonicity constraint within the time interval $[2,3]$, since its value goes from $2$ down to $0.5$. \item \lit{c\_overshoots}$_4$: The signal shows a dual behavior: it undershoots,  going below the value $\synt{v}_1$, and remains below that value instead of overshooting. 
For instance, signal
$\beta_4$ goes (and remains) below the value of $3$. It reaches value $2$ at timestamp $3$ and takes, right after then, values ranging over $[0.5,2]$. \end{itemize}
\paragraph*{\textbf{Diagnoses}}
The diagnoses associated with the four violation causes above are the following:
\begin{itemize}
\item \lit{d\_overshoots}$_1$ and \lit{d\_overshoots}$_2$ include the records in which the signal shows a maximum and a minimum value.
The reported diagnosis allows engineers to understand the range of values taken by the signal.
For instance, for signal $\beta_1$ in \figurename\ref{fig:osh}, the diagnosis is 
$\langle (5,0.8), (6.7, 2.5)\rangle$.
\item \lit{d\_overshoots}$_3$ includes two consecutive records 
from a signal that overshoots, but violates the monotonicity constraint.
This diagnosis shows the time interval over which the signal violated the monotonicity constraint within $[t_l,t_u]$.
For instance, for signal $\beta_3$ in \figurename\ref{fig:osh}, the diagnosis is 
$\langle (2, 2), (3, 0.5)\rangle$.
\item \lit{d\_overshoots}$_4$ includes two consecutive records in which the signal shows a dual behavior (i.e.,
it undershoots instead of overshooting).
More specifically, we report the last record at  which the signal value is delimited by $[\synt{v}_1, \synt{v}_{1}+\synt{v}_{2}]$, followed by the next-seen record at which it undershoots, going below $\synt{v}_1$. 
This diagnosis allows engineers to understand the  interval over which the signal shows a dual behavior within the time interval $[t_l,t_u]$.
For instance, for signal $\beta_4$ in \figurename\ref{fig:osh}, the diagnosis is
$\langle (2, 3.8), (3, 2.1)\rangle$.
\end{itemize}

\subsubsection{\textbf{Order relationship}}
\paragraph*{\textbf{Violation causes}}
A property with an \lit{if-then} construct is based on two patterns, each of which represents one of the pattern constructs we support in \dslname. According to the construct syntax \lit{if} $p_1$ \lit{then} $p_2$,  $p_1$ is referred to as a cause pattern and $p_2$ as an effect pattern. We consider two possible violation causes of a property with the \lit{if-then} construct. 
\begin{itemize}
\item \lit{c\_if-then}$_1$: the cause pattern $p_1$ holds at some time interval $[t_1, t_2]$ within the time interval $[t_l,t_u]$, but then, the effect pattern $p_2$ fails to hold until the last timestamp ($t_u$) of that time interval.
\item \lit{c\_if-then}$_2$: the cause pattern $p_1$ holds within a time interval $[t_1,t_2]$ but since then, whenever the effect pattern $p_2$ holds (after $p_1$) within a time interval $[t_3,t_4]$, the time distance ($t_{3}-t_{2}$) between the occurrences of two patterns $p_1$ and $p_2$ is violated. 
\end{itemize}

\paragraph*{\textbf{Diagnoses}}
The diagnoses associated with a violation of an expression with an \lit{if-then} construct  
are the following:
\begin{itemize}
\item \lit{d\_if-then}$_1$ includes the time interval delimited by the last timestamp ($t_2$) of the last occurrence of pattern $p_1$ and the last timestamp $t_u$ of the time interval $[t_l,t_u]$, showing the exact interval over which the effect pattern $p_2$ failed to hold.
The corresponding diagnosis is then the following: 
$\langle [t_2,t_u] \rangle$.
\item \lit{d\_if-then}$_2$ includes the time interval delimited by the last timestamp ($t_2$) in which the cause pattern $p_1$ holds, and the first timestamp ($t_3$) in which the effect pattern $p_2$ holds. 
The diagnosis also includes the violated time distance ($t_{3}-t_{2}$) between the occurrence of patterns $p_1$ and $p_2$. The diagnosis is the following:
$\langle [t_2, t_3], t_{3}-t_{2} \rangle$.
\end{itemize}

\subsection{\emph{Scopes}}
Since the same syntactic constructs of \dslname (e.g., the keyword \lit{before}) can be used to define both absolute and event scopes (see \figurename~\ref{tab:syntax}), 
we use the identifiers $\lit{a\_}$ and $\lit{e\_}$  before the violation cause name depending on whether it refers to an absolute or an event scope. 
Additionally, we use $\lit{bef}$,  $\lit{aft}$, and $\lit{bet}$ as shortcuts for $\lit{before}$, $\lit{after}$, and $\lit{between}$, respectively. 
For example, $\lit{c\_a\_bef}_2$ denotes a violation cause for the \lit{before} absolute scope.
\paragraph*{\textbf{Violation causes}}
\begin{itemize} 
\item \lit{c\_a\_at}$_1$, \lit{c\_a\_bef}$_1$ and \lit{c\_a\_aft}$_1$ are violations related to absolute boundaries (i.e., timestamps) that are not within the time interval of the trace over which the property is evaluated.
\item  \lit{c\_a\_bet}$_1$ indicates that either at least one of the scope boundaries is outside the  time interval $[t_i,t_e]$ or the left boundary (which is supposed to be smaller than the right one) is greater than or equal to the right boundary.
\item \lit{c\_e\_bef}$_1$ states that scope pattern $p_1$ holds in the execution trace, whereas the property pattern $p$ fails to hold sometime before $p_1$ held.
\item \lit{c\_e\_aft}$_1$ indicates that scope pattern $p_1$ holds in the execution trace, whereas the property pattern $p$ fails to hold after that.
\item \lit{c\_e\_bet}$_1$ states that  scope patterns $p_1$ and $p_2$ hold in the execution trace, whereas the property pattern $p$ fails to hold between the last timestamp where $p_1$  held and the first timestamp in which $p_2$ held.
\end{itemize}
\paragraph*{\textbf{Diagnoses}}
The diagnoses associated with scope-based violations are the following:
\begin{itemize} 
\item \lit{d\_a\_at}$_1$, \lit{d\_a\_bef}$_1$ and \lit{d\_a\_aft}$_1$ include the time interval $[t_i,t_e]$ that delimits the execution trace, as well as the absolute boundary $t$ that is not within the range delimited by that time interval.
\item  \lit{d\_a\_bet}$_1$ includes the time interval $[t_i,t_e]$ of the execution trace over which the property is evaluated, as well as the left and the right absolute boundaries of the scope (i.e., timestamps $n$ and $m$, respectively; see  \figurename~\ref{tab:syntax}).
\item \lit{d\_e\_bef}$_1$ includes the time interval $[t_1,t_2]$ in which the scope pattern $p_1$ held and before which the property pattern $p$ failed to hold. 
\item \lit{d\_e\_aft}$_1$ includes the time interval $[t_1,t_2]$ in which the scope pattern $p_1$ held and after which the property pattern $p$ failed to hold.
\item \lit{d\_e\_bet}$_1$ includes the time interval $[t_2,t_3]$ where $t_2$ represents the last timestamp in which pattern $p_1$ (the left event-boundary) held and $t_3$ (the right event-boundary) is the first timestamp in which pattern $p_2$ held throughout the execution trace. 
\end{itemize}

\subsection{\emph{Atoms}}
\paragraph*{\textbf{Violation causes}}
The violation cause $\lit{c\_not}_1$ for the construct ``\lit{not}\ \synt{sc}'' requires \synt{sc} to be satisfied, since for ``\lit{not}\ \synt{sc}'' to be violated, \synt{sc} must be satisfied (see the semantics in \figurename~\ref{tab:propertypatterns}). 
\paragraph*{\textbf{Diagnosis}}
As depicted in~\figurename~\ref{tab:diagnosticInformationAtomsScopes},  many diagnosis are associated with the violation cause $\lit{c\_not}_1$.
Indeed, for $\lit{c\_not}_1$ the diagnosis should explain why the
violation cause $\lit{c\_not}_1$ (see \figurename~\ref{tab:diagnosticPatterns}) holds, i.e., why \texttt{sc} is satisfied. 
The reasons that lead to the satisfaction of \texttt{sc} depend on the \dslname scope and the pattern used to define \texttt{sc}.
When \texttt{sc} is satisfied, both the scope and the pattern are satisfied.
Our diagnosis explains \emph{why the pattern used to define \texttt{sc} holds}. 
For this reason, the name of the diagnosis is obtained by adding the string ``$\lit{d\_not}$'' before the name of the pattern used to define \texttt{sc}. 
In the following, we explain each of the diagnoses w.r.t the pattern defining \texttt{sc}:
\begin{itemize}
\item diagnosis $\lit{d\_not\_assert}$ includes a timestamp $t$ in which one or more signals ($s_1, s_2, \dots, s_n$) defined in the related property satisfy the corresponding condition $c$ as well as the corresponding value(s) taken by each of these signals.   
\item diagnosis $\lit{d\_not\_becomes}$ includes the first timestamp $t$ that satisfies the property condition, as well as the value of the signal recorded at $t$.
\item diagnosis $\lit{d\_not\_spike}$ includes the first and the last records  of a spike instance that occurred within the time interval $[t_l,t_u]$.
\item diagnosis $\lit{d\_not\_oscillation}$ includes the first and the last records of an oscillation instance.
\item diagnosis $\lit{d\_not\_rises}$ includes the first  record at timestamp $t$  in which (1) the signal defined in the property rises, reaching the property threshold $v$, and (2) the monotonicity constraint (if defined in the pattern) is satisfied within the time interval $[t_l,t]$.
\item diagnosis $\lit{d\_not\_overshoots}$ includes the first record at which the signal defined in the property overshoots (i.e., reaching a value that ranges between values $\synt{v}$ and $\synt{v}_1 + \synt{v}_2$) and satisfies the monotonicity constraint, if defined in the pattern.
\item diagnosis $\lit{d\_not\_if-then}$ includes two time intervals delimiting where the cause pattern and the effect one of the property hold throughout the execution trace.
\end{itemize}

 \section{\NAME at work}
\label{TDAtWork}
In this section, we illustrate how \NAME works by applying
algorithm~\ref{tdApp} to three example properties, each of them with different
constructs.

\subsection{Property with a single atom}
Let us consider property P1, checked on the trace shown in \figurename~\ref{fig:spk}:
\[
  \begin{array}{ll}
     \text{P1} \equiv &  \lit{after}\ 7\ \lit{exists}\  \lit{spike}\
                        \lit{in}\ \beta_1 \\
    &\lit{with}\ \lit{width} < 0.5\  \lit{amplitude} < 90.\\
  \end{array}
\]
Based on the \dslname grammar in \figurename~\ref{tab:syntax}, this
property is made of a single atom of the form $\lit{after}\  \synt{t}\
\synt{p}$, i.e., it consists of an \lit{after} scope construct
(delimited by an absolute time instant, parameter $\synt{t}=7$)
constraining a pattern \synt{p} of type  \lit{spike}.

Given the presence of only one atom in the property,
algorithm~\ref{tdApp} first determines whether the atom itself is
violated by the trace  (line~\ref{alg:subpropviolation}). Since
the trace violates the specification defined by the atom,
algorithm~\ref{tdApp} continues by computing the associated diagnosis
using function~\textsc{TD-Atom} (algorithm~\ref{tdAtom}).

Algorithm~\ref{tdAtom} relies on the auxiliary function
\textsc{getViolationCauses}, which analyzes the syntactic structure of the atom
and determines the possible violation causes associated with it.
In this case, the possible violation causes are the one  associated with
the  \lit{after} scope construct with an absolute boundary, i.e.,
\lit{c\_a\_aft}$_1$, and the five ones associated with the  \lit{spike}
construct, i.e., \lit{c\_{spike}}$_{i}$ with $1\leq i \leq 5$.
This means that function \textsc{getViolationCauses} returns the list 
$\mathit{vcs} = [\lit{c\_a\_aft}_1$, \lit{c\_{spike}}$_{1}$,
\lit{c\_{spike}}$_{2}$, \lit{c\_{spike}}$_{3}$, 
\lit{c\_{spike}}$_{4}$, \lit{c\_{spike}}$_{5}]$.

The algorithm continues by looping through the violation
causes in $\mathit{vcs}$, to determine the first violation cause that
holds on the trace; it will then return the corresponding diagnosis. 
In this example, the violation cause $\lit{c\_a\_aft}_1$ holds on the
trace since the value of parameter \synt{t} (7) is outside the time interval $[0,6]$.
The corresponding diagnosis $\lit{d\_a\_aft}_{1} = \langle [0,6], 7
\rangle$ shows  the interval $[0,6]$ and the absolute boundary $7$.

\subsection{Property with a single atom and negation}
Let us consider property P2, checked on the trace shown in \figurename~\ref{figures/fig:sDA}:
\[
  \begin{array}{ll}
     \text{P2} \equiv &  \lit{not}\ \ \lit{globally}\  \beta_3 \ \lit{becomes} < 3 \\
  \end{array}
\]
Based on the \dslname grammar in \figurename~\ref{tab:syntax}, this
property is made of a single atom of the form \lit{not}\  \synt{sc}, 
where $\synt{sc} \equiv \lit{globally} \ \synt{p}$ consists of a
\lit{globally} scope construct constraining a pattern \synt{p} of type
\lit{becomes}.

As in the previous example, with
only one atom in the property,
algorithm~\ref{tdApp} determines whether the atom itself is violated
by the trace (line~\ref{alg:subpropviolation}). 
Since the trace violates the specification defined by the atom, algorithm~\ref{tdApp} continues by computing the associated diagnosis using function~\textsc{TD-Atom} (algorithm~\ref{tdAtom}).

During the execution of algorithm~\ref{tdAtom},  the auxiliary function
\textsc{getViolationCauses} returns the list 
$\mathit{vcs} = [\lit{c\_not}_1]$, since, in this example,
 the only possible violation cause is associated with the  \lit{not}
 construct.

Algorithm~\ref{tdAtom} will then compute the diagnosis corresponding
to the only violation cause included in list $\mathit{vcs}$, using 
the auxiliary function \textsc{getDiagnosis}.
In this example, the violation cause $\lit{c\_not}_1$ holds on the
trace since there exists a time instant (timestamp 3) in which the value of 
signal $\beta_3$ decreases from value 4.5 to 0.9 (at timestamp 4).
The corresponding diagnosis
$\lit{d\_not\_becomes} = \langle 4, 0.9\rangle$ shows the first record
(at timestamp 4) in which the predicate associated with the
\lit{becomes} pattern holds, as well as the value of the signal.

\subsection{Property with a conjunction of two atoms}
Let us consider property P3, checked on the trace shown in \figurename~\ref{figures/fig:rt}
\[
  \begin{array}{ll}
     \text{P3} \equiv & \lit{globally}\ \beta_3\  \lit{rises}\ \lit{monotonically} \ \lit{reaching}\ 3 \\
    & \lit{and}\ \lit{between} \ 2 \ \lit{and} \ 6 \ \lit{assert} \ \beta_3 <= 4 
  \end{array}
\]
Based on the \dslname grammar in \figurename~\ref{tab:syntax}, this
property is made of a single clause that consists of a conjunction of
two atoms $\delta_1$ and $\delta_2$, where 
$\delta_1 \equiv  \lit{globally}\  \beta_3 \ \lit{rises} \
\lit{monotonically} \ \lit{reaching}\ 3$
and $\delta_2 \equiv  \lit{between} \ 2 \ \lit{and} \ 6 \ \lit{assert}
\ \beta_3 <= 4$.
Atom $\delta_1$ consists of a \lit{globally} scope construct constraining a pattern \synt{p} of type \lit{rises};
atom $\delta_2$  consists of a \lit{between} scope construct (delimited by two absolute 
time instants, parameters $\synt{t}\textsubscript{1}=2$ and $\synt{t}\textsubscript{2}=6$) constraining a pattern \synt{p} of type  \lit{assert}.

Given the presence of two atoms in the property, the loop at
lines~\ref{alg:getsubpropertiesloop}--\ref{alg:diagnosticinf} of
algorithm~\ref{tdApp} is executed twice.
More in details, algorithm~\ref{tdApp} first determines whether atom
$\delta_1$ is violated by the trace. Since
the trace violates the specification defined by the atom,
algorithm~\ref{tdApp} continues by computing the associated diagnosis
using function~\textsc{TD-Atom} (algorithm~\ref{tdAtom}). During the
execution of the latter, the auxiliary function
\textsc{getViolationCauses} returns the list 
$\mathit{vcs} = [\lit{c\_rises}_1, \lit{c\_rises}_2, \lit{c\_rises}_3,
\lit{c\_rises}_4]$, since four possible violation causes are
associated with the \lit{rises} pattern construct. Function
\textsc{checkViolationCause} will then determine that  the first
violation cause (among those in $\mathit{vcs}$) that
holds on the trace is  $\lit{c\_{rises}}_3$, since signal $\beta_3$
  violates the monotonicity constraint in two time instants (at
  timestamp 2 with value 2 and at timestamp 3 with value 0.5).
The corresponding diagnosis, computed by function
\textsc{getDiagnosis}, is $\lit{d\_{rises}}_{3} = \langle \langle 2, 2
\rangle, \langle   3,0.5 \rangle  \rangle$, consisting of the tuples
(each with a timestamp and the corresponding signal value)  that violate the monotonicity constraint.

A similar process is followed for atom $\delta_2$, which is also
violated by the trace. In this case, function
\textsc{getViolationCauses} returns the list
$\mathit{vcs} = [ \lit{c\_{a}\_{bet}}_1, \lit{c\_{assert}}_1 ]$, since
the possible violation causes are associated with the \lit{between}
scope construct and the \lit{assert} pattern construct. The first
violation cause that holds on the trace is $\lit{c\_{assert}}_1$,
since signal $\beta_3$ violates the predicate associated with the
assertion at timestamp 5, when its value reaches 4.9.
  The corresponding diagnosis
$\lit{d\_assert}_1 = \langle 5,4.9 \rangle$ shows the timestamp and
the signal value.
  Algorithm~\ref{tdApp} then ends by returning the set of the diagnoses
instances, containing $\lit{d\_{rises}}_{3}$ and $\lit{d\_{assert}}_1$.

 \section{Evaluation}
\label{sec:evaluation}

Recall that, in CPSs, temporal properties are often complex, since they are typically expressed as constraints on different signal behaviors.
Although we support characterizations of individual signal behaviors, these can be and are often considered together, to report violations within a single property.
As a result, we are interested in assessing the \emph{applicability} of \NAME, that is to which extent and how efficiently \NAME is able to report diagnoses
of industrial properties violated by industrial traces.

\subsection{Datasets}
To the best of our knowledge, there is no public dataset containing traces and properties suitable for investigating the diagnosis of signal-based temporal
properties expressed in \dslname. For example, existing works on the topic of trace diagnostics, that use
different specification languages (e.g.,
STL~\cite{10.1007/978-3-319-89963-3_18,ferrere2015trace}), have not released their traces and properties.
Moreover, 
the lack of standardized benchmarks in the field of runtime
verification is a well-known issue~\cite{DBLP:journals/sttt/BartocciFBCDHJK19},
hindered by the
diversity of the tools' specification languages~\cite{DBLP:conf/rv/Reger17}.
Existing specification-based generators for synthesized traces target
a particular specification language, such as
MFODL~\cite{krstic2020benchmark}, MLTL~\cite{li2018mltl}, and MTL~\cite{ulus2019timescales}.
No trace generator exists for SB-TemPsy-DSL or for other languages for
signal-based properties (like STL and HLS~\cite{9402030}).

In light of this, to investigate the applicability of \NAME, we considered two different sources for obtaining traces and getting access to properties of interest:
\begin{itemize}
    \item an industrial system from the satellite domain (hereafter referred to as PROP-SAT), provided by our industrial partner;
    \item the  \emph{Fuel Control of an Automotive Powertrain} (referred to as AFC) benchmark model~\cite{10.1145/2562059.2562140} and its requirements used in the ARCH competition~\cite{ernst2021arch}, a competition for the falsification of temporal logic specifications written in STL.
\end{itemize}

\subsubsection*{PROP-SAT dataset} 
This dataset was defined as follows. 
We considered
\numsimulationtraces traces provided by our industrial partner;
each of these traces logs the in‐orbit operations of a satellite. 
The number of records in the traces ranges from \num{\minnumberofentries} to \num{\maxnumberofentries} ($\mathit{avg}=\num{\averageumberofentries}$, $\mathit{StdDev}\approx \num{\stdevnumberofentries}$),
and the recording interval ranges from
\SI{\minsimulationminutes}{\minute} to
\SI{\maxsimulationhours}{\hour}.\SI{\maxsimulationminutes}{\minute} ($\mathit{avg}=\SI{\avgsimulationhours}{\hour}.\SI{\avgsimulationminutes}{\minute}$, $\mathit{StdDev} = \SI{\stdDevsimulationhours}{\hour}.\SI{\stdDevsimulationminutes}{\minute}$). 

We considered \numproperties properties defined with our industrial partner and expressed in \dslname.
These properties were first elicited (and defined in English) through a series of meetings with a group of system and software engineers of our industrial partner. 
The corresponding \dslname properties were then written by the first author and validated by the engineers. This task  cumulatively lasted about $80$ hours.

The number of occurrences of each scope and pattern construct of \dslname in the properties is the following. 
For scopes: \lit{globally}   73, \lit{before}  1, \lit{after}  8, \lit{at} 3, \lit{between} 15;
for patterns: \lit{assert}  111, \lit{becomes}  13, \lit{spike} 5,
\lit{oscillation}    23, \lit{rises}  3, \lit{falls}  7,
\lit{overshoots}  4, \lit{undershoots}  4, \lit{if-then} 23.
All the listed scopes and pattern constructs are used in the definition of at least one property; 
we remark that none of the properties used the  $\lit{not}\ \synt{sc}$ construct for defining atoms.

We considered $\numsimulationtraces\times\numproperties=\num{\totalNumberOfCombinations}$ trace-property combinations, each obtained from one of the $\numsimulationtraces$ traces and one of the $\numproperties$ properties.

We removed \num{\numtracesremoved} trace-property combinations for which the trace did not log (in any of its records) any  variables used in the property and therefore did not enable the verification of its satisfaction.
Such a situation occurred because some properties are supposed 
to be checked \emph{only} at a certain operational stage  (e.g., only during the launch phase and not during the operational phase). As a result, system engineers chose not to instrument the system, at certain stages of operation, when properties were not meant to be verified.

Then, we iteratively considered each of the remaining \num{\numpreprocessedtraces} trace-property combinations. 
Due to the sampling strategy used by our industrial partner, 
two records of the same trace may log different variables, i.e., a value may not be present in every record for some of the variables.
Therefore, to ensure that the traces have the format described in section~\ref{sec:motivating}, we proceeded as follows. 
For each trace-property combination, we 
(a)~removed entirely from the trace all the  records that only contain
variable values that do not refer to any of the variables used in the
considered property, since these records do not affect its
satisfaction;
(b)~removed, from each of the remaining records, the values of the variables that were not used in the property, while preserving the rest of the record;
(c)~generated missing values for the remaining variables by using various interpolation functions~\cite{szabados1990interpolation}.
We considered different interpolation functions depending on the type of the signal, as commonly done in the literature (e.g.,~\cite{icse2021}).

Then, we analyzed each of the resulting \num{\numpreprocessedtraces} trace-property combinations.
Since \NAME aims to support engineers in detecting the source of property violations, we are interested in selecting the trace-property combinations that lead to such violations.
We executed \checkname by setting a timeout of \SI{\timeoutpreprocessing}{\minute}, thus enabling us to consider all the trace-property combinations in approximately $\SI{\timeoutpreprocessing}{\minute}\times \numpreprocessedtraces$=15 days of computation.

Out of the \num{\numpreprocessedtraces} trace-property combinations, $\SBTemPsyCheckTimeout$ of them timed out ($\approx \percentagetimeout$). 
The main reason behind such a timeout is the known scalability issue of the trace checking tool~\cite{boufaied2020trace}, especially when the property to check is defined with an order relationship pattern or an event scope.

Among the remaining \num{\SBTemPsyCheckNoTimeout} ($\approx \percentageNOtimeout$) trace-property combinations that did not timeout,  \num{\numfinaldataset} combinations represent traces that violate a property ($\percentageViolationsAmongNonTimeouts\%$). 
Though this may appear surprising at first, some of the properties only refer to specific phases of the satellite life cycle (e.g., satellite launch, deployment). We nevertheless checked these properties by considering all the traces provided by our industrial partner, including the ones that refer to the actual regular operations of the satellite. These trace-property combinations naturally led to a property violation\footnote{A property that does not refer to the regular operations of the satellite is expected to be violated if checked on a trace recording such regular operations.}.

Our final dataset contains \num{\numfinaldataset} trace-property combinations leading to a property violation. In this dataset,
the number of records in the traces ranges from \preprocminnumberofentries to \num{\preprocmaxnumberofentries} ($\mathit{avg} \approx \preprocaverageumberofentries$, $\mathit{StdDev} \approx \preprocstdevnumberofentries$), the recording interval ranges  from
\SI{\minsimulationminutesprec}{\second}, for traces with a single record, to
\SI{\maxsimulationhoursprec}{\hour}.\SI{\maxsimulationminutesprec}{\minute} 
($\mathit{avg}=\SI{\avgsimulationhoursprec}{\hour}.\SI{\avgsimulationminutesprec}{\minute}$, $\mathit{StdDev}= \SI{\stdDevsimulationhoursprec}{\hour}.\SI{\stdDevsimulationminutesprec}{\minute}$). 
Notice that the number of records and the recording intervals  of the traces of the final dataset are significantly smaller than those observed for the original traces. 
This is due to the fact that, for each trace-property combination, we
removed from the trace all the records that only contained variable
values that did not refer to any of the variables of the considered
property (see step~(a) above).

\subsubsection*{AFC dataset} 
The AFC benchmark model~\cite{10.1145/2562059.2562140} used in the ARCH competition~\cite{ernst2021arch} comes with three properties (namely, AFC27, AFC29, and AFC33). We excluded property AFC27 because it could not be expressed 
with \dslname, since the language does not support nested operators\footnote{We refer the reader to our previous work~\cite{boufaied2020trace} in which we discuss the expressiveness of \dslname}.
Properties AFC29 and AFC33 have the same formula structure and differ only in terms of their parameters. We considered only one of them (property AFC29) and expressed it in \dslname. The property states that \emph{``between 11 and 50 seconds, signal $\mu$ shall be lower than 0.007 ''}, corresponding to the \dslname specification:
\[
 \begin{array}{ll}
     \phi_{29} \equiv & \lit{between}\ 11 \  \lit{and}\ 50 \ \lit{assert} \ \mu \ < 0.007
  \end{array}
\]

As we are interested in reporting a diagnosis corresponding to the violation of AFC29, we used the ARIsTEO tool~\cite{aristeo}, a plugin for S-Taliro~\cite{annpureddy2011s}, to generate \afclogs\ traces that falsify the property, sampled over \SI{50}{\second} (i.e., a time horizon of $[0,50]$), with simulations configured to use a variable sample step.

We therefore obtained \afclogs\ trace-property combinations; the number of records in the generated traces ranges from \num{\afcMinRec} to \num{\afcMaxRec} 
($\mathit{avg} = \num{\afcAverage}$, $\mathit{StdDev} =  \num{\afcStd}$).

This is admittedly a much smaller dataset than the PROP-SAT one, if we consider the number of trace-property combinations it contains and the fact that we only considered one property.
Moreover, given the simplicity of property AFC29, considering more traces would not lead to different diagnoses and conclusions (i.e., the catalogue of violation causes and corresponding diagnoses in \NAME includes a single violation cause $\lit{c\_assert}_1$  and one diagnosis $\lit{d\_assert}_1$ for the \lit{assert} construct supported by \dslname). 
We remark that, out of the eight STL properties included in the benchmark description~\cite{10.1145/2562059.2562140}, only three of them were included in the ARCH competition benchmark~\cite{ernst2021arch}, and thus were known to be falsifiable. 
Nevertheless, this dataset is adequate for our goals, which are (a) to demonstrate that we can obtain similar results on a different dataset obtained from a publicly available benchmark model; (b) to support open science, using a non-proprietary dataset that can be made publicly available.

\subsection{Applicability}
\label{sec:rq1}
We assessed the applicability of \NAME by considering the
\numfinaldataset\ trace-property combinations in the PROP-SAT dataset 
as well as the \afclogs\ trace-property combinations in the AFC one. Applicability entails the capacity to report diagnoses within reasonable time.

We remark that we could not perform a comparison between \NAME and
state-of-the-art tools, for a number of reasons. First, some
alternative approaches~\cite{dawes2019explaining,dou2018model,10.1007/978-3-319-11164-3_24}
do not support signal-based temporal properties, thus making any comparison impossible.
The only alternative that supports signal-based properties is AMT2.0~~\cite{10.1007/978-3-319-89963-3_18, ferrere2015trace}. However, the tool is no longer publicly available\footnote{\url{https://www-verimag.imag.fr/AMT-2-0.html}} and its successor \emph{rtamt}~\cite{nivckovic2020rtamt} has dropped support for diagnostics capabilities, rendering impossible any experimental comparison.

\subsubsection{PROP-SAT dataset}

\subsubsection*{Methodology} 
We executed \NAME on each trace-property combination in the PROP-SAT dataset,  with 
a timeout of \SI{\diagnosticsTimeoutInMinutes}{\minute}, leading to approximately $\SI{\timeoutpreprocessing}{\minute}\times \numfinaldataset$=10 days of computation.
For each execution, we recorded whether \NAME finished within the timeout and the diagnoses (if any) it yielded.
To assess the applicability of \NAME, we analyzed the number of combinations in which \NAME finished within the timeout and whether it yielded a diagnosis, i.e., whether at least one violation cause was applicable. 
We conducted our evaluation on a high-performance computing platform, using nodes equipped with
Dell C6320 units (2 Xeon E5-2680v4@\SI{2.4}{\giga\hertz}, \SI{128}{\giga\byte}).

\subsubsection*{Results} 
\NAME finished within the timeout for $\approx\diagnosticsNoTimeoutsPercentage\%$ of the combinations (\num{\diagnosticsNoTimeouts} out of \num{\numfinaldataset}).

For the remaining \num{\diagnosticsTimeouts} combinations that \emph{timed out}, $\approx\ordertimeoutPercentage\%$ of these combinations ($\ordertimeout$ out of $\diagnosticsTimeouts$) come from properties using the \lit{if-then} construct, 
$\approx \eventBoundariestimeoutPercentage\%$ of the combinations ($\eventBoundariestimeout$ out of $\diagnosticsTimeouts$) come from properties using the event scope constructs,  and $\approx \complexPropTimeoutPercentage\%$ of the combinations ($\complexPropTimeout$ out of $\diagnosticsTimeouts$) come from properties that used the \lit{or} and \lit{and}  operators to combine properties  and clauses defined using the aforementioned constructs.
For these combinations, the computational overhead to compute the diagnosis led to timeouts. Note that, in practice, engineers are likely to use larger timeouts  than the one selected here, which is due to experimental constraints, and, therefore, we expect the percentage of combinations that time out to decrease.

For the \num{\diagnosticsNoTimeouts} trace-property combinations that \emph{finished within the timeout},
\NAME always returned a diagnosis.
We recall that a diagnosis is made by one or more diagnosis instances that are generated by \NAME for the different atoms of the formula (see Section~\ref{sec:approach}). These instances describe why the scope and pattern constructs of \dslname used for the definitions of the atom are violated by a trace.
\NAME produced one diagnosis instance for each atom of the formula (corresponding to the input property) for $\supportedViolationsOnlyPercentage\%$  of the combinations (\num{\supportedViolationsOnly} out of \num{\diagnosticsNoTimeouts}).
For the remaining $\mixedViolationsPercentage\%$ of the combinations (\num{\mixedViolations} out of \num{\diagnosticsNoTimeouts}), some atoms of the formula did not lead to any diagnosis instance.
In total, the \num{\diagnosticsNoTimeouts} combinations returned \num{\occurrencesOfsupportedViolationsOnly} diagnosis instances. 

The top part of Table~\ref{tab:diagnosticprevalence} shows the number of diagnosis instances (column \#\textbf{N}) computed by \NAME for each scope and pattern construct of \dslname (as used in the properties of the PROP-SAT dataset).
These results suggest that a relatively high percentage of the diagnosis instances ($1002+294+660+2640=\scopesOccurrences$ out of \num{\occurrencesOfsupportedViolationsOnly}, $\approx \scopesOccurrencesPercentage\%$) is related to scope constructs.
Indeed, since we considered all the possible trace-property combinations in the dataset,  there are many combinations for which the time instant values used to define the scope operators exceeded the maximum timestamp recorded in the trace. 
The remaining $\approx \patternsOccurrencesPercentage\%$ of the diagnosis instances ($7098+460+321+11=\patternsOccurrences$ out of \num{\occurrencesOfsupportedViolationsOnly}) is related to patterns constructs. 
\NAME returned diagnosis instances for the \lit{assert}, \lit{spike}, \lit{oscillation}, and \lit{if-then} constructs, though with different prevalence. 
\NAME did not report any diagnosis instances for the \lit{becomes}, \lit{rises}, \lit{falls}, \lit{overshoots}, and \lit{undershoots} constructs. 
We further analyzed the properties containing these constructs and noticed that,
in all these cases, \NAME detected a violation of the corresponding scope. 
In such cases, 
the diagnosis instances returned by \NAME are only related to the scope constructs.

\begin{table}[t]
\caption{Number (\#\textbf{N}) of diagnosis instances generated by \NAME for each scope and pattern construct  of \dslname (as used in the properties of our datasets).}
    \label{tab:diagnosticprevalence}
\scriptsize
    \centering
    \begin{tabular}{@{\hspace{5pt}}l@{\hspace{5pt}}l@{\hspace{5pt}}l@{\hspace{5pt}}l@{\hspace{5pt}}l@{\hspace{5pt}}l@{\hspace{5pt}}l@{\hspace{5pt}}}
    \toprule
      \textbf{Type} &  \textbf{Construct}  &  \#\textbf{N}  &  \textbf{Construct}  &  \#\textbf{N} &  \textbf{Construct}  &  \#\textbf{N}  \\
    \midrule
    \multicolumn{7}{c}{PROP-SAT dataset}\\ \midrule
Scope & \lit{globally}  & 0
& \lit{before} & 294 & \lit{after} & 2640  \\
&  \lit{at} & 1002 & \lit{between} & 660 & \\
    \midrule
Pattern & \lit{assert} & 7098 
& \lit{becomes} & 0 & \lit{spike} & 321 \\
& \lit{oscillation}  &  460 & \lit{rises} & 0 & \lit{falls} & 0 \\
& \lit{overshoots} & 0 & \lit{undershoots} & 0 & \lit{if-then} & 11\\ \midrule
\multicolumn{7}{c}{MD1 dataset}\\ \midrule
Pattern  & \lit{becomes} & 656 &  \lit{overshoots} & 942  & \lit{undershoots} & 940 \\ \midrule
\multicolumn{7}{c}{MD2 dataset}\\ \midrule
Pattern  &   \lit{rises} & 90 & \lit{falls} & 90 & &
\\ \midrule
\multicolumn{7}{c}{AFC dataset}\\ \midrule
Pattern  &   \lit{assert} & 10  & & & & \\
    \bottomrule
    \end{tabular}
    
\end{table}

To guarantee a complete applicability assessment of \NAME, covering all \dslname constructs, we built two additional datasets (MD1 and MD2), derived from the PROP-SAT one, using the following strategies:
\begin{asparadesc}
\item [MD1 (replacing the property scope).]
We considered all the $\combinationsNotCoveredByRealProps$ trace-property combinations where the properties are defined using only one single pattern of type \lit{becomes}, \lit{overshoots}, or \lit{undershoots} within a scope operator. 
We changed the scope of the patterns to \lit{globally} in order to avoid any scope violations, thus making the detection of violations of these property patterns possible. As a result, we obtained  $\combinationsNotCoveredByRealProps$ additional trace-property combinations which constitute the MD1 dataset.
\item [MD2 (changing the pattern definition).] 
The patterns \lit{rises} and \lit{falls} were not used in any property containing only one single pattern, but were always used within the \lit{if-then} construct.
Therefore,
we considered the $\risesfallsCombinations$ trace-property combinations where the properties contained the \lit{rises} and \lit{falls} patterns, and then we extracted from the \lit{if-then} construct the subproperties that were using these patterns.
This led to $\risesfallsCombinations$ additional trace-property combinations, which constitute the MD2 dataset. 
\end{asparadesc}

We executed \NAME on the MD1 and MD2 datasets with a timeout of \SI{\diagnosticsTimeoutInMinutes}{\minute}. 
In the case of MD1, \NAME yielded  diagnosis instances for $\approx \violatedCombinationsWithGloballyPercentage\%$ of the combinations ($\violatedCombinationsWithGlobally$ out of $\combinationsNotCoveredByRealProps$), with no timeout;
the distribution of these instances is 
shown in the second block (from the top) of Table~\ref{tab:diagnosticprevalence}.
For MD2, \NAME yielded  diagnosis instances for all  $\risesfallsCombinations$ combinations; the distribution of these instances is 
shown in the third block (from the top) of Table~\ref{tab:diagnosticprevalence}.

\subsubsection{AFC dataset}
\subsubsection*{Methodology}
We executed \NAME on each of the \afclogs\ trace-property combinations in the AFC dataset, with a timeout of \SI{\diagnosticsTimeoutInMinutes}{\minute}. The total time to execute \NAME on all these trace-property combinations  was  $\approx \SI{14}{\s}$; hence, no timeouts occurred. 

Also in this case, we assessed the applicability of \NAME by 
analyzing all the \afclogs\ trace-property combinations processed by \NAME, checking whether it yielded a diagnosis.

\subsubsection*{Results}
\NAME yielded diagnosis instances for all the
\afclogs\ trace-property combinations in the AFC dataset;
all these diagnoses were of type $\lit{d\_assert}_1$.
\begin{figure}[tb]
   \hspace*{-1cm}
   \centering
    \includegraphics[scale=0.6]{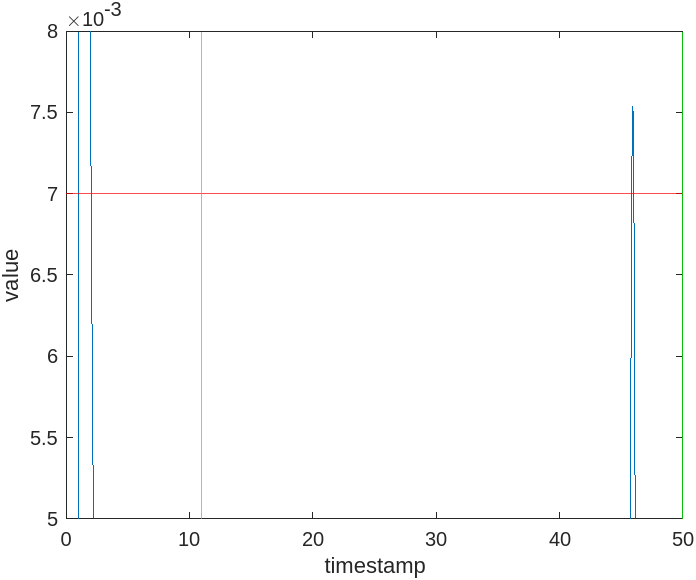}
    \caption{A trace (from the AFC dataset) with a signal violating the expression ``\lit{between}\ 11 \  \lit{and}\ 50 \ \lit{assert} \ $\mu$ \ < 0.007''.}
\label{muViolation}
\end{figure}

For instance, the signal in \figurename~\ref{muViolation} violates property $\phi_{29}$, showing at least one timestamp in which the signal violates the condition ($\mu\ < 0.007$). More specifically, the signal violates the condition for $99$ records within the time interval $[45.85,46.014]$, which lies within the scope interval $[11,50]$ in  $\phi_{29}$ (delimited by green vertical lines in the figure), 
with values ranging between $0.0070103$  and $0.0075373$. According to the definition of 
the diagnosis $\lit{d\_assert}_1$, \NAME reports the first record (i.e., timestamp and the corresponding signal value) in which the condition was violated. The diagnosis is then as follows: $\langle 45.85,0.0070278\rangle$. The choice of reporting the first record that violates the condition is motivated by the fact that we are interested in detecting and reporting the root cause(s) of the property violation (See section~\ref{patternAssertion}).

\subsection{Discussion}

The results show that \NAME was widely applicable in the context of the two datasets we considered, including one based on an industrial case study.  

Indeed, when considering the PROP-SAT dataset as well as MD1 and MD2, \NAME was able to finish within the small timeout of \SI{1}{\minute} for 
$\num{\diagnosticsNoTimeouts}+\combinationsNotCoveredByRealProps+\risesfallsCombinations=\num{\totalNonTimeouts}$
out of 
$\num{\numfinaldataset}+\combinationsNotCoveredByRealProps+\risesfallsCombinations= \num{\totalCases}$
($\approx~\totalPercentageA\%$) of the trace-property combinations in this group of datasets,
returning a diagnosis for $\num{\diagnosticsNoTimeouts}+\violatedCombinationsWithGlobally+\risesfallsCombinations=\num{\totalSupported}$ combinations ($\approx~\totalSupportedPercentageA\%$ of the cases).
Moreover, in the case of the AFC dataset, \NAME processed all the \afclogs\ trace-property combinations well below the timeout, yielding a diagnosis for all of them.

These results suggest that, in practice, our set of violation causes provide sufficient coverage of observed violations.

\subsubsection{Threats to Validity}
In terms of \emph{internal validity}, the choice of a timeout of \SI{\diagnosticsTimeoutInMinutes}{\minute}, justified by the high computational time (15 days) for the experiments on the PROP-SAT dataset, led to a number of trace-property combinations for which \NAME did not finish its execution.
Considering a larger timeout would further increase the applicability of \NAME. 
Moreover, we assumed that the traces in the PROP-SAT dataset (provided by our industrial partner) were correctly collected after the satellite deployment. 
The possible presence of erroneous records might lead to a different number of trace-property combinations leading to a violation, and to different diagnosis instances. 
Furthermore, we have assumed that the verdicts reported by \checkname were correct.

In terms of \emph{external validity}, the trace-property combinations in the PROP-SAT dataset may be a threat for the generalization of our results, as other datasets may differ in terms of (a)~the constructs used for expressing the properties, (b)~the type of property violations. 
We mitigated this threat by selecting an industrial case study in the satellite domain that is representative of complex CPS, with large traces and many complex properties elicited with experts. Further, we modified scopes and patterns in the  properties of the PROP-SAT dataset to expand our analysis such as to consider more trace-property combinations.  
Moreover, we also considered an additional dataset (AFC) from the benchmark used for a popular competition for the falsification of temporal logic specifications
over CPS.

Regarding \emph{conclusion validity}, the trace-property combinations in our datasets did not trigger all the 34 violation causes in our catalogue (see \figurename~\ref{tab:diagnosticPatterns}).
More in details, 
the trace-property combinations in the PROP-SAT and AFC datasets cover $9$ out of the $34$ violation causes. If we include the two additional datasets MD1 and MD2, the total number of covered violation causes reach $17$ out of $34$ ($8$ new ones). 
These $8$ new covered violations causes are distributed as follows: $2$ \lit{becomes}, $2$ \lit{rises} (and its dual \lit{falls}) and $4$ \lit{overshoots} (and its dual \lit{undershoots}).

\subsection{Data Availability}
We cannot publicly release the traces and properties used in the experiments for the PROP-SAT because they are subject to a non-disclosure agreement. 
We make the raw output of \NAME, the traces generated as part of the AFC dataset,
and the script used for the analysis of the evaluation data available as supplementary material in a permanent repository~\cite{Boufaied2023-supp}.
\NAME is available under the Apache 2.0 license at \url{https://github.com/SNTSVV/TD-SB-TemPsy}; a permanent record is also available on Figshare~\cite{Boufaied2023-TDSBTemPsy}.

 \section{Practical Implications}
\label{sec:practical}

\subsection{Usefulness of the diagnoses}

When engineers use a run-time verification tool that only yields Boolean verdicts, if a property is violated on a trace, engineers have to inspect the trace to understand the cause of the violation.

Such an inspection is not necessarily trivial, especially for the more complex types of properties (e.g., those involving spike or oscillatory behaviors), and cannot rely on a simple visualization of the signals. This problem is even more noticeable when dealing with huge execution traces, containing thousands of records.

For example, let us consider property $\phi_{29}$ from the AFC dataset. 
\begin{figure}[tb]
   \hspace*{-1cm}
   \centering
    \includegraphics[scale=0.5]{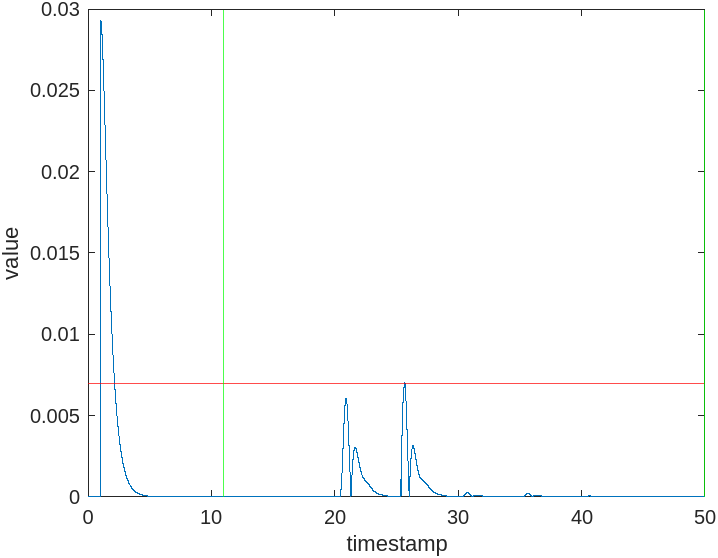}
    \caption{A trace (from the AFC dataset) with a signal violating the expression ``\lit{between}\ 11 \  \lit{and}\ 50 \ \lit{assert} \ mu \ < 0.007'' (with no zoom factor in the plot).}
\label{muViolationNoZoom}
\end{figure}
Figure~\ref{muViolationNoZoom} shows one of the execution traces considered  in our evaluation. 
Given the small order of magnitude used in the numeric parameters of the property (e.g., the threshold $0.007$) and the shape of the signal, detecting the violation through a visual inspection is not immediate, even if a large zoom factor is used in the visualization.

On the other hand, if an engineer uses our approach, she can immediately look up the important record in the trace, since it is indicated in the diagnosis  $\langle 45.85,0.0070278\rangle$. 

Note that the example above is based on a simple assertion property. Automating the diagnostics of violation is even more  important for more complex types of properties (e.g., those involving spike or oscillatory behaviors).

The actual effort savings are specific to individual case studies and can only be estimated through a user study. We plan to conduct one as part of future work. 

\subsection{Extending the catalogue of violation patterns and diagnoses}

As discussed in section~\ref{sec:introduction}, our catalogue of 34 violation causes, each associated with one diagnosis, \emph{is not complete}. 
Nevertheless, following the methodology illustrated in section~\ref{sec:methodology}, users can add new violation causes depending on their specific needs or on the requirements of particular domains.

We remark that this extension of the catalogue is a one-time effort. It can be performed by  engineers following the three steps: behavior analysis, definition of violation causes, and definition of diagnoses.
In particular, step ``Definition of violation causes'' requires new violation causes to be checked (for correctness) by verifying whether a formula (obtained from the formal specification of the violation cause semantics) is unsatisfiable (see section~\ref{sec:definition-violation-causes}). This check can be performed with state-of-the-art constraint solvers like Z3.
Overall, fulfilling this requirement prevents the users from introducing errors in the definition of violation causes.

Finally, even if the original catalogue of  34 violation causes is not complete, we remark that it results from an extended industrial case study and relies on a taxonomy~\cite{DBLP:journals/corr/abs-1910-08330} of pattern-based constructs that have been identified through a thorough review of the literature, whose completeness has been validated in an industrial context. 
Based on this, we expect our catalogue to be widely reusable in different CPS domains,
provided that the requirements to be checked using a run-time verification tool can be expressed
using \dslname.

 \section{Related Work}
\label{sec:related}
The problem of enriching Boolean verification verdicts with additional information that supports reasoning on the causes of such verdicts has been widely studied in the literature. 
This section discusses related work in the trace-checking and model-checking areas. We included the latter since a trace can be seen as a model made by a sequence of consecutive states, each representing one trace record, with  transitions connecting the consecutive records.

In the \emph{trace-checking} area, there are two main strategies (see Section~\ref{sec:introduction}) that aim to provide additional information on the causes of a property violation: (i)~isolating slices of the traces that explain the property violation (e.g.,~\cite{ferrere2015trace,mukherjee2012computing,beer2009explaining,10.1007/978-3-319-89963-3_18}); and (ii)~checking whether the traces show common behaviors that lead to the property violation (e.g.,~\cite{dou2018model,10.1007/978-3-319-11164-3_24,dawes2019explaining}). 
The first strategy produces large explanations for complex properties since the size of the explanation increases with the number of operators of the formula expressing the property of interest. 
Existing approaches based on the second strategy do not support complex signal-based temporal properties (as the ones considered in this work) and are not complemented by a precise methodology that describes how to add new causes that support more complex properties.
Therefore, in this work we have proposed a novel, language-agnostic methodology for defining \emph{violation causes} and \emph{diagnoses}, and applied it in the context of signal-based temporal properties expressed in \dslname.

\begin{table}[tb]
\caption{Comparison of trace diagnostic approaches}
\label{relWTable}
\centering
\begin{tabular}{l@{\hskip 5pt}c@{\hskip 5pt}c@{\hskip 5pt}c@{\hskip 5pt}c}
\toprule
\textbf{Approach} & \textbf{SBTP} & \textbf{Lang.} & \textbf{Method.} & \textbf{Eval.} \\ \midrule
\citet{ferrere2015trace} & + & TL & - & T \\
\citet{mukherjee2012computing} & - & TL & - & P \\
\citet{beer2009explaining} & - & TL & - & P \\
\citet{10.1007/978-3-319-89963-3_18} & + & TL & - & P \\
\citet{dawes2019explaining} & - & TL & - & P \\ 
\citet{dou2018model} & - & DSL & - & S \\ 
\citet{10.1007/978-3-319-11164-3_24} & - & DSL & - & P
\\
\NAME & + & DSL & + & I,P \\ 
\bottomrule
\end{tabular}
\end{table}

Table~\ref{relWTable} provides a comparison, in terms of 
trace diagnostics support, of the aforementioned trace checking approaches. 
Column \emph{SBTP} indicates whether the approach supports signal-based temporal properties. Column \emph{Lang.} indicates --- using the symbols DSL and TL --- whether the approach supports, respectively, a high-level, DSL-like specification language or a low-level, temporal-logic language. Column \emph{Method.} indicates whether the diagnostic approach supports a methodology to add new violation causes.
Column \emph{Eval.} indicates the type(s) of benchmarks used in the evaluation of the approach (\textit{I}: industrial case study, \textit{P} public benchmark, \textit{S}: synthetic benchmark, \textit{T}: toy example).  

As shown in the table, \NAME is the only approach that supports 
signal-based temporal properties expressed in a DSL-like specification language, that is complemented by a methodology allowing users to define new violation causes, and that has been evaluated using datasets derived both from a complex industrial case study and from a public benchmark.

In the \emph{model-checking} area, some approaches (e.g.,~\cite{fase2020,schuppan2012towards,DBLP:journals/corr/abs-1109-2656,zheng2021flack,10.1007/978-3-540-31984-9_17,5634313,griggio2018certifying,10.1007/978-3-030-45190-5_18,timm2020model,10.1007/3-540-36577-X_12})   extract information from the model (e.g., model slices) to explain the model checking verdict. 
Typically, these approaches have limited scalability and therefore are not easily applicable to the trace-checking scenario. 
\NAME relies on a conceptually different technique, which leverages violation causes and diagnoses to explain trace checking verdicts. 
Moreover, its implementation uses existing technologies that showed encouraging scalability results in previous works; our evaluation confirms the applicability of our solution.
Other approaches (e.g.,~\cite{peled2001model,10.1007/978-3-319-66197-1_4,peled2001falsification,mebsout2016proof,basin2018optimal,pnueli2002deductive,balaban2010proving}) rely on deductive reasoning techniques to explain model checking verdicts.
Different from the approach proposed in this work, they usually provide an exhaustive explanation for a verdict by considering some initial assertions
(e.g., simple conditions on the values assumed by the variables in the states of the model) and examining how logical operators can be applied to reach a specific logical conclusion. However, the proofs produced by deductive reasoning approaches are usually difficult to understand for non-experts.
Besides, their size significantly grows with the size of the model to analyze~\cite{Grebing2020}. Therefore, when the model represents a trace, which is typically large in practice (i.e., because of a large number of records), the generated proofs are likely to be extremely large and difficult to understand by engineers.

 \section{Conclusion}
\label{sec:conclusion}
In this paper, we proposed \NAME, a trace-diagnostic approach for signal-based temporal properties, based on violation causes and diagnoses.
We defined a methodology for defining  violation causes and diagnoses that provides formal soundness guarantees.
We proposed a catalog of 34 violation causes, each associated with one diagnosis, for properties expressed in \dslname.
We evaluated \NAME by assessing its applicability on 
on two datasets, including one based on a complex industrial case study.

For the latter, \NAME finished within the stringent timeout  of \SI{1}{\minute}
for $\approx\totalPercentageA\%$ of the trace-property combinations,
yielding a diagnosis in $\approx\totalSupportedPercentageA\%$ of these cases; 
moreover, it also yielded a diagnosis, within the same timeout, for all the trace-property combinations in the other dataset.
In practice, outside of experimental settings, longer timeouts can be considered.

In the future, we plan to perform a large-scale, systematic evaluation to assess (a) the scalability of \NAME with respect to the trace size
and (b) its applicability when dealing with different violation causes. 
This evaluation requires  the use of synthesized traces, which enable
varying the trace size and controlling the causes of property
violations.
Furthermore, we are going to assess the applicability of \NAME on 
a diverse set of CPS case studies (e.g.,  unmanned aerial vehicles).
Moreover, we expect to revisit the implementation of \NAME and \checkname, to support a tighter integration between the two tools, so that some intermediate outputs for the trace diagnostics procedure could be computed during the execution of the trace checking one.
In addition, we intend to conduct a user study to assess the usefulness
of the diagnoses provided by \NAME, for example in the context of
fault localization.

\section*{Acknowledgments}
Part of this work was supported by the Natural Sciences and Engineering Research Council of Canada (NSERC), through its DG and CRC programs;  
by European Union's Horizon
2020 Research and Innovation Programme under grant agreement
No. 957254 (COSMOS).

\bibliographystyle{IEEEtranN}

\end{document}